\documentclass[sigconf]{acmart}

\setcopyright{none}
\renewcommand\footnotetextcopyrightpermission[1]{} 
\usepackage{etoolbox} 
\newbool{SUBMISSION}
\booltrue{SUBMISSION} 

\DeclareGraphicsExtensions{.pdf,.png,.jpg}

\usepackage{comment}

\usepackage[most]{tcolorbox}
\usepackage{url}

\usepackage[color=pink]{todonotes}
\usepackage{fancyvrb}
\usepackage{xcolor}    
\usepackage{framed}    

\colorlet{shadecolor}{yellow!20}
\usepackage[T1]{fontenc}
\usepackage{listings}
\usepackage{makecell}
\usepackage{pifont}
\usepackage{subcaption} 
\usepackage[framemethod=tikz]{mdframed}

\usepackage{tikz}
\usetikzlibrary{mindmap,trees,arrows.meta,shadows}

\usepackage{forest}
\usepackage{hyperref}
\usepackage{adjustbox}

\usetikzlibrary{arrows.meta,shadows}

\newcommand{\TODO}[1]{\todo[inline]{#1}}
\newcommand{\GVL}[1]{{\begin{blue} #1}}

\usepackage{smartdiagram}

\forestset{
  skan tree/.style={
    for tree={
      drop shadow,
      text width=3cm,
      grow'=0,
      rounded corners,
      draw,
      top color=white,
      bottom color=blue!20,
      edge={Latex-},
      child anchor=parent,
      anchor=parent,
      tier/.wrap pgfmath arg={tier ##1}{level()},
      s sep+=2.5pt,
      l sep+=2.5pt,
      edge path'={
        (.child anchor) -- ++(-10pt,0) -- (!u.parent anchor)
      },
      node options={ align=center },
    },
    before typesetting nodes={
      for tree={
        content/.wrap value={\strut ##1},
      },
    },
  },
}
\usepackage{smartdiagram}

\usepackage{forest}

\definecolor{BrickRed}{rgb}{0.8, 0.25, 0.33}

\lstdefinestyle{python}{
  language=Python,
  basicstyle=\scriptsize\ttfamily,
  keywordstyle=\color{blue},
  commentstyle=\color{green!50!black},
  stringstyle=\color{Bittersweet},
  showstringspaces=false,
  breaklines=true
}

\lstdefinestyle{sh}{
  language=sh,
  basicstyle=\scriptsize\ttfamily,
  keywordstyle=\color{blue},
  commentstyle=\color{green!50!black},
  stringstyle=\color{Bittersweet},
  showstringspaces=false,
  breaklines=true,
  keywords={singularity,echo,cms,export,cd,mkdir,nvidia-smi,python,seff}
}

\newcommand{\YES}{\ding{51}}
\newcommand{\NO}{$\times$}

\author{Gregor von Laszewski}
\affiliation{%
  \institution{Biocomplexity Institute, University of Virginia,}
  \city{Charlottesville, VA 22911}
  \country{USA}}
\email{laszewski@gmail.com}

\author{Wesley Brewer}
\affiliation{%
  \institution{Oak Ridge National Laboratory}
  \city{Oak Ridge, TN 37831,}
  \country{USA}}

\author{Sean R. Wilkinson}
\affiliation{%
  \institution{Oak Ridge National Laboratory}
  \city{Oak Ridge, TN 37831,}
  \country{USA}}

\author{Andrew Shao}
\affiliation{%
  \institution{Hewlett Packard Enterprise Canada}
  \city{Victoria, British Columbia}
  \country{Canada}}

\author{J.P. Fleischer}
\affiliation{%
  \institution{$^{5}$ University of Florida}  
  \city{Gainesville, FL 32611}
  \country{USA}}

\author{Harshad Pitkar}
\affiliation{%
  \institution{Cummins}
  \city{Columbus, IN 47201}
  \country{USA}}

\author{Christine R. Kirkpatrick}
\affiliation{%
  \institution{San Diego Supercomputer Center, University of California}
  \city{San Diego, La Jolla, CA 92093}
  \country{USA}}

\author{Geoffrey C. Fox}
\affiliation{%
  \institution{Biocomplexity Institute, University of Virginia,}
  \city{Charlottesville, VA 22911}
  \country{USA}}

\newcommand{\TITLE}{Towards Experiment Execution in Support of Community Benchmark Workflows for HPC
}

\makeatletter
\begingroup\endlinechar=-1\relax
       \everyeof{\noexpand}%
       \edef\x{\endgroup\def\noexpand\homepath{%
         \@@input|"kpsewhich --var-value=HOME" }}\x
\makeatother

\def\overleafhome{/tmp}
\ccsdesc[cssClassifiers^500]{}

\begin{document}


\onecolumn

%
%

\setcounter{tocdepth}{4}






%
%

\title{\TITLE}






\newcommand{\PARGRAPH}[1]{\parindent 0 {\bf #1}}

\begin{abstract}


    Over many decades, High Performance Computing systems have been made available to the research community through research organizations and also recently made available through commercially available cloud solutions. The use of such systems has traditionally been restrictive due to the high costs of computing resources as well as the complex software to offer them efficiently to the community. Over time, we have also seen the utilization of federated resources such as Grids, Clouds, and today's hyperscale data centers. Still, despite the many software systems and frameworks in support of these resource infrastructures, their utilization has been a challenge, especially in the academic community educating the next generation of scientists. We also found that using limited benchmarks on various machines, even if they are not federated, pose a significant hurdle in demonstrating compute resource capability to the many communities with their highly varied computational needs. While popular frameworks such as Gateways and, on the other spectrum, Jupyter notebooks promise usage simplification, it is even more important that a wide variety of scientific benchmarks are available that outline how such machines can be best utilized and align with the scientists application-specific computational challenge. We found that this is best done in the form of workflow templates that are designed for a specific problem and can be adapted to a specific scientific application.

  Within this paper, we focus first on identifying common usage patterns that outline which workflow templates we have found most useful based on our experiences over many decades. Naturally, there are many other patterns available that may be addressed by other frameworks. However, we focus here on templates that have been used by us, based on decades of use of HPC systems dating back to the early parallel computers. Recently, we have enhanced and expanded our experience by participating in  the MLCommons Science working group. We found that focusing on simple tools addressing what we call {\em experiment management} as part of the the more general computational workflow improves adaptability in the  educational community. Hence, they can become valuable aspects into what we term \textit{benchmark carpentry}.

We have verified this approach based on the experiences of two independently developed software tools and frameworks that upon closer inspection provide a large amount of overlap in functionality. These systems are the Experiment Executor which is part of a larger bag of services distributed as part of Cloudmesh that has been used for more than a decade, as well as SmartSim developed by Hewlett Packard Enterprise which similarly addresses experiment management. These frameworks have been tested on various scientific applications. In our earlier work, this was done on two scientific applications: conduction cloudmask and earthquake prediction. More recently this work was extended to include experiment executions that involve the interaction of simulation and AI/ML. Lastly, we focus on how these frameworks are also simplifying the development of a surrogate for computational fluid dynamics.



\end{abstract}

\maketitle
\pagestyle{plain}
\thispagestyle{empty}

\section{Introduction}
\label{sec:introduction}

Benchmarks are useful for comparing the performance of computing resources and their suitability for applications that may be executed with them. In particular, application users can benefit when benchmarks have analogues with their applications allowing them to assess and potentially predict feasibility of their expected scientific workloads. Typical benchmarks include measuring the performance of CPUs, GPUs, data storage, data transfer, and energy consumption.

The most challenging application problems require HPC-scale computing resources. These applications are influenced by machine-specific performance, but it is often not clear how a specific application performs when benchmarks may not relate closely enough to the applications. Furthermore, the shared nature of systems with hundreds of users  adds another dimension of complexity: the execution of the benchmark may differ if deployed on hardware reserved for a single user versus shared hardware and in cases with multi-step jobs, scheduling on crowded machines introduces additional delay. These issues make predicting real-time end-to-end performance benchmarking very challenging. Hence, in many systems a benchmark is run as a single user and the queue wait time is often ignored.

Arguably, Linpack performance is the most well-known HPC benchmarks forming the basis for the published list of the top 500 HPC machines~\citep{www-top500}. Recently, a benchmark using High-Performance Conjugate Gradient (HPCG) has been added to complement the Linpack benchmark so that an additional understanding of the machine's performance can be achieved~\citep{www-top500} through a different application. With increasing awareness of the electrical requirements for these machines, the energy consumption of watts per flop is the figure of merit for the Green500 benchmark~\citep{green500}.

From such benchmarks, one can only derive the {\em potential} of the entire HPC machine, whereas application users need to develop their own benchmark experiments representative of the application's needs. These benchmarks are often run at smaller scales and introduce scaling factors based on theoretical assumptions to then predict the performance of larger problems. In some cases, the application needs to be rewritten to fit within the constraints of available resources and compute time to access them. In other cases, it is not the hardware of the machine that leads to lower-than-expected performance, but logistical policies. For example, HPC platforms usually have scheduling policies that maximize overall machine utilization while allowing a balanced number of jobs for users. While the HPC policies can be changed for special circumstances, it is often not a permanent option because the individual benchmark user impacts negatively the larger user community. Therefore, realistic application benchmarks often need to distinguish between performance based on single-user optimal policies vs. production policies. 

The increasing using of machine learning has led to the development of benchmarks focused on training and inference from neural networks. The MLCommons group tries to make the use of machine learning easier while creating typical benchmarks for supercomputers. While raw performance is measured by most working groups in MLCommons measuring the performance of many well-known AI applications, the science working group also tries to identify the best algorithms or even data sets to obtain benchmarks based not only on performance but also on the accuracy of the problem solution.

Clearly, these multiple objectives preclude the use of a single benchmark, but rather require many different experiments with potentially different hyperparameters and datasets. Setting up a workflow that supports such experiments is often complex especially if they exceed the center's policies and need to be modified accordingly. Therefore, the capability of coordinating many experiments, their workflows, and aggregating the results is key to this type of benchmark while staying within the limits associated with the HPC user defined by the datacenter.

Throughout the paper we use the general term {\em computational workflow} or simply {\em workflow} to indicate the chain of tasks that are needed to produce from a set of inputs the outputs \citep{def-workflow} as has been used for decades. To distinguish separately executed workflows that may lead to a set of very distinctive results we use the term experiment execution. To be more specific, we use the term {\bf\em experiment execution} to indicate a specific workflow that includes the execution of a benchmark experiment while considering the provisioning of data, the execution of the algorithm on the data to achieve a result, and the variation of the hyperparameters as part of the many different single executions run on the infrastructure to obtain the result. We distinguish this very specific definition of this workflow from the rather overloaded term of {\em workflow} by many different communities. As our focus is the experiment execution as part of benchmarks, we also use the term {\bf\em Experiment Executor} in this paper referring to the execution engine to conduct such experiments. The coordination of computational task placed on the available compute infrastructure including different sites we term {\bf\em compute coordinator}. The different terms have been introduced as the task of benchmarking includes a pipeline often with repeated executions. 

This is subset of a general workflows, which by many in the community have been 
predominantly used to express them as direct acyclic graphs (DAGs). However for our experiments it is essential that we have an easy way to integrate multidimensional loops iterating over hyperparameters, which are much easier to understand and formulate than DAGs. Nevertheless, we obviously also need to support DAGs and not only do iterations.

The complexity of the compute infrastructure and the applications requires that workflow toolkits provide sufficient flexibility to support the experiment execution. Thus we need to support a bottom-up approach as well as a top-down approach. Through the bottom-up approach we need APIs, components and scripts that can be adapted by integrating new applications. In addition the top-down approach also allows the conceptual integration of multiple experiments run on various compute resources including those hosted on different sites. The results of these experiments is then consolidated and a consistent  eport can be generated from them while promoting Open Science and the FAIR Principles \citep{wilkinson2016fair}. We also need to be able to support a top-down approach where we learn from the application users what features their benchmarks need to include and be able to utilize lower-level libraries to support them.

Two software libraries Cloudmesh and SmartSim were developed indepently and contains apsects of these bottom-up and top-down approaches while offering similar functionality. This gives us the confidence that what we describe here has general applicability and is useful to the community. Both provide Python-based libraries used to describe the components of an experiment execution. While Cloudmesh also provides a compute coordinator to integrate heterogeneous experiments, SmartSim allows users to deploy an in-memory specialized datastore which is also capable of performing AI inference and exchange data between components of the workflow. In Cloudmesh \citep{cloudmesh-ee} community-developed reusable tools can be used for this.

The paper is structured as follows. After a brief introduction in Section \ref{sec:introduction}, we outline some requirements in Section \ref{sec:requirements} that we found important as part of our activities in the area motivated by hardware, user, software, and data management requirements. In  Section \ref{sec:executors} we focus on presentation an overview of two independent implementations addressing many of the requirements presented, namely SmartSim and Cloudmesh Experiment Executor and Compute Coordinator. The overview includes also a comparison between these systems and provide evidence how the requirements we identified are fulfilled by them. Furthermore, we present a recent extension to our work making it possible to utilize cloud resources.
We have tested the system on use cases that we very briefly list in Section \ref{sec:use_cases}.
To position our work we also added a related work Section \ref{sec:related}.
 Finally, we conclude the paper with Section \ref{sec:conclusion}.

\section{Workflow Requirements}
\label{sec:requirements}

In this section, we make some important observations that have a direct impact on workflow requirements particularly in the context of scientific computing done and their benchmarks at national laboratories and academia. These workflows are largely distinguished from commercial workflows by their reuse of community-shared resources and the fundamental requirement to share results externally. Additionally, a strong trend towards open science and scientific reproducibility has led to a push towards making the tools, software, and applications (e.g. simulation code, execution scripts, etc.) open source.  These requirements are not a comprehensive list, but they highlight key aspects we needed to address when developing software to design experiment executors, run benchmark experiments, collect results, and perform comparisons. Importantly,  these requirements listed had a direct impact in the development of the experiment executors for {\em SmartSim} and {\em Cloudmesh}. For pointers to additional requirements we refer to our related research (see Section \ref{sec:related}).

\subsection{Compute Systems Requirements}\label{sec:hw-requirements}

In the U.S., the HPC flagship computing resources have traditionally been offered by national scale computing centers, most notably the Department of Energy (DOE) and the National Science Foundation (NSF). In addition, we see NASA, NSF ACCESS, NAIRR, and others additionally providing HPC resources for specialized mission and scientific objective efforts serving particular communities. Others such as the DoD-related facilities are not available in general to the open science communities without restrictions, hence, we exclude them from our discussion. Other resources include commercial computing resources offered through cloud providers as part of hyper-scale computing centers. For the latter, the pricing is discussed in more detail in Section \ref{sec:cloudcluster}.

Beyond these systems, we see additional regional or topical shared resources to smaller scale (e.g. many universities support such HPC resources). Such tiered levels of HPC resources are necessary to serve the various communities by allowing granularity of customization of computational resources based on topic, scale, and budget. These smaller systems also serve as an important on-ramp for training in preparation for effective usage of the leadership classsystems. Some organizations define these tiers more specifically, for example the European Union's Partnership for Advanced Computing in Europe (PRACE) \cite{www-prace,prace-fact} associates {\em Tier-0:} with European centers with petaflop machines, {\em Tier-1:} with National centers, and {\em Tier-2:} with Regional centers.
  
However, it is essential to recognize that there are additional tiers that are of importance when requirements are gathered to support all of them. Hence, we suggest the use of a five-tiered model that classifies resources accordingly:

\begin{itemize}
\item {\bf Tier-0:} Leadership Class machines with worldwide leading
  performance (Listed at the top of the Top 500 list). Such machines also include large specialized data file systems to allow serving the many computational nodes. Recently, such systems include a large number of GPUs. They are typically served by batch queuing systems and serve the most challenging scientific applications. An allocation request is typically needed to get access to such machines.
\item {\bf Tier-1:} Machines in the upper portion of the Top-500 list which may be part of National centers, Regional Centers, or Universities. Such systems are similar to those in Tier-0, but of significantly lower capabilities.
An allocation request is typically needed to get access to such machines.
\item {\bf Tier-2:} Machines whose performance is similar to machines
  in the rest of the Top 500 list. These machines are either smaller systems or if still operated older HPC clusters that have been replaced by newer machines.
  An allocation request is typically needed to get access to such machines. In the case of universities, the HPC is shared based on internally set policies.
\item {\bf Tier-3:} Smaller scale HPC clusters funded by a university or entity
  that are no longer listed in the Top500 list. Many universities have their own small clusters that are not as powerful, but serve their individual university or research communities, or are operated by specialized  labs. They may or may not run batch systems and at times use other software such as OpenStack, or more recently Kubernetes. 
\item {\bf Tier-4:} Privately owned machines supporting development
  and debugging. These are machines operated by individual researchers that may include powerful GPUs or CPUs, often performing on a small scale individually faster than even those offered by servers operated by universities. They provide an excellent cost-performance option for many researchers to develop and debug their programs quickly if the scale of the targeted application allows. These systems obviously are not targeting large parallel computing jobs. Although these machines do not typically represent an HPC machine as they are mostly single-node computers they can provide valuable input in performance expectations, development, and debugging.
\end{itemize}

\begin{figure*}[htb]
    \centering
    \begin{tabular}{cc}
        \vtop{\null\hbox{\includegraphics[width=0.45\textwidth]{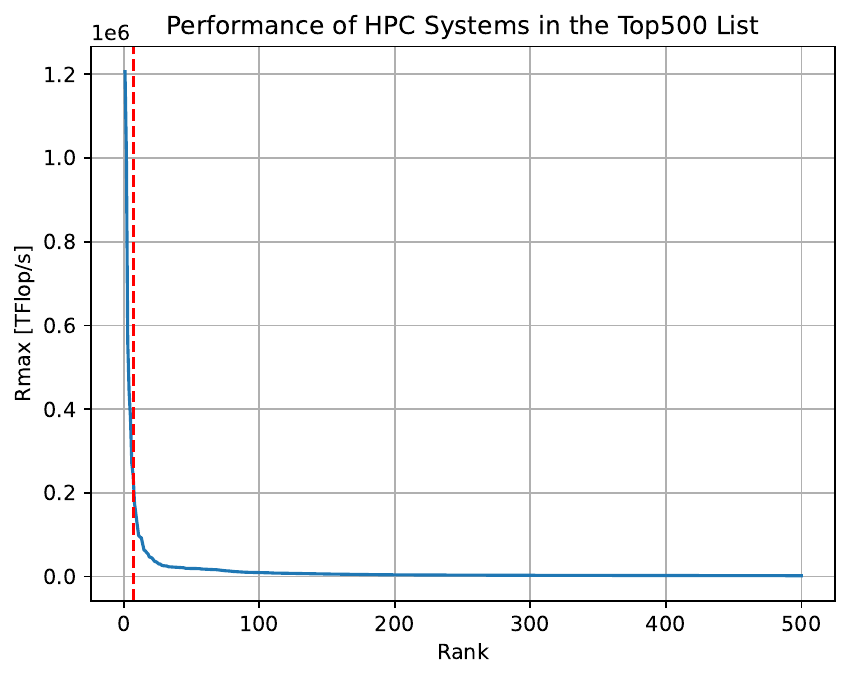}}} &
        \vtop{\null\hbox{\includegraphics[width=0.45\textwidth]{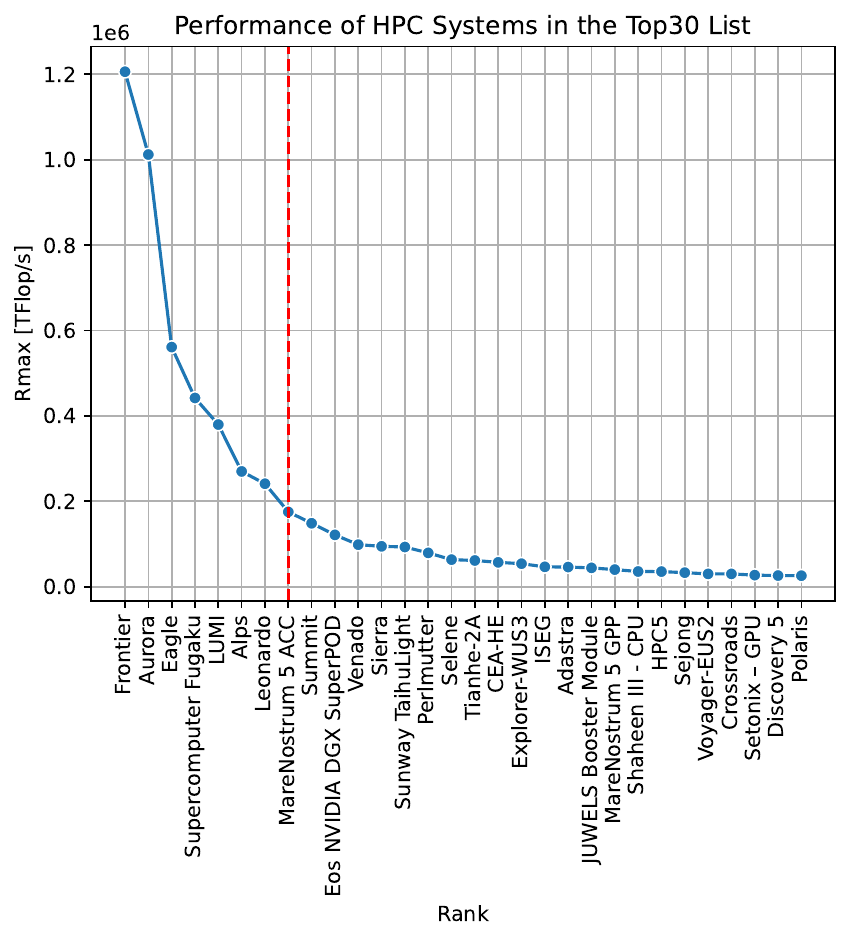}}} \\
        {\bf a.} Top500. & {\bf b.} Top30. \\
    \end{tabular}
    
    \caption{Top500 List Comparison with the index equilibrium at about 7.}
    \label{fig:top500comparison}
\end{figure*}

Using this distinction, we note that the Tier-0 (the top machines in the Top500 list) dominate the overall computational capacity. This is evident as the Index Equilibrium is at about 7, that is the Rmax [TFlops/s] for
the first 7 resources of the list are equal to the sum of all other
493 resources, where Rmax is the maximal LINPACK performance achieved (see Figure \ref{fig:top500comparison}).

\newcommand{\myPARAGRAPH}[1]{
    \parindent 0pt \textbf{#1}
}

\myPARAGRAPH{Resource Sharing.}
The Tier 4 has especially become important as it provides potentially individually small scale computational power for developing a subset of scientific applications. As they typically deployed as single user controlled systems, they are immediately available during the development and debugging cycles of the scientific workflow.

Accessing resources in the other Tiers is more structured and managed through well-established queuing/batch systems to allow shared usage among the community accessing the resources. On larger systems especially when large scale experiments are needed this may result in significant runtime delays.

\myPARAGRAPH{Potential Heterogeneity in Resource Integration.}
Scientific workflows must be able to interface with such systems easily. However, especially at the university level, we see significant complication in the use of such resources, as machines often integrate different heterogeneous resources into the batch system, which makes intrinsic knowledge about the design and availability of specific resources necessary by its users. This is introduced when the university rolls out multiple generations of infrastructure that is integrated over time in a shared heterogeneous HPC cluster. Thus, such systems may include older hardware, or have file systems not well integrated leading to a significant slowdown in the potential performance as we have discovered in \citep{las-frontiers-edu}.

\myPARAGRAPH{Authentication and Authorization.}
One of the major requirements is the authentication and authorization. In general, all resources are accessible through remote access. It is well established that two concepts are used to access them. First is the use of SSH, and second, some systems also provide additional security through VPNs and 2-factor authentication. Hence, if such systems are used in concert, we must be able to integrate with the various authentication mechanisms. Grids were supposed to solve this issue, and at least for some US-based systems, is continued in the use of inCommon \citep{incommon}. However, anecdotally it is often practically easier to gather the authentication mechanisms and have the client directly authenticate to the various systems utilizing ssh key authentication when available. This is based on experience dating back even 30 years ago \cite{las-99-loosely}.

\myPARAGRAPH{Operating Systems requirements.}
On the Top500  list, 100\% of the operating system family is Linux-based. We also observe that all but the last Tier are also most likely Linux-based. The exception is Tier 0, where client machines are using Windows, Linux, and MacOS. As we strive not only to run workflows on the clusters but also to control them from client machines, it is important to support a wide variety of operating systems. This may be easy to do as MacOS can interface with the various tiers through shells or API calls. Even on Windows we can leverage Git and Bash (or other POSIX-adjacent terminals) to emulate a Linux environment, and thus shell scripts 
can be ported to even Windows. Alternatively, on Windows one could use Windows Subsystem for Linux (WSL). This allows us to restrict our development environment choices. However, in some cases it is beneficial to also support native Windows on the client side, especially as some scientists and students may not be familiar with the Linux capabilities recently made available in Windows.


\tcbset{
    colback=yellow!20,
    colframe=yellow!50!black,
    title=Colored Box,
    breakable, 
}

\newenvironment{BOX}[1][title]
{
    \ifx\homepath\overleafhome
        \bigskip
        \begin{quote}
        \begin{mdframed}[hidealllines=true,backgroundcolor=yellow!20]
        \noindent\rule{\linewidth}{0.4pt}
        {\bf #1}
    \else
        \begin{tcolorbox}[title=#1]
    \fi
}
{
    \ifx\homepath\overleafhome
        \noindent\rule{\linewidth}{0.4pt}
        \end{mdframed}
        \end{quote}
    \else
        \end{tcolorbox}
    \fi
}



\begin{BOX}[Implications from Section \ref{sec:hw-requirements} Compute Systems Requirements]

\begin{itemize}
\item {\bf Hardware at wide scale:} {\it The hardware that we need to
    support includes a wide range of large HPC systems down to the individual researcher's computer.}
\item {\bf Integration of GPUs:} {\it One of the important aspects especially with the advent of integration of AI-based algorithms and the need for faster calculation is that GPUs must be easily accesible.}
\item {\bf Interface to workload managers:} {\it As different systems
    have different workload managers such as SLURM, LSF, and others, the workflow system must be expandable and allow easy
    integration of workload managers by the community.} 
\item {\bf Simple uniform access through shells:} As shells are supported on all machines, a framework can leverage shells uniformly. 
\item {\bf Minimal support for access via authentication and authorization:} {\it Although it is not
    necessary to support a fully-federated resource infrastructure,
    client-based workflows must support the integration of
    heterogeneous resources with potentially different workload
    managers. Access could be supported through multiple keys or
    services that are specifically set up by the user to allow
    interfacing with the hardware.}
\item {\bf Batch access and direct access:} {\it As many of the
    workload managers are batch queuing systems, we need to support
    them in general. However, as we also have access to machines that
    may not be controlled by batch queues, we need to be able to
    potentially simulate such batch queues or provide mechanisms to
    install them so that such resources can simulate the same
    interfaces as those provided by HPC centers.}
 \item {\bf Cloud HPC resources:} {\it Most recently several cloud
     vendors are also supporting the provisioning of HPC resources,
     but the complexity of managing them is beyond those present by the
     typical application users. Workflow systems should support
     easier provisioning of such HPC resources so they can be readily
     integrated into the scientific research and benchmark efforts.}
\item {\bf Minimal support for virtualization in the cloud:} {\it Although we
    previously spent a lot of time interfacing with virtualized
    resources and cloud-based resources, we recently have shifted our
    focus on the more traditional approach to interface with queueing
    managers. This is motivated by the fact that many of the
    most complex state-of-the-art projects are conducted on the most
    capable machines (the first 7 machines in the Top500 provide the
    same compute power as the remaining 493).}
 \item {\bf Container and virtual machine support:} {\it
     Being able to support containers helps abstract the application from the underlying hardware. Thus, a potent workflow system ought to
     support both virtual machines and containers. In the case of containers, this includes Docker, Kubernetes, and Apptainer (the latter being representative of the Docker-like container solutions that work in the heightened security environments on HPC platforms that isolate the runtime environments).}
  \end{itemize}
\end{BOX}

\subsection{User Requirements}
\label{sec:user-requirements}

To identify the user requirements, we first ask: who are the users \citep{wilkinson2022-2}? Based on our experience, we characterize the following six user personae:

\begin{itemize}
    \item {\bf Application users} are users with focus on application experiment workflow usage. Often they are supported by Graphical User Interfaces (GUIs), Gateways, or even customized application-specific frameworks. In many cases, the complex workflows to utilize sophisticated cyberinfrastructure, including hardware and software, are hidden from the users so they can focus on their applications \citep{lee2021}. However, these abstractions may come with the problem that such users are divorced from the actual cost of a particular workflow. Thus especially when evaluating the cost-benefit ratio of various applications, being able to project in easy-to-understand terms the expected runtime (the estimate prior to running the experiment) costs for experiments is needed to help plan and prioritize. These cost metrics may include not only the dollar cost but other factors such as availability, wait time, and energy costs. Thus users can be more informed about the cost impact of their workflows.
    \item {\bf Application Scripters} are users that may not have GUIs or Gateways available or even prefer to use scripts to formulate their experiment workflows. This may include shell scripts or programming languages such as Python to coordinate the experiments. The requirements for such users include that scripting must be simple enough so that the application is still the main focus of their workflow. In some cases, workflow templates that feed scripts promote reuse and can accelerate the repeated execution of experiments. When using HPC systems, users typically have to learn how to use a batch queuing system, as well as have minimal understanding of the command shell \citep{wilson2021}. This include the specific policies that are imposed by the organizations offering the resources. 
    \item {\bf Application developers} are developing specialized applications as part of their scientific workflows. They either develop the workflows from scratch as part of the regular programming or reuse libraries that interface either with the application domain or the cyberinfrastructure so that through reuse the experiments they target can be simplified through sophisticated but easy-to-use APIs, libraries, experiment management workflow components that coordinate one or multiple experiments. It is important that the libraries that are developed for this community can be integrated in some fashion into the preferred programming language or framework. This may go beyond the availability of Python frameworks that are very popular with AI experiments. 
    \item {\bf DevOps Engineers} tasks include the management of a software development life cycle and enabling the integration of cyberinfrastructure to allow workflows that integrate automated provisioning, testing, and release management. They can be essential in the better utilization of the infrastructure in general but also support large-scale experiments with complex workflows that are these days more common while utilizing large-scale cyberinfrastructure. Hence, Experiment workflows need not only be able to be defined by application users for large-scale HPC, but it is advantageous to consult with DevOps Engineers to fine-tune experiments before they are placed on the infrastructure or are refined throughout their lifetime. A whole set of tools have been developed in support of DevOps, which is beyond the scope of this paper.
    \item {\bf System Administrators and Support Staff} support experiments while maintaining the systems designed for a user community. They will provide support and help to any users utilizing the infrastructure. In many cases, application users do not need or access DevOps Engineers but interface directly with the System Administrator to define strategies to utilize the infrastructure for their experiment workflows. In all organizations we used HPC resources for experiment workflows, dedicated support was available to address questions on how to improve the runtime experiments as well as application performance improvements.
    \item {\bf Organization and Funding agencies} are an often overlooked part of the scientific experiment workflow. They provide, in many cases, access to the often costly infrastructure and need to be informed how they are used. This may include not only an overall breakdown for the entire organization, but it can also help the individual experimenter to understand their own demands placed on computing resources (see \citep{las-15-tas}). 
\end{itemize}

From this diverse set of users that we encounter in support of repeated experiments executions, it is obvious that the requirements vary by user group. While the application user is satisfied with a high-level interface, an application developer and scientific researcher may need access to much more sophisticated tools and libraries. In many cases, they could also benefit from a standardization of libraries that supports their and other researchers' experiments even across domains. The education of the users may play a very important role. While we have seen in some projects users have been educated by system staff, the users as well as the system staff may not have known tools that simplify certain processes in the experiment execution management.

\subsubsection{Benchmark Carpentry Requirements}
\label{sec:carpentry}

In \citep{las-frontiers-edu} we introduced the concept of {\em benchmark carpentry} in relationship to educational activities as we believe that based on our own experience this topic is much needed but is not adequately addressed. There, we identified that educational efforts to enable benchmark carpentry includes diverse tasks related to  (a) installing software, (b) reserving compute resources for exclusive use, (c) preparing experiments (potentially using a large number of batch jobs) and executing them, (d) evaluating and validating the resulting performance including computational power of CPUs, GPUs, data I/O, networking, and energy, (e) record the results in a uniform format so that comparisons can be made (f) and submit results to the community to allow others to contribute, either through publications or efforts such as promoted by MLCommons. All of them should ideally be integrated in a well-defined mechanism allowing to support the FAIR principles \citep{wilkinson2016fair}.

Due to the wide variety of potential user communities involved targeted educational material must be available. This goes beyond material that has been typically distributed as part of software carpentry \citep{software-carpentry} efforts while significantly expanding them with the focus on benchmarking. Hence the term benchmark carpentry is appropriate for such efforts.

\begin{BOX}[Implications from Section \ref{sec:user-requirements} User Requirements]

\begin{itemize}

\item {\bf Wide Variety of Users:} {\it To support a wide variety of users, experiment execution management needs to be available from the lowest to the highest level of interfaces targeting the specifics of the user community. This has a significant impact on the software in support of these communities which we explain in the next section.}
\item {\bf Ease of Use:} {\it In order for the user community to utilize experiment execution management, whatever tool and software is supported must be easily used by the targeted user community.}
\item {\bf Experiment Automation:} {\it Users strive for replication of their experiments. This includes experiments that can be replicated by different users, but also experiments that can be replicated on different hardware.}
\item {\bf Experiment Reporting:} {\it As experiments are recorded at a particular time under a selection of software and hardware utilization, it is important that results encompass reporting of the environment. This will help the reproducibility of the experiment and if the underlying system has changed the repetition of the experiments with minimal changes.}
\item {\bf Portability:} {\it Users that conduct benchmark experiments also require portability which allows them to compare and contrast different experiment setups on different systems.}
\item {\bf Cost Considerations:} {\it One important factor in conducting benchmark experiments is the ability to understand cost considerations prior to running a large-scale or time-consuming experiment. Having the ability to scale and predict performances from a small scale to the target scale is an important need. Tools and software should be provided that assist in this often complex endeavor.}
\item {\bf Benchmark Carpentry:} {\it As benchmarking is often not enough covered in educational activities, there is an opportunity to engage in a specific Benchmark Carpentry effort. As part of this can be the introduction to workflows, and tools that coordinate the experiment management of benchmarks.}
\end{itemize}
    
\end{BOX}

\subsection{Workflow Specification Requirements}
\label{sec:definition}

Due to the diverse user communities, a one-fits-all solution cannot be delivered. In particular, the software requirements need to address each community. However, we can identify common patterns as part of the definition that overlap between the communities. This includes, in particular, arrays and loops of experiments iterating over hyperparameters or specific machine configurations to be provisioned or used as they are common to define benchmarks. Although DAGs help coordinate certain experiments, loops and arrays often provide simpler and clearer ways to define repeated or iterative experiments. This insight was already available in previous work where we described allowing iterations and dynamically changing workflow graphs \citep{las07-workflow} in addition to DAGs. We like to emphasize that the the inclusion of hyperparameter searches in deep learning benchmarking experiments projects the requirement to integrate not only Graphs but also more importantly iterations or goal-oriented searches with termination or adaptation conditions. 
In general the definition of the experiments need to be simple so that they can easily be generated executed, and their results uniformly gathered. 

When working with the communities we are most closely related to, we found that specifications using YAML was accessible and rich enough to express a hierarchical structure to represent graphs, but also iterations were easy to understand by others provides a simple way to define experiments. Others may refer to more elaborate workflow languages as indicted by the efforts collected in \citep{workflow-list}. The definition of workflows must also be programmable so that more sophisticated experiments can be built that integrate efforts conducted by others. Thus, it is important to also ensure that the APIs encourages reuse in other frameworks. The functionality ought to be exposed on various levels. This includes availability to access the queueing system, the availability of APIs accessbile to other frameworks or even programming languages, or providing high-level abstractions such as YAML formulations that allow the definition of iterations or DAGs in with easy-to-understand specifications. The latter is based on our long-term experience, present even in our earliest work that included a simple workflow language \citep{las-96-ecwmf} followed by an XML based language \citep{las07-workflow}. However, our recent experiences with the educational community make us believe that the definition using YAML and JSON is more straightforward and presents an opportunity to simplify the workflow definition requirements \citep{cloudmesh-cc,cloudmesh-ee}.

The ability to integrate checkpointing/restoring, result recording, and backup is another important requirement that ought to be supported either implicitly or explicitly during the experiment definition. Furthermore, through the making workflow experiments availability  on various abstraction levels via specification, we can flexibly integrate multiple supporting frameworks, leveraging community activities.  One additional way of exposing and integrating with other frameworks is to provide interfaces in OpenAPI. This will allow in many cases a sufficiently detailed integration in other languages also through the wide support of OpenAPI. However, due to the dominant use of Python in academia as part of newer developments, we suggest to promote the use of a native Python API. This has the advantage that many built-in libraries and tools can be leveraged to simplify the development of integrated workflows. We also can leverage the language support for loops and existing libraries to support graphs which we have demonstrated successfully in \citep{cloudmesh-cc}.

\myPARAGRAPH{Workflow Template Requirements.} A closely related requirement is the ability to clone a workflow and to utilize it for new experiments. While the workflow definition is an instantiation, the template refers to a potential instantiation of a workflow. We found that providing such templates has two benefits. First, it allows users to learn from previous experience, second it allows easy adaptation to utilize the workflow in other infrastructures if properly adapted. 

\myPARAGRAPH{Workflow Template Repositories.} Using templates allows us to address two concerns: (a) the reuse of sophisticated infrastructure while learning from templates targeting such machines, and (b) the reuse of specific application templates formulating experiments that ar similar to other applications. This could be facilitated by an experiment management template repository shared with the community. As the infrastructure and knowledge about the application changes over time, such a repository can also include evolving templates. As we also have to consider different scales, templates to serve HPC, Hyperscaler, federated, cloud, and quantum computing \citep{bieberich2023} resources ought to be integrated. Being able to specify them in a uniform format is beneficial. However, such repositories could also include various formats as long as the metadata with them includes information about the origin and which tools can be used to execute them producing data following the FAIR \citep{wilkinson2016fair} principles.

\subsection{Runtime Requirements}
\label{sec:druntime}

In this section we gather some common runtime requirements from various use cases from scientific HPC applications.

\myPARAGRAPH{Cyclic and non-cyclic Experiment Execution Requirements.} \label{subsubsection:cyclic}
Workflow execution has typically focused on workflows that can be represented by a pipeline or more generally a directed acyclic graph (cycles with predefined loop limits can be unrolled). 



Another example stemming from deep learning is an exhaustive set of trials for every combination of the provided hyper-parameter values also termed gridsearch. 
Hence experiment executors need mechanisms that support such cyclic experiment execution while integrating with or without conditional runtime executions as part of an experiment.

\myPARAGRAPH{Queuing Policy Requirement Implications.}
Besides needing to be authorized to use a particular machine, we especially note that often we need to address policy-based restrictions to the resources restricting authorization. This includes addressing queuing system policy restrictions set up by the organization and managed by the administrators. Although such restrictions could be changed, they are often not scalable as they need to be changed back. However, in many cases, reformulating the experiment workflow can avoid such restrictions. For example, a job that is terminated due to exceeding time restrictions could be split into multiple jobs while allowing checkpointing at the end of the individual jobs and restarting them into the next phase can help. In other cases, instead of creating a loop over all possible calculations needed to conduct the overall experiment, submitting multiple jobs with the experiments split up between them can be used. 

Supporting this effort requires also that results are being stored in a coordinated fashion following the FAIR principles, the analysis of the final result while combining the results of the many individual experiments can be split up; experiment management can even deal with outages of the resources. Another example is where a resource allows the utilization of thousands of CPUs or GPUs, but only for a small amount of time. In such cases, the application user ought to be encouraged to parallelize their algorithms, and the experiment framework needs to support such a modality. This, however, is outside of the scope of our work and needs to be addressed by the application developers. Small-scale runtime experiments could be used to project how the runtime or the design of an algorithm is impacted by such queuing policies. In all cases workflow templates could be used to communicate to users how to deal with such limitations.

Furthermore, we note that such policies differ widely between HPC systems and need to be integrated into the planning of a heterogeneous experiment across resources. Obviously, the availability of time and space limits at runtime that could be queried dynamically can help improve the deployment of dynamic experiments that deal with the limits if they arise. Guidelines and APIs to checkpoint an application will be of importance. However, although automatic check-pointing is desired, for many applications only a fraction of data is needed and not the entire state of the running application. Therefore it is best to identify data that is needed in consecutive runs and only checkpoint those.

\subsection{Authentication and Authorization Requirements}

The desire to execute benchmark workflows on multiple HPC resources to compare them will bring up the topic of federated resources using the same security context. However, although this would be highly desirable in organizations such as DOE \citep{antypas2021} and maybe even achievable, we will always have resources that are outside a single federated organization. This is not only due to the independence of many research organizations and universities but also due to policies in place by the countries in which they operate. Hence, it is important that other means are used to allow using resources from a wide variety of organizations from which we know that federation can not be achieved. To allow maximal progress with minimal effort we did not further explore the use of more formal federation capabilities. When using cloud resources such as HPC resources provided by major cloud providers, federated security becomes even more complex. Our strategy to stay within the cloud-based security context for such resources will ideally be integrated within such resources in a wider benchmark effort.

It is beneficial to strive for security requirement simplification which is motivated by our own experiences utilizing HPC and other compute resources into benchmark experiments. SSH is the de facto solution for accessing HPC platforms, spanning DOE, NSF, and university resources, potentially allowing all experiments to leverage SSH to create heterogeneous experiments across these resources. This potential is complicated when organizations require the use of a VPN. For example, consider a case with one of our universities. In this case, the university mandated VPN redirected all traffic to the universities IP infrastructure, causing a network bottleneck to other resources. In response, we have developed a toolkit that supports a split-VPN functionality, simplifying our authentication and authorization requirements to gain access to the resources. Hence, our primary requirement was SSH availability, and if SSH is only accessible behind a VPN server, allow support for split-VPN.

Furthermore, it is important to note that that policies are controlled by superuser access to HPC platforms which is often limited to a very small subset of support staff. Thus any workflow software managed by the users must be able to be run with basic user privileges. Software designed for containers and/or for deployment in the cloud tends to assume that they have the ability to register system-level services and/or kernel-level access. This is not the case on many of the machines available to the academic public. As such, some container technologies commonplace in the cloud are simply not available on major HPC platforms. Hence restrictions of using container technology (if deployed) is limited to those allowed by the various internal policies and may be different from site to site. 

One of the other important requirements is that the workflows do not hardcode any credentials into it either through workflow specification files, nor programming. The credentials must be kept separate.

\subsection{Data Management Requirements}
\label{sec:data-requirements}

In order to support data benchmark requirements, we need to consider the wide variety of data needs served by the benchmarks. This includes statically generated data which may not involve any data storage, and reaches hundreds of petabytes, such as  the training of large language models (LLM). As the full training of such LLM can exceed the availability of the resource for the benchmark, it is often necessary to reduce the benchmark either in size of the data or the duration of the runtime. 
We see in some use cases only a few files, but in others a plethora of files incorporated into an application benchmark. The inclusion of such complex data needs can reveal shortcomings of the hardware design, leading to potential issues that despite the availability of modern CPUs and GPUs the file system is not designed to leverage them efficiently as the workflow to utilize them can not keep them busy and the data management becomes a bottleneck. We have shown recently that at a university cluster, this was the case and the machine despite excellent GPU capabilities performed subpar due to data management issues of the file system. Hence, benchmarks should not only project the potential of the sheer computational power of a system but also integrate the data I/O performance. Furthermore, benchmarks need to define the needs for required space, the type of the storage, such as blocks, file, or object store which then have an impact for the utilization and design of a storage system suitable for the benchmark.

\myPARAGRAPH{Requirement for a Federated Results Repository.}
To compare multiple experiment results across diverse HPC and other resources a federated results repository is required. This is typically done by a lead organization, and results are loosely contributed to the organization's case (such as demonstrated by MLCommons). Often it can be useful to include official organization-level submission, and also results gathered by different users from the same organization. The later has been especially useful to us, as it allowed us to avoid policy restrictions governing the execution of a large number of experiments needed to complete the overall benchmark on some sites.

In each of these experiment results, the underlying structure allows unique and easy identification within the repository. This is emphasized when designing adaptive benchmarks. This is important to not only store timing information for a fixed benchmark, but all metadata needed to replicate such a benchmark. This can even include a variation in the application input data if variation is needed by the application to for example obtain better results or update the benchmark with timely data and rerun it on resources.

Organizing data for later reuse also allows for later application of machine learning to uncover additional insights that can lead to further optimization, e.g., subtle performance gains related to specific software and hardware device combinations. This leads to a potential complication in organizing and interpreting the data. Instead of creating a single repository, multiple repositories could gather the information, but scripts should be provided that can merge results from multiple repositories.

\myPARAGRAPH{Support for FAIR.}
The FAIR Principles provide a framework for aligning research outputs. FAIR is an acronym with 15 underlying principles that relate to Findability, Accessibility, Interoperability, and Reusability (FAIR) \citep{wilkinson2016fair}. More specifically, the FAIR Principles call for the use of well-described, standardized metadata, clear licensing and usage information, open or freely available protocols and methods for access, and the use of globally unique identifiers. The actual implementation of the FAIR Principles varies by domain, data architecture, and resources available to sustain data management practices and infrastructure \citep{jacobsen2020fair}.

In the case of this work, several practices were identified that relate to making HPC benchmark results more FAIR~\citep{kirkpatrick2025}. These include ensuring controlled vocabularies are used when describing system information. Where these are not available, it is preferable for the software to extract information from other sources, such as system configuration, libraries, or registry settings, especially over the use of free-form text. Ideally, globally unique identifiers would be assigned to each benchmark output, just as they would be for other digital objects \citep{garcia2025}. Other implementations of FAIR can include writing provenance information about how the benchmark was created in the results file \citep{souza2023}. The FAIR Principles in this work are also exemplified in structured abstraction, e.g., the use of YAML, the use of standardization, NIST standards, and (machine) accessibility via an API.

\subsection{License Requirements} 
\label{sec:license}

As part of our long-term commitments to support scientific workflows we have seen a number of adverse effects when licensing of a software library or tool have changed.
Examples include MongoDB, Redis, and conda. Hence, one of the requirements must be that the software chosen to implement workflows support not only open-source development but are usable on a large scale so that deployment costs do not effect usability. 

At the same time, the software to manage the workflows should be distributed under a well-known and established open-source license allowing others to also easily contribute and reuse. 

\begin{BOX}[Implications from Sections \ref{sec:definition} to \ref{sec:license} ]

\begin{itemize}

\item {\bf Workflow specification:} {\it It is beneficial to have a workflow specification that includes more then DAGs and addresses emphasis on iterative/cyclic experiments.}
\item {\bf Workflow Templates:} {\it To ease and learn from previous experiments it is important to be able to formulate templates for specific infrastructure, but also applications. They can be used to be adapted by others.}
\item {\bf Template Repositories:} {\it Collections of workflow templates ought to be collected in a template repository with enough meta data so the FAIR principle can be leveraged and new clones can be developed easily by the users.}
\item {\bf Experiment Reporting:} {\it As experiments are recorded at a particular time under a selection of software and hardware utilization, it is important that results encompass reporting of the environment. This will help the reproducibility of the experiment and if the underlying system has changed the repetition of the experiments with minimal changes.}
\item {\bf Runtime support for Cyclic, pipeline, and DAG Executions:} {\it To execute a workflow specification that may be derived from a template all major execution paradigms must be supported at runtime. This includes the cyclic execution of experiments, the formulation as an execution pipeline, but also  the execution formulated as DAG.}
\item {\bf Runtime Batch Queuing Integration:} {\it As many resources utilize batch queues an easy abstraction to access them must be available. This must also integrate special software that can deal with queuing policies that may limit workflows. }
\item {\bf Authentication and Authorization:} {\it The system must at least be able to deal with authentication and authorization and can leverage existing frameworks for it. At minimum SSH and split-VPN ought to be supported.}
\item {\bf Data Management:} {\it All related data management aspects ought to follow the FAIR principles. The results projected by the experiments ought to be gathered in a single repository that federates correlated results. Alternatively scripts should be provided that can merge results from multiple repositories.}
\item {\bf Licensing:} {\it licensing of all parts of the benchmark efforts benefit from using open source and open access licenses. However, care must be taken if an open source project changes their license in such a form that it is not allowed to be used in scalable fashion.}
\end{itemize}
    
\end{BOX}

\section{Overview and Implementation of the Experiment Executors}
\label{sec:executors}

The requirements discussed previously were distilled from our experience as principal developers of two workflow libraries Cloudmesh and SmartSim. One thing to note particularly is that these two projects were developed completely independently without knowledge of the other until the writing of this paper. Despite this, the abstractions, responsibilities, and even terminology are often very similar and have overlap. This remarkable convergence and their demonstrated use across a wide variety of novel use cases suggests that the requirements discussed here are fundamental to the types of emerging computational paradigms in the exascale era. Both systems started with a bottom-up software design approach, but whose target user base aligns more the application-oriented subset of users described in Section \ref{sec:user-requirements}. In addition to the similarities, the use cases and target userbase has differed between the two of them, resulting in divergences between the implementation of some abstractions and additional featuresets.

While a number of other workflow systems also solve many of the same problems, the authors are not privy to the software engineering and design history of those. We thus limit the discussion of how workflow requirements have manifested in SmartSim and Cloudmesh, describe where the disjoints between the two come from, and discuss where the commonality arises. By doing so, we aim to engage with developers of other workflow engines and also to provide a basis for a shared taxonomy.

\subsection{Overview of Cloudmesh and SmartSim}

In the following sections, we provide a brief overview of each package to discuss the underlying design philosophies and show simple examples of their usage. 



\subsection{SmartSim}
\label{sec:smartsim}

SmartSim is a Python-based, open-source library developed by Hewlett Packard Enterprise (HPE) that allows users to describe and execute hybrid AI/ModSim workflows using an in-memory datastore to exchange data. The original intention was to provide domain scientists on traditional, on-prem HPC platforms a way to 1)  describe and execute large ensembles of simulations, 2) incorporate in-the-loop inference capabilities for simulations, and 3) provide a solution for storing and consuming data for online visualization, analysis, and surrogate model training. It has a sibling library SmartRedis that provides C, C++, Fortran, and Python clients for simulation and analysis codes to enable components to communicate with the datastore.

To define a workflow, users write a Python driver script that imports the \textit{Experiment} object. This object has a variety of factory methods used to define the essential components of the workflow: \textit{Model} ( actual applications to be run), \textit{Ensemble} (configurable collections of the \textit{Models}), and \textit{Orchestrator} (the in-memory database used to store data from workflow components). As with any Python script, the user can introduce their own branching logic and code at various points of the execution. When the script is executed, additional internal entities and code are created and called, interacting with the HPC platform's filesystems and workload manager.

As a brief overview of a SmartSim workflow, we show a simple (but representative) driver script in Listing \ref{lst:smartsim-ensemble} and describe briefly what happens at each step (for further detail refer to \citep{smartsim-repo}).  The instantiation of the {\em Experiment} object defines the name of the experiment and the workload manager specify which workload manager (SGE, SLURM, PBSPro, LSF, or a local executor) during the instantiation of this object. Next, the {\em RunSettings} are created that allow define how a {\em Model} will be executed and what resources are required. To actually create an {\em Ensemble} users specify how the application should be parameterized:  

\begin{enumerate}
    \item specifying parameter values in templated input files
    \item defining different collections of input files
    \item modifying runtime arguments passed to the executable
\end{enumerate}

In this particular example, we assume that the user has provided a templated configuration file which will be parsed and modified to adjust the {\em foo} and {\em bar} parameters.

The call to {\em generate}, creates the actual run directories where each ensemble member will be run and configuring files as necessary. {\em start} actually attempts to launch the ensemble, interacting with the workload manager to queue and execute jobs. As each ensemble member executes, its standard error and output are captured and archived for later inspection. SmartSim also captures additional experiment telemetry (e.g. the timestamp for when an application runs) that the user can examine.

\begin{lstlisting}[language=python,caption={Configuration of an ensemble in 
  SmartSim},captionpos=b,label=lst:smartsim-ensemble,basicstyle=\small]
  from smartsim import Experiment

  exp = Experiment("example-ensemble-experiment", launcher="slurm")
  rs = exp.create_run_settings(exe="path/to/example_simulation_program")
  params = {
    "foo": [2, 11],
    "bar": [1.0, 1.5]
  }
  ensemble = exp.create_ensemble(
    "example-ensemble", 
    run_settings=rs, 
    params=params, 
    perm_strategy="all_perm")
  ensemble.attach_generator_files(
    to_configure=["/path/to/templated/config/file"],
    to_copy=["/path/to/input/files"]
  )
  exp.generate(ensemble)
  exp.start(ensemble)
\end{lstlisting}


\subsection{Cloudmesh}
\label{sec:cloudmesh}

\myPARAGRAPH{Origin.}
The origin of Cloudmesh is based in the support of providing an easy to use interface for clients to cloud resources, especially virtual machines. Cloudmesh has many contributors and is organized on the concept of plugins that can enhance cloudmesh so that users can develop their own specialized plugins easily to address their specific needs.  Cloudmesh is based on Python and allows integration of other frameworks and API. The architectural design principle is based on a bottom-up approach where simple functionality is implemented first that is then expanded upon to deliver API, components, programs, and services  useful for the end-user.

Recently, we added plugins that target HPC infrastructure, but refrained from renaming the project due to the large number of plugins related to the project name Cloudmesh. This additionally reflects the support of traditional on-premise HPC clusters from DOE, NSF, universities and also cloud HPC resources that can be used for experiments.  This extends to resource aggregation in the same or different data centers. To simplify the execution on such infrastructures, we developed a hybrid multi-cloud and HPC analytics service framework that was created to manage heterogeneous and remote workflows, queues, and jobs. Relevant services can be accessed through a Python API, the command line, and a REST service. It is supported on multiple operating systems like macOS, Linux, and Windows 11.  

Cloudmesh provides a number of interfaces for various user communities. This includes Python-APIs, Python-plugins, services, commandline tools, a Cloudmesh commandshell and templates to use them.

While cloudmesh has over the years included lots of contributors developing over 100 plugins, we focus here on a very small subset of plugins enabling {\em experiment execution} and {\em compute coordination} as discussed earlier. 
The examples are detailed enough to understand basic features, but we refer to the Cloudmesh GitHub for more detailed explanations through code examples, code templates, and plugins \citep{www-cloudmesh-org}.


\myPARAGRAPH{Simple Cloudmesh Plugin Management}
Cloudmesh is built around the concept of Python plugins, that are integrated automatically through Python namespaces. This is facilitated by a tool that allows developers to create a plugin template that can then easily be deployed with pip install. These plugins can also be hosted on GitHub and PyPI for deployment to other users. Plugins can be dependent on other plugins, but their dependencies are included in the relevant setup instructions. The plugin framework allows the definition of a new command with user defined parameters that can be called on the command line as well as in the Cloudmesh shell.

\myPARAGRAPH{Cloudmesh Providers}
To simplify the integration with various computational backends, Cloudmesh uses the concept of providers to achieve a portable integration to Cloud providers such as AWS, Azure, or Google. The same principle can be utilized when interfacing with HPC queuing systems to conduct job management. Alternatively, scripts can be derived from templates that can be executed directly on local or remote resources including HPC batch queues. Cloudmesh comes also with an easy to use configuration file written in YAML allowing to define resources. Similar features exist when working with file storage systems. 

\myPARAGRAPH{Automated Service Generation}
Furthermore, Cloudmesh contains a prototype that can automatically generate REST services using OpenAPI specifications using Python functions and classes.  More information about this part of Cloudmesh is provided at \citep{www-cloudmesh-org}.

\myPARAGRAPH{Experiment Management}
To manage experiments we have implemented two components. The {\em Cloudmesh Experiment Executor} (EE) executes workflows on a single HPC cluster or compute resource, and the Cloudmesh Compute Coordinator (CC) manages tasks on remote resources that may include those started by EE \citep{las-2022-hybrid}\citep{las-2022-templated}.  

An experiment using EE is specified via YAML files. To coordinate them across resources CC can either use YAML files or DAGs that can even be rendered in a simple GUI. Hence, this thus provides the following functionality 

\begin{itemize}
\item {\bf (a) Heterogeneous System Integration:} the placement of the workflow onto a number of different compute resources including HPC, cloud, and local computing while also providing adaptations to various batch queues
\item {\bf  (b) Heterogeneous Compute Coordination:} the coordination of task-based parallelism to execute the workflow on the various resources, and 
\item  {\bf (c) Heterogeneous Experiment
Execution:} the coordination of hyperparameter sweeps as well as infrastructure parameters used in the application through the experiment coordinator.  

\end{itemize}

\begin{figure*}[htb]

  \ifbool{SUBMISSION}{  
      \centering\includegraphics[width=1.0\textwidth]{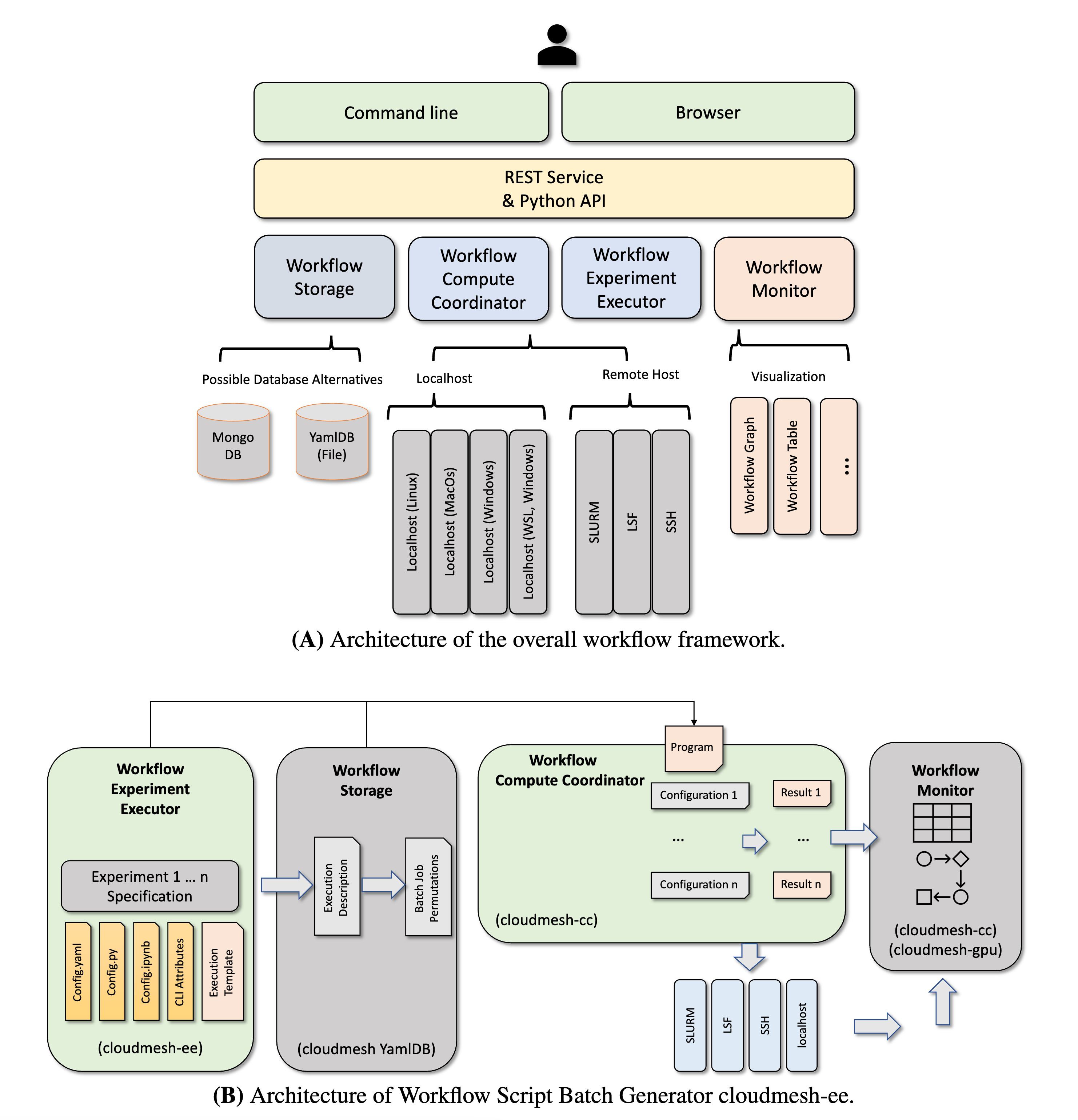}
  }{
    \centering\includegraphics[width=0.70\textwidth]{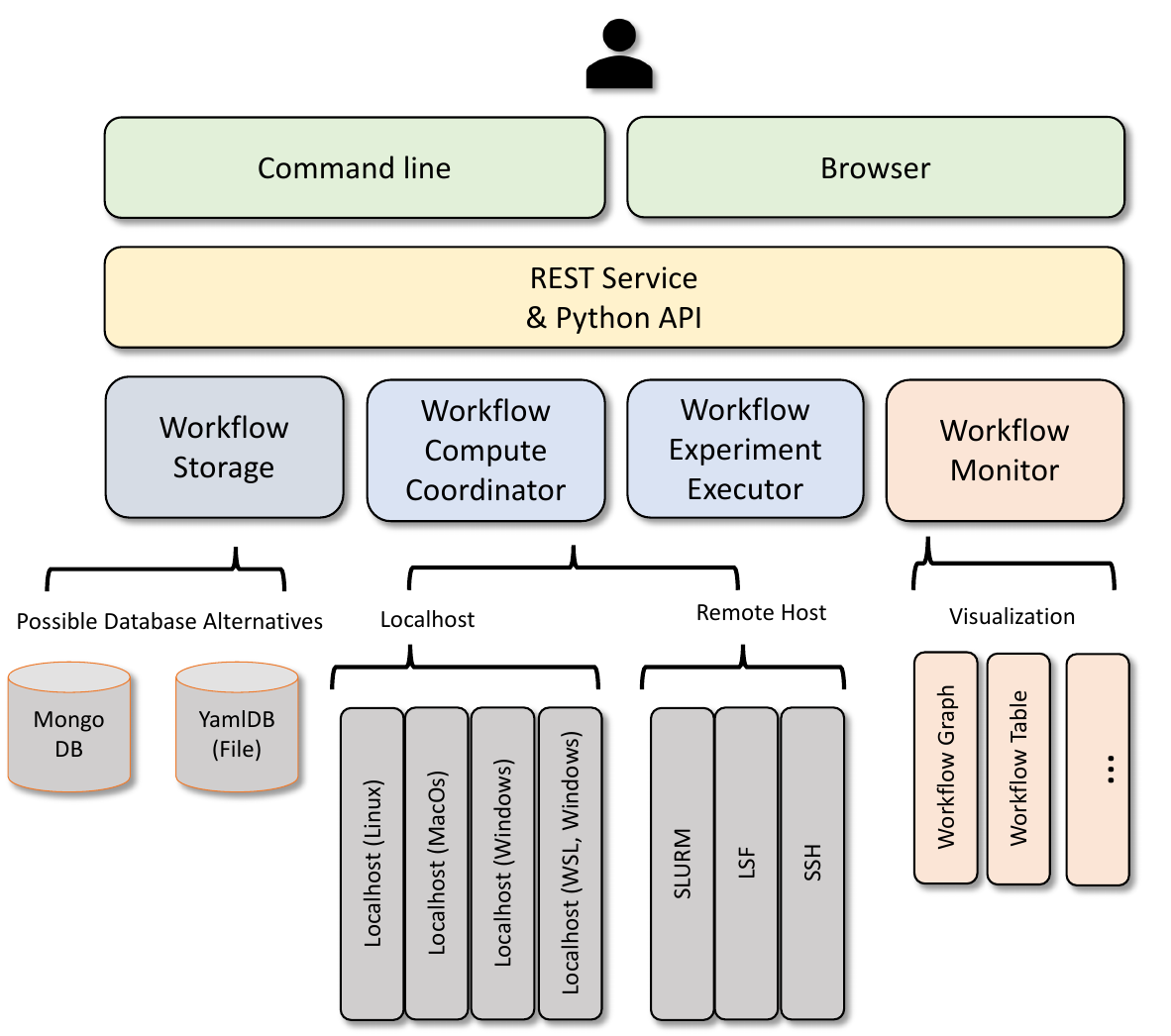}
    
    {\bf (A)} General architecture of the integrated Cloudmesh {\bf Compute Coordinator} and {\bf Experiment Executor} \citep{las-frontiers-edu}.

\bigskip\bigskip
    
    \centering\includegraphics[width=1.0\textwidth]{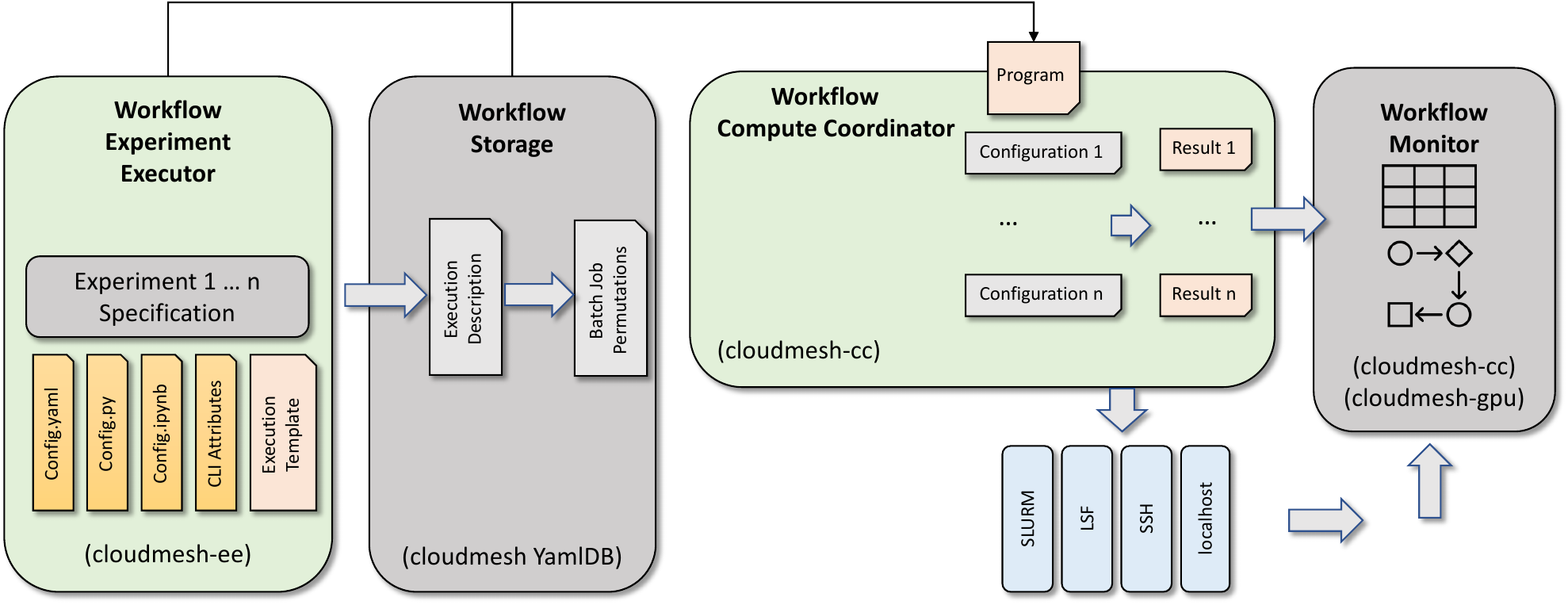}

    {\bf (B)} General architecture of Cloudmesh Experiment Executor interfacing with abstract schedulers \citep{las-frontiers-edu}.
  }

    \caption{Architecture of the Cloudmesh Workflow Service Framework.}
    \label{fig:cc-2}

\end{figure*}

The architecture of the framework is depicted in Figures \ref{fig:cc-2} A and B.  The framework is based on a layered architecture so that it can be improved and expanded on at each layer targeting developers and end users. The system can also be used on a uniform HPC infrastructure, but can be extended by the the user to integrate multiple systems into the workflow as needed. This is the reason we use the term {\em heterogeneous}. Next, we describe the two components in more detail.

\subsubsection{Compute Coordinator, a Cloudmesh Plugin}
\label{sec:workflow-cc}

The role of the Compute Coordinator (CC) is to execute tasks on compute resources. Tasks can be scheduled on a variety of schedulers operating on compute resources. Examples are LSF, SLURM, and SSH.

The experiment workflows are defined with human-readable YAML and can be stored in various formats on databases such as Cloudmesh file-based YamlDB \citep{yamldb}. The concepts have been influenced by our earlier work \cite{las-94-ecwmf} but have been updated with current technologies and the use of YAML as specification language.

It is easy to create a variety of add-ons to CC such as monitoring components that can be part of a Web browser-based implementation to display the Workflow and its status as a graph, table, or even as a continuous log file.  Tasks and Jobs report their status at runtime into a database which can also be just a file in a filesystem.  To provide uniformity,  we have introduced an abstract job class that is integrated into a workflow class that allows us to define jobs, start them, and cancel them, to name only the most important management methods. Internally, each job creates a status file in which the actual progress of the job is recorded.  This status file is managed directly on the compute resource on which the job is run and is queried through pull requests on demand to return the status to the client. This way, the status of all jobs can be monitored easily. As we strive not to run jobs that execute in milliseconds but rather in the multiple-second or hour range, such status reporting and propagation is well-suited for us because they are typically long-running tasks as the particular benchmark applications we work with require.  As our status progress update specification is universally applicable via integration into notifications through files (including stdout and stderr) they can, also be issued by bash scripts, SLURM scripts, Python programs, Jupyter notebooks, or any frameworks written in other computing languages. The workflow status updates are implicitly and uniformly augmented with timestamps, the name of the HPC resource, the compute resource within the HPC, and additional messages are appended to be sent to the monitoring component.  The workflow allows the specification of dependencies between tasks and supports a DAG. The code is compatible with Windows, macOS, and Linux.

The workflow specification plays an important role in not only defining a workflow but also in simplifying status updates that update an instantiation of a workflow. As we have completely separated the status of the workflow from the responsibility of obtaining status updates, this component can be shut down while the underlying jobs as part of the system integration are still executed and updating their statuses on the remote resources. Once the system is started again on the user's local machine, it self-synchronizes its status from the system integration services that query the status of the appropriate resources. To summarize, the client is stateless and fetches the state of the submitted jobs on demand. It will return the latest state found as reported by the job execution service.

The workflow definition for CC is rather simple and
intuitive and has been introduced in \cite{las-2022-templated}. An example is presented in  Figure~\ref{fig:workflow-example} depicts. We find it important to display this information in this paper rather then pointing to a manual as it showcases the simplicity of the framework and some of its unique features such as dynamic labels. In the example, a graph
with the tasks ($start \rightarrow f\!etch\!-\!data
\rightarrow compute \rightarrow analyze \rightarrow end$) representing a typical minimalistic use case for Deep Learning (DL) is shown. The workflow
 executes three scripts ([fetch-data,compute,analyze].sh) while the dependencies are specified in a human-readable format using the names of the nodes. The nodes contain easy-to-manage information such as the name of the node, a label that is used to print the node's progress, and can contain parameterized variables such as any value defined as part of a particular node, or specially formatted time stamps. To demonstrate the easy use our label contains the {\em name} and {\em progress} of the workflow which is rendered by the graph or table monitoring components. One can also use variables accessible from Python including operating system or batch system variables to name only a few. Selected examples of values usable in the nodes are listed in \citep{las-2022-templated}.
 
\begin{figure*}[htb]
\begin{minipage}[t]{1.0\columnwidth}
\centering  \includegraphics[width=1.0\columnwidth]{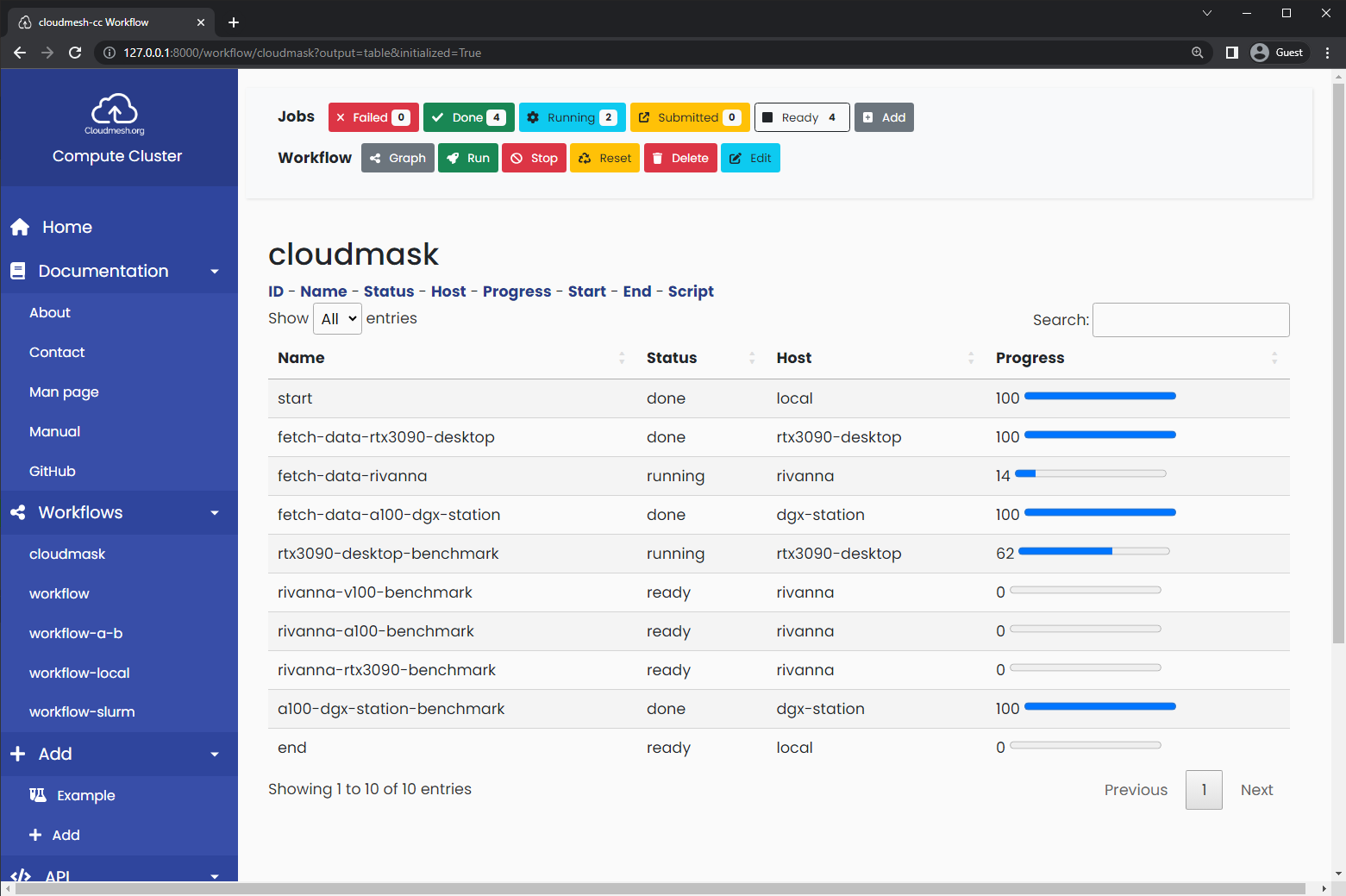}
\end{minipage}
\hfill
\begin{minipage}[t]{1.0\columnwidth}
  \includegraphics[width=1.0\columnwidth]{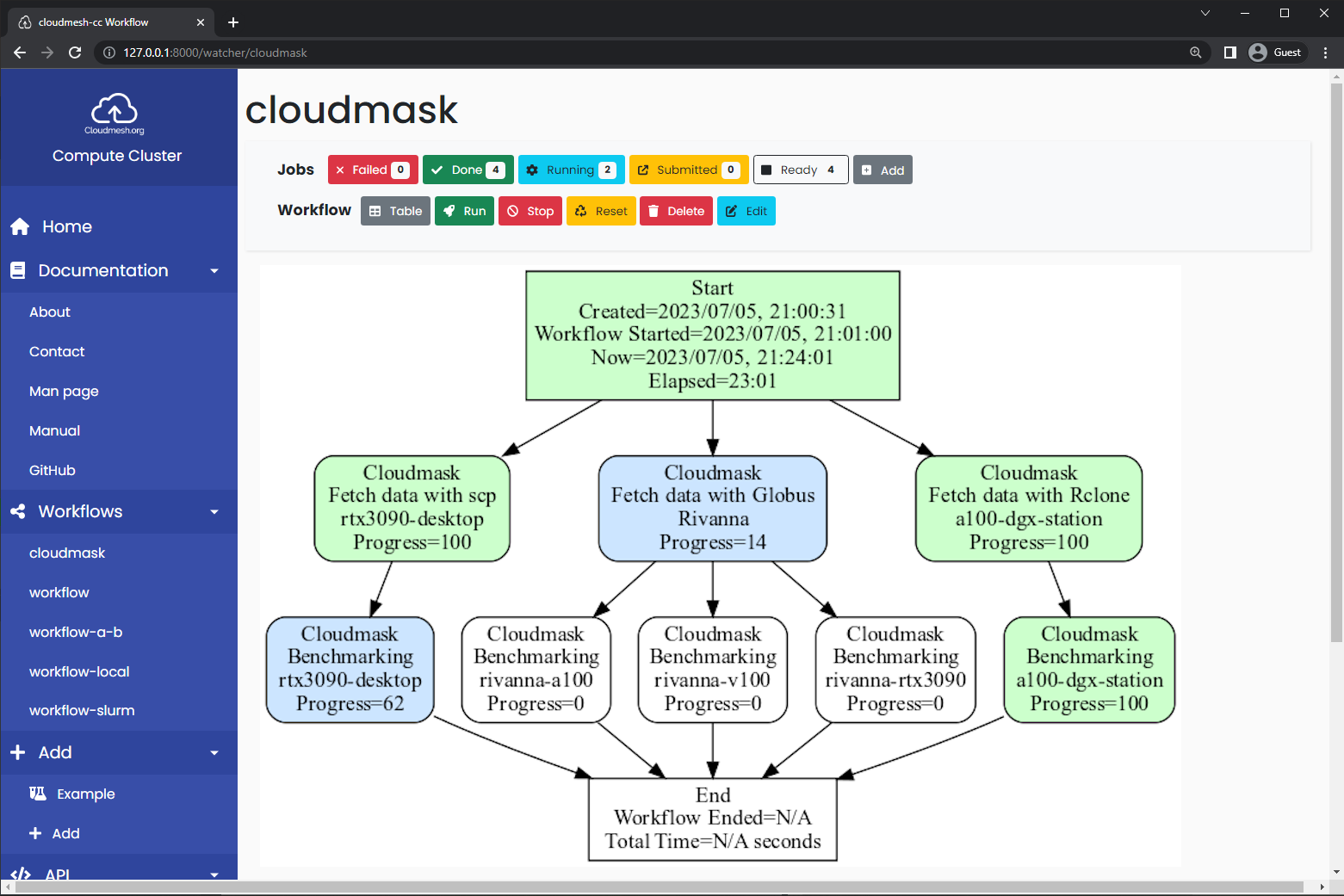}
\end{minipage}
 \caption{Table and Graph view of CC experiment}
\label{fig:gui-view}
\vspace{-0.5cm}
\end{figure*}

\begin{figure}[htb]
    \centering\includegraphics[width=1.0\columnwidth]{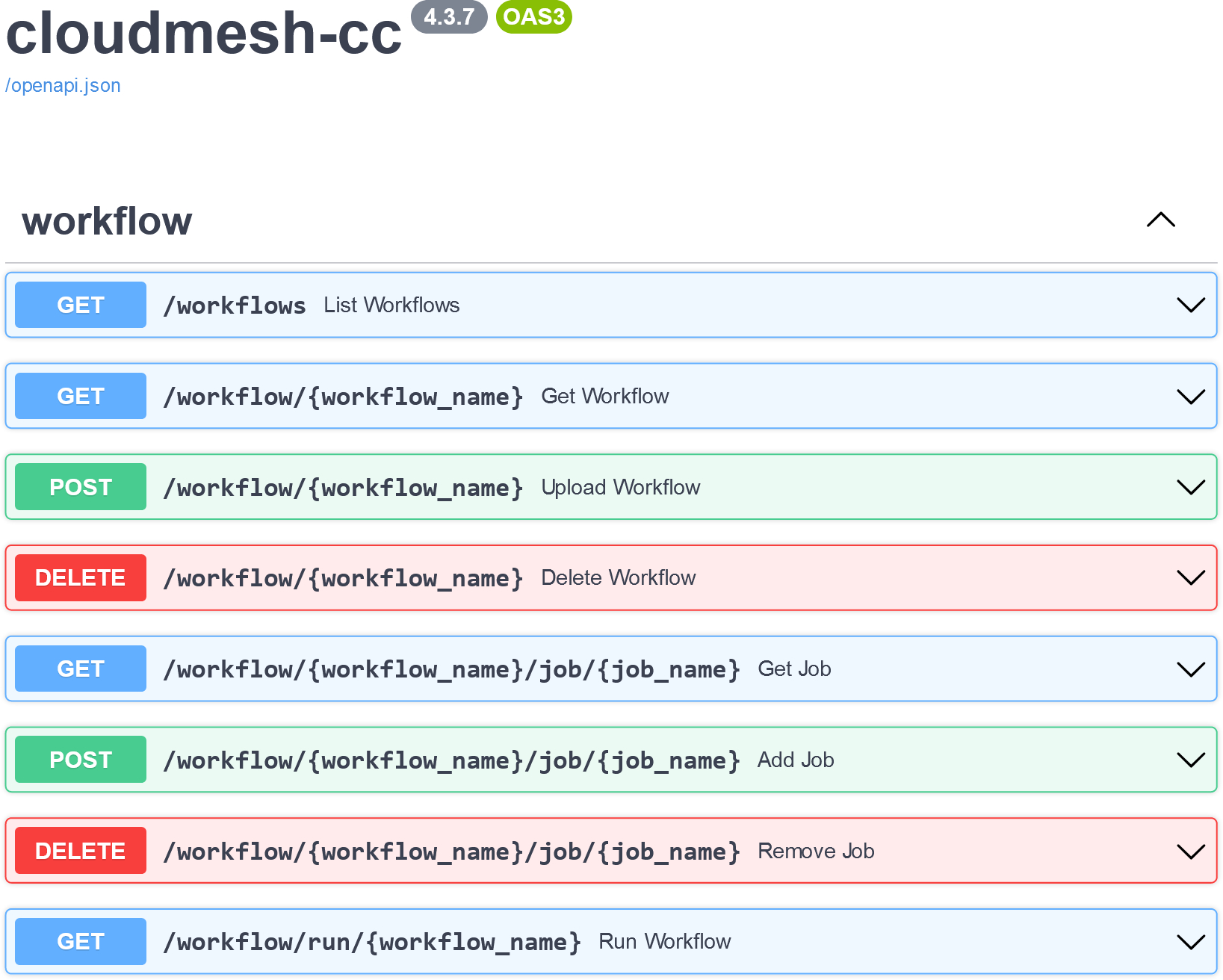}
      
    \caption{OpenAPI workflow interfaces.}
    \label{fig:cc-3}
\end{figure}

Figure \ref{fig:gui-view} shows a prototype  example graphical view of the status through a  web browser-based interface that renders a workflow in either table or graph format. Another, feature is to utilize the Cloudmesh OpenAPI generation which is utilized to produce 
Figure \ref{fig:cc-3}. Through OpenAPI we can integrate this functionality also into REST services.

\begin{figure}[htb]
\vspace{-0.4cm} 
\begin{lstlisting}[breaklines=true,basicstyle=\small]
(*\bfseries workflow:*)
  (*\bfseries nodes:*)
    start:
       name: start
    fetch-data:
       name: fetch-data
       user: gregor
       host: localhost
       status: ready
       label: '{name}\nprogress={progress}'
       script: fetch-data.sh
    compute:
       name: compute
       user: gregor
       host: localhost
       status: ready
       label: '{name}\nprogress={progress}'
       script: compute.sh
    analyze:
      name: analyze
      user: gregor
      host: localhost
      status: ready
      label: '{name}\nprogress={progress}'
      script: analyze.sh
    end:
       name: end
  (*\bfseries dependencies:*)
    - start,fetch-data,compute,analyze,end
\end{lstlisting}
\vspace{-0.4cm}
\caption{Cloudmesh Experiment Specification Example.}\label{fig:workflow-example}
\label{fig:yaml-file}
\end{figure}

\subsubsection{Experiment Executor, a Cloudmesh Plugin}
\label{sec:workflow-ee}

The Cloudmesh Experiment Executor (EE) \citep{las-frontiers-edu} allows the execution of experiments described by experiment parameters. It includes two kinds of parameters. In traditional machine learning workflows and benchmarks, hyperparameter tuning and configuration are key elements in assessing and optimizing the performance of models. However, scaling hyperparameters for highly parallel execution with heterogeneous hardware is complex. EE is used to generate many parameter combinations in a gridsearch (an exhaustive set of trials for every combination of the provided hyper-parameter values). Besides hyperparameters, EE also allows the specification of resource-specific parameters determining hardware and even software properties when an experiment is executed. The architecture of the EE framework is depicted in Figure \ref{fig:cc-2}B.

EE experiments can utilize various queuing systems such as SLURM, and LSF, but also sequential and parallel SSH jobs. The order of scheduling the tasks generated by EE could be customized. Besides static gridsearches, EE can also leverage dynamic functions and introduce through them dynamically changing searches at runtime.

As a result, the output structure of the experiment includes the hyperparameter values, providing a unique identifier for each experiment, the results from different computing systems can be merged into the overall combined results. Thus, EE supports the creation of coordinated results while allowing the generation of cooperating, selective, and distributed result generation. To showcase the simplicity of integrating iterative experiments  \citep{cloudmesh-cc}, we present a specification template that generates experiments tasks for epochs 1, 30, and 60 on A100 and V100 GPUs, repeating it 5 times. For more details on how to utilize EE we refer to the GitHub repository \citep{cloudmesh-ee}.

{\scriptsize
\begin{lstlisting}[breaklines=true,basicstyle=\small]
    application:
        name: cloudmask
    data: "/scratch/{os.USER}/{application.name}"
    experiment:
        epoch: "1,30,60"
        gpu: "a100,v100"
        repeat: "1,2,3,4,5"
\end{lstlisting}
}

This specification template is then used and instantiated via the commandline or shell to produce a template for a specific machine and its queuing system policies. The result is formulated as templated batch script that gets executed through EE while filling out runtime parameters automatically based on  prior defined user specifications. As the templates are in principle user-independent, they can also be executed via different user accounts and even organizations if desired. Through the integration with EE one can also use an energy monitor to create energy traces at runtime. As the overall experiment can be designed in independent chunks representing a variety of independent parameter searches it is possible to create first experiments with a smaller runtime in order to estimate the impact larger experiments have on the runtime.  

While practically working with the system, we observed that students (as part of research experiences) not using our experiment executor spend a significant amount (weeks-months) of a semester on setting up a benchmark and replicating only a fraction of the functionality provided by the EE. However, we tested the system out while other students used EE and we observed that the applications for which a template and configuration file has been designed reduced the on-ramp time to less than a day. Not only that, instead of needing a team including graduate students, the work could be performed by a single undergraduate student.

EE differentiates itself from other approaches as gridsearches can trivially be formulated either as API calls or as displayed here through an easy-to-understand YAML file. Through this mechanism, thousands of independent experiments can be run as part of a large-scale experiment workflow.

EE takes two configuration files. The first is a YAML file that includes all parameters used by the benchmark including an experiment section that defines the Cartesian product (or dynamic changing values in case a function is defined and used). The second is a SLURM template. From these files, it will create through commandline, for example, SLURM scripts, while at the same time 

\begin{enumerate}
  \item using a unique directory for the experiment,
  \item taking a parameter set from the Cartesian product of the experiment parameters,
  \item creating from a batch job template an instantiation of the template while replacing all variables from the configuration file and replacing the specific experiment parameters,
  \item creating an instantiation of the configuration file while replacing all experiment parameters with the one for the current experiment.
\end{enumerate}

An example of a configuration file \verb|config.yaml| where we iterate over epochs, GPUs, and repeat it 5 times is shown next, while more elaborate examples can be found in the manual:

\begin{lstlisting}[language=sh,basicstyle=\small]
    #!/bin/bash

    #SBATCH --job-name={experiment.repeat}-{application.name}
    #SBATCH --nodes=1
    #SBATCH --gres=gpu:{experiment.gpu}:1
    #SBATCH --time=02:00:00
    #SBATCH --mem=64G
    #SBATCH -o {experiment.gpu}-{application.name/{experiment.repeat}-%j.out
    #SBATCH -o {experiment.gpu}-{application.name}/{experiment.repeat}-%j.err
    #SBATCH --partition=bii-gpu
    #SBATCH --account=bii_dsc_community

    echo {cloudmesh.vesrion}
    echo {os.name}
    export USER_SCRATCH=/scratch/$USER
    cd USER_SCRATCH
    mkdir -p $USER_SCRATCH/{experiment.gpu}-{application.name}/%j.out
    nvidia-smi

    cms gpu watch --gpu=0 --delay=0.5 --dense > outputs/gpu0.log &

    python earthquake.py --config config.yaml

    seff $SLURM_JOB_D
\end{lstlisting}

This example is needed to point out some additional features of our specification language. This includes variable names using dot notations which define a context. These variables obtain their values either from our YAML specification file as shown earlier, but also operating system variables (starting with \lstinline|os.|) , and variables stored in the cloudmesh database (starting with \lstinline|cloudmesh.|). 

Users now can use common variables value expansion to create experiments and integrate system specific values. Examples include graphics processing units, memory, file systems used, versions of Python, versions of TensorFlow, epochs, learning rate, and many other important parameters that can influence the benchmark.  More details can be found in \citep{las-2022-templated}. One of the implicit features is that policies that may be restrictive to run such long running jobs as a single executable or job submission, cloudmesh creates separate jobs out of them that are run independently. This has for our usecases shown that we were able to run them on or available infrastructure. The same mechanism can be applied to in SmartSim.

\subsubsection{Cloudmesh Timers}
\label{sec:monitoring}

We have observed that many of our students spend too much time augmenting their code with timers in an uncoordinated fashion. Therefore, we have provided a simple Python library that can be installed with pip so students can augment their codes in a most simple fashion. The important part is that EE and CC also use their library and thus an experiment has a consistent reporting function throughout.  Furthermore, we have made an extension so that it also directly returns in addition to the Cloudmesh timer format, timers used by MLCommons in mllog format. The advantage is that the timers in Cloudmesh format are humanly readable, while when also exporting mllog such benchmarks fulfill the MLCommons logging conventions. Besides these formats, the StopWatch also produces summary tables in txt, csv, HTML, JSON, and YAML. Furthermore, this includes the automatically detected specification of operating systems parameters, which comes in handy when an experiment is to be replicated or further analyzed.




In addition, we developed a simple command line tool to augment batch scripts to monitor the GPU performance characteristics such as energy, temperature, and other parameters \citep{cloudmesh-gpu}.



Monitoring time and system GPU information can provide considerable insights into the application's performance characteristics. Hence, it is significant for planning a time-effective schedule for parameters while running a subset of planned experiments.

\subsection{Provisioning Cloud Clusters with the Cloudmesh Plugin}
\label{sec:cloudcluster}

We report next on a new development we started over the last several months to contrast with the primarily on-premise focused cases described previously. In particular, this focuses on an increasingly common cost analysis: whether an HPC application would be more cost-effective to run in the cloud instead of an on-premise machine. We are providing some starting points for this discussion.

A new plugin to Cloudmesh extends the high-level API and abstractions to provision a High Performance Computing  (HPC) cluster in the cloud. It currently utilizes AWS Parallel Computing Service (PCS) ~\citep{awspcs:online} as the supported computational backend. The plugin simplifies the infrastructure deployment of an HPC by creating an abstraction layer for the end user who intends to run their experiments on HPC resources in a cloud. It simplifies access, deployment, and management of the infrastructure in the cloud. Unfortunately, deploying a cluster with PCS is still too complex for many users. Hence, we developed a sophisticated but easy-to-use plugin for Cloudmesh called {\em create}, which as its name suggests, creates a cloud cluster. It provides the necessary automation to deploy clusters within minutes from the command line or via an API (a fully functional HPC cluster using AWS can be built in about 6 minutes).  

In AWS PCS, a cluster consists of compute nodes configured by EC2 instances. The cluster is managed by a controller that provides batch processing to its users via SLURM (Simple Linux Utility for Resource Management).  The total cost ($H$) of running a PCS cluster in AWS can be calculated using Equation \ref{eq:aws},

\begin{equation}
 H = C + (N * (M + I)) \label{eq:aws}
\end{equation}

where 
$C$ is the controller fee per hour, 
$N$ is the number of nodes, 
$M$ is a fixed cost called the node management fee which is charged for each node per hour, and
$I$ is the node cost per hour for each node stemming from the instance type used.

The advantage of using a cloud is motivated by scaling hardware features to the exact count necessary by the application's computational needs with on-demand timing. Through auto-scaling based on workload, users can define a minimum and maximum number of nodes, as well as adding and deleting nodes on-demand when they are not needed.

In many cases, researchers have very limited budgets to conduct experiments. For this reason, researchers tend to utilize freely available HPC clusters through project proposals in nationally or university-funded clusters. Cloud HPC provides an alternative cost structure following the pay-as-you-go model with the hope that the cost of executing the experiment is reasonably cheap.  Some suggest  \citep{munhoz_performance_2023} that cloud-based HPC clusters can be a more economical option for many users, particularly for those with fluctuating workloads or limited budgets. 

A detailed pricing structure for PCS can be found at \citep{www-aws-pricing}. It is interesting to compare the cost of running a benchmark on an on-premise cluster versus running a benchmark on PCS which we highlight on some examples focusing on GPU usage. To obtain a better understanding, we have mapped out the pricing based on typical MLCommons benchmark with A100 GPUs in Table \ref{tab:aws-1}. Each node has 8 A100 GPUs. We find that the cost per GPU per hour is about \$4.18.

To further quantify the cost, we compared benchmarks for DeepCAM and CosmoFlow run on a cluster with Nvidia GPUs \citep{mlperf-nvidia-benchmark} as reported by MLCommons and show the price for running them on similar clusters in the cloud with A100 GPUs in Table \ref{tab:aws-1}. From the MLPerf Training HPC Benchmarks \citep{mlcommonsBenchmarkMLPerf} and training policies documented in the GitHub repository \citep{mlperftrainingpolicies}, we are able to derive the cluster type as well as the runtime for repeated experiments.

\begin{table*}[h]
\caption{AWS PCS Pricing Information, examples for $n$ GPU (p4d.24xlarge) Nodes value as of Mar. 2025.
p4d.24xlarge offers 96 CPUs, 1TB memory, and 8 NVIDIA A100-SXM4-80GB GPUs per node.}
\label{tab:aws-1}
\smallskip
\resizebox{\textwidth}{!}{%
\centering
\begin{tabular}{|r||r|r|r|r|r|r|r|}
 \hline
 \makecell[t]{\bf Slurm \\ \bf Controller \\\bf Size} &
 \makecell[t]{\bf Number of \\ \bf AWS Nodes } & 
 \makecell[t]{\bf Number of \\ \bf GPUs} & 
 \makecell[t]{\bf Node cost \\ \bf per node \\ \bf hour } &  
 \makecell[t]{\bf Controller \\ \bf Fee per hour} & 
 \makecell{\bf Node \\ \bf Management \\ \bf Fee per hour} & 
 \makecell[t]{\bf Total Cost  \\ \bf per hour \\ } &
 \makecell{ \bf Cost per GPU\\ \bf per hour \\ \bf (approx.)}
 \\
   & \makecell{$N$} & \makecell{~} & \makecell{$I$} & \makecell{$C$} & \makecell{$M$} & \makecell{$H = C + (N * (M + I))$} & \\
 \hline
 Small   & 64 & 512     & \$32.77 & \$0.60  & \$0.67 & \$2,140  & \$4.18 \\
 Medium  & 128 & 1024   & \$32.77 & \$3.34  & \$0.67 & \$4,283  & \$4.18 \\
 Large   & 256 & 2048   & \$32.77 & \$6.71  & \$0.67 & \$8,567  & \$4.18 \\
 \hline
\end{tabular}
}

\caption{MLPerf Benchmarking Results from MLCommons using NVIDIA A100-SXM4-80GB (400W) GPU model \citep{mlcommonsBenchmarkMLPerf}}
\smallskip
\label{tab:aws-2}
\resizebox{\textwidth}{!}{%
\centering
\begin{tabular}{|r||r|r|r|r|r|r|r|}
 \hline
 \makecell[t]{\bf System\_Configuration \\ \bf Name} & 
 \makecell[t]{\bf Total \\ \bf Nodes} & 
 \makecell[t]{\bf Total \\ \bf GPUs} & 
 \makecell[t]{\bf Benchmark} & 
 \makecell[t]{\bf Average Duration \\ \bf of execution \\ \bf in Minutes} &
  \makecell[t]{\bf Number of \\ \bf Repeated\\ \bf Experiments} &
 \makecell[t]{\bf Total Duration \\ \bf of execution \\ \bf in Minutes} &  
 \makecell[t]{\bf Projected \\ \bf Total Cost \\ \bf on AWS for all\\ \bf Experiments} 
 \\
    
 \hline
 dgxa100\_n64\_ngc21.09\_pytorch & 64  & 512  & DeepCAM & 2.65 &  5 & 13.25 & \$473 \\
 dgxa100\_n128\_ngc21.09\_mxnet  & 128 & 1024  & CosmoFlow & 8.04 &  10 & 80.4 & \$5,740\\
 dgxa100\_n256\_ngc21.09\_pytorch  & 256  & 2048  & DeepCAM & 1.67 &  5 & 8.35 & \$1,192\\ 
 \hline
\end{tabular}
}

\bigskip
\end{table*}

Although, Table \ref{tab:aws-1} provided prices per hour (AWS charges by minute), AWS offers also a long term charge for 1 year for \$5,098, or for 3 years for   
 \$3,140 per hour. 
 
As cloudmesh can easily provision such clusters, the expected costs must be communicated to the user as otherwise unexpected costs may occur. This can be achieved by adding in future an interactive question such as alerting the users of the overall cost, or by defining limits for cost when such a cluster is provisioned.

Next, we list some advantages and disadvantages. Advantages include easy deployment no administrator costs, the Cluster can be updated to utilize new hardware once it is made available by the cloud provider, utilizing spot instances reduces cost, and a cluster is only needed when the demand arises. As for the disadvantages, it is essential to understand the cost model so as to not be charged unexpectedly, although deployment is easy users still have to know more than just the interface to a queuing system (this is simplified by cloudmesh), community on-premise clusters may be ``free'' once the scientific need has been approved, and community clusters have dedicated support staff helping to port scientific applications.

In this paper, we have not answered the question if a cloud cluster is cheaper than an on-premise cluster. This we hope to address in a follow up paper. The included cost outline gives a pretty good understanding of what a typical cost arises when we identify the needs for selected MLCommons benchmarks.

However, we note that the cost of computation is often not taken into consideration when running applications like benchmarks for traditional HPC platforms because the user seldom bears the cost directly. In contrast, the need to minimize the cost associated with developing and executing experiments in the cloud does add an additional requirement to what might be expected of experiment executors.

\subsection{Comparing Cloudmesh and SmartSim against technical requirements and features}
\label{sec:compare}

Table \ref{tab:features} shows a number of technical features based on the development of SmartSim and Cloudmesh Experiment Executor and Compute Coordinator have been identified as useful and provide overlap between the systems. The purpose is not to perform bean-counting to identify which system is better than the other, but to show the similarities and differences that arise from shared philosophies but different user-driven requirements  as shown in Table \ref{tab:req-features}. One thing that we like however to project that if we were to develop a new system it ought to combine also the unique aspects of each of the systems.

\subsection{Similarities between Cloudmesh and SmartSim}

Both solutions abstract the complexities of scheduling systems of the target platform from the user. This was identified early on because an individual user has very little influence on what platform is available to them. Often, the user simply desires to specify a node count, number of tasks, and distribution of tasks. In both Cloudmesh and SmartSim, the users define that at a higher level and each system generates the batch request and launching commands on execution. This helps simplify the porting of a workflow to a new machine because only minimal modification is required and the user does not tend to need any deep understanding of the particular semantics of a new workflow.

Both systems are built in Python, provide an extensive API, and significant documentation. This is a direct result of knowing that the target users come from a domain science perspective. Python serves both developer and domain users well and provides a common language and tooling. The need for an expressive API additionally maps onto a level of expressiveness and abstraction that allows the user to create their own custom workflows. Lastly, while good documentation is a fundament of software engineering, having a wide variety of examples through templates on many different platforms has helped both user communities rapidly prototype their own workflows. This is particularly true in the emerging HPC/AI motifs \citep{brewer2024ai} where the expression of each components unique, but the overall structures are similar.

The configuration and parameterization of workflow components are a primary concern in each system. Regardless of whether they are defined in a YAML file (Cloudmesh) or by a user-provided way through a program (SmartSim), the key piece is the encapsulation of the parameters in a single location. This provides a single source of truth for the parameters that define the workflow and a way for users to modify it for their own purposes. 

The licensing of both systems align with permissive open-source licenses. This is key, particularly for science users funded by public funds. Important to note is that neither follows a freemium type model, where the open-source library may have a restricted featureset or limitations based-on the scale of the workflow and the utilization of undelaying respurces. Again, this a key requirement due to the difficulties in a priori knowing what scales are needed for scientific advancement, logistic challenges when administering licenses for individual users on shared platforms, and barriers to distribution and reproducibility that paid licenses introduce.

\subsection{Contrasting Cloudmesh and SmartSim}

Cloudmesh and SmartSim differ in how the primary functionality are extended. Cloudmesh system provides a very sophisticated but easy-to-use plugin system allowing extensibility and integration of new functionality through add-on packages that can easily be installed with pip. Furthermore, it includes an extension to allow new components to not only be integrated via command line, but it contains an extensible command shell. Internally all components are written as Python code exposed through APIs bound into a single namespace. SmartSim prefers to extend functionality by incorporating new features into a main library. It offers a number of ways, especially during the execution of a workflow, where users can inject their own code. In general however, SmartSim is deliberately more opinionated to help streamline how users can interact with the library direclty and avoid user pitfalls.

Experiments/Ensembles can be formulated in both systems as YAML files and not only pure Python code utilizing Python language constructs and integrating with the API calls from each system. Through them, they can conduct customizable grid searches as needed in many AI applications. Moreover, the Cloudmesh experiment YAML file has the ability to implicitly use multi-valued variables instead of just using lists. This also includes the integration of Python functions that can be executed at the time of creation of the experiments. Such functions can be dynamic.

Cloudmesh has a strong emphasis on allowing for submission from a machine to another with support for SSH-based remote execution. In particular, because some platforms require a VPN for security, Cloudmesh provides a plugin for split-VPN that can use multiple VPNs and, based on the target organization in which the resource is hosted, automatically chooses the correct one. Thus, workflows across different organizational boundaries can be defined as shown in \citep{las-frontiers-edu}. Both systems manage credentials through existing frameworks separately from the experiment workflow specifications. It could be possible to integrate other libraries such as CILogon to enable easier integration with NSF ACCESS. However, this was not our current focus as we worked predominantly with VPN protected and DOE resources allowing us to use SSH. 

As previously mentioned, the open-source nature of the libraries are key, however Cloudmesh focuses more on ensuring that its dependencies also are not restricted and are under an open-source license. SmartSim still has dependencies on Redis and RedisAI which must be compiled from source due to its license restrictions. This has added considerable complexity to the distribution process. Additionally the recent move of Redis itself to a more restrictive license represents an external risk that affects a key component of SmartSim.

One activity that Cloudmesh has recently started is not just to look into using virtual machines in various cloud providers, but also looking into provisioning and utilizing clusters based on HPC and Kubernetes as provided by them. Together these features will provide an even more powerful system extending the capabilities of both. 

To recap, we like to refer to our two tables. We summarize the technical similarities and differences in \ref{tab:features}. As we introduced a number of requirements in Section \ref{sec:requirements} we have added the Table \ref{tab:req-features} that lists which of them are being fulfilled by the systems.

\begin{table*}[htbp]
\caption{Comparison between the workflow-related features of SmartSim and Cloudmesh}
\label{tab:features}
\centering
\resizebox{\textwidth}{!}{%
\begin{tabular}{|l|l|l|}
\hline
\bf Feature &  \makecell{\bf Cloudmesh \\ \bf Experiment Executor and \\\bf Compute Coordinator} &  \bf SmartSim \\
\hline
\multicolumn{3}{|l|}{\bf\em Scheduler}\\
\hline
Queue & SLURM, LSF, SSH, others possible &  SLURM, PBS, LSF, SGE \\
Batch Submission & \YES & \YES \\
Within Allocation & \YES & \YES \\
DAGs  & \YES  & \YES \\
Inferencing Capabilities & \NO & \YES \\
In-memory data exchange & \NO & \YES \\
Experiment/Ensemble & \YES & \YES \\
\hline
\multicolumn{3}{|l|}{\bf\em Interface}\\
\hline
Python API & \YES & \YES \\
Command line & \YES & \NO \\
Command shell & \YES & \NO \\
GUI  & (\YES) & \YES \\
\hline
\multicolumn{3}{|l|}{\bf\em Parameters}\\
\hline
Native YAML configuration & \YES & \NO  \\
multi-value YAML & \YES & \NO \\
evaluative YAML ($f(\vec{x}$)) & \YES & \NO \\
Gridsearch & \YES & \YES \\
Customizable Strategies & \YES & \YES \\
\hline
\multicolumn{3}{|l|}{\bf\em Federation}\\
\hline
SSH & \YES & \NO \\
Split VPN support & \YES & \NO \\
Build in parallel multi resource experiments & \YES & \NO \\
Combine results by multiple users & \YES & \NO \\
Combine results from multiple resources & \YES & \NO \\
\hline
\multicolumn{3}{|l|}{\bf\em Expandable}\\
\hline
Plugin Manager & \YES & \NO \\
\hline
\multicolumn{3}{|l|}{\bf\em Distribution}\\
\hline
only pip & \YES & Without Redis \\
pip with compile & N/A & With Redis \\
Container & \YES & \YES \\
Singularity & \YES & \YES \\
Docker & \YES & \YES \\
Licence  & Apache 2.0 & BSD-2-Clause license \\
\hline
\multicolumn{3}{|l|}{\bf\em Other Deployments}\\
\hline
AWS Parallel Cluster & \YES & \NO \\ 
AWS Kubernetes  & in progress & \NO \\
\hline
\end{tabular}
}
\end{table*}


\begin{table*}[htbp]
\caption{Requirements addressed by the two experiment executors.}
\label{tab:req-features}
\centering
\resizebox{\textwidth}{!}{%
\begin{tabular}{|p{6cm}|l|l|}
\hline
Requirements & Cloudmesh& SmartSim \\
\hline
\hline
\multicolumn{3}{|l|}{\bf\em Implications from Section 2.1 Compute Systems Requirements}\\
\hline
\hline
Hardware at wide scale &  used at DOE, NSF, university, private   & used at DOE, NSF, university, private    \\ 
\hline
Integration of GPUs &   \YES  & \YES   \\ 
\hline
Interface to workload managers &  SLURM, LSF, SSH, others possible    &  SLURM, PBS, LSF, SGE  \\ 
\hline
Simple uniform access through shells &  \YES  &    \\
\hline
Minimal support for access via authentication and authorization &  \YES   & \YES   \\ 
\hline
Minimal support for virtualization in the cloud & \YES    &    \\ 
\hline
Batch access and direct access &  \YES   &  \YES  \\ \hline
Cloud HPC resources &   \YES  &    \\ \hline
Container and virtual machine support &  \YES   & singularity   \\ \hline
\hline
\multicolumn{3}{|l|}{\bf\em Implications from Section 2.2 User Requirements}\\ \hline
\hline
Wide Variety of Users &  developer, user  & developer    \\ \hline
Ease of Use (trhough) &  Python, OpenAPI, scripts, templates, YAML  &  Python  \\ \hline
Experiment Automation &  \YES   &  \YES  \\ \hline
Experiment Reporting &   filesystem, integratable into database  &  filesystem \\ \hline
Portability &  \YES   & \YES   \\ \hline
Cost Considerations &  prototype Plugin  &    \\ \hline
Benchmark Carpentry &  Manual, examplesin git repo   &  Manual, examples in git repo \\ \hline
\hline
\multicolumn{3}{|l|}{\bf\em Implications from Sections 2.3 to 2.7}\\ \hline
\hline
Workflow specification &  YAML, Python, scripts   & Python   \\ \hline
Workflow Templates &  \YES   &  \YES  \\ \hline
Template Repositories & self-managed    &  self-managed  \\ \hline
Experiment Reporting &  self-managed   & self-managed   \\ \hline
Runtime support for dynamic, Cyclic, pipeline, and DAG Executions &     \YES & \YES \\ \hline
Runtime Batch Queuing Integration &  \YES   & \YES   \\ \hline
Authentication and Authorization &  SSH, split-vpn, extensible   & focus on single hardware executions  \\ \hline
Data Management & filesystem, prototype cloudstorage interfaces   & filestystem, Redis   \\ \hline
Licensing & Apache 2.0. replaced mongodb due to license 
issues    & Redis, will likely replace redis due to license change \\ \hline
\end{tabular}
}
\end{table*}


\section{Use Cases}
\label{sec:use_cases}

\subsection{Open Surrogate Model Inference (OSMI) Benchmark}

Most AI workflows on HPC can be categorized into six different execution motifs: Steering, Multistage Pipeline, Inverse Design, Digital Replica, Distributed Models, and Adaptive Training \cite{brewer2024ai}. One component that shows up across multiple motifs is machine-learned surrogate models. Such models typically are used in hybrid ModSim/AI workflows, where traditional simulations are used for a large part of the workflow, and then particular aspects of the simulation, such as a turbulence or radiation model, are replaced by digital surrogates, e.g., \cite{partee2022using, martinez2022roam, bhushan2023assessment}. Because of the challenges of integrating the simulations with the AI model in a highly scalable manner, developing a benchmark was necessary to assess the performance of various configurations. Initial developments of a surrogate model benchmark, called {\em OsmiBench}, were studied by Brewer et al. \cite{brewer2021production}. The studies showed that using a separate load balancer on each compute node, which round-robins the inference requests across multiple GPUs on the node, and also using the maximum batch size that the GPU memory allows yields optimal inference performance. This study was followed by a secondary investigation by Boyer et al. \cite{boyer2022scalable}, which investigated performance implications of the full coupling between the surrogate inference and the simulation code, and showed that using a concurrency level of two batch inference requests was optimal. 

\begin{figure}[htb]
    \centering
    \includegraphics[width=\linewidth]{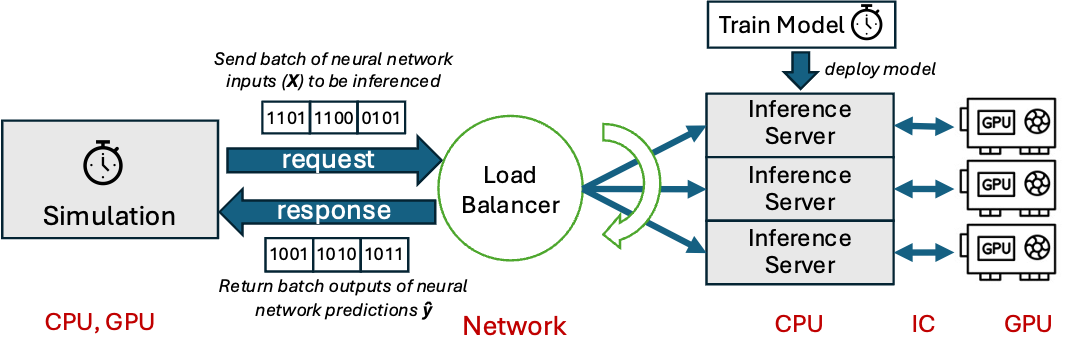}
    \caption{Architecture of OSMI benchmark.}
    \label{fig:osmi}
\end{figure}

The Open Surrogate Model Inference (OSMI) benchmark was developed as an open-source community benchmark founded upon these principles. The architecture of OSMI is shown in Fig. \ref{fig:osmi}. The benchmark supports either TensorFlow or SmartSim/PyTorch-based frameworks as shown in Table \ref{tab:osmi}. Inference requests are initiated from within the simulation using a client API call (e.g., SmartRedis or gRPC API), the requests are then sent to a load balancer (e.g., HAProxy), which distributes the requests in a round-robin fashion to multiple inference servers, each bound to a single GPU. Benchmark timings are able to be measured at multiple places in the architecture, but the primary measurement of interest is how long it takes from the time an inference request is initiated from the simulation until the response is returned back to it. As opposed to chip-level benchmarks such as MLPerf \cite{reddi2020mlperf}, OSMI is able the measure system-level performance, which includes the performance of the CPU, GPU, network, and interconnect (IC), giving a holistic performance representation of the system. This same approach was used to benchmark a wide range of HPC systems, revealing significant performance differences between seemingly similar machines, often due to factors such as different interconnect performance \cite{brewer2020inference}.

Work is in progress to incorporate this type of in-the-loop inference into MLCommons. The initial effort implements the configuration and execution of the benchmark with both SmartSim and Cloudmesh. As expected, because the distilled requirements for OSMI are fulfilled by both solutions, no changes needed to be made to either package to implement. Notably, the key requirements were the ability to interface with the workload manager, configure variations of the benchmark, and execute the benchmark on HPC resources. The portability is being tested by execution on both Department of Energy HPC platforms and academic clusters.

\begin{table}[htb]
\centering
\renewcommand{\arraystretch}{1.5}
\begin{tabular}{llll}
\hline
\textbf{AI framework} & \textbf{Inference server} & \textbf{Client API} & \textbf{Protocol} \\ \hline
TensorFlow & TF Serving  & TF Serving API  & gRPC  \\
PyTorch    & RedisAI     & SmartRedis      & RESP  \\ \hline
\end{tabular}
\caption{OSMI-supported AI frameworks.}
\label{tab:osmi}
\end{table}

\begin{BOX}[Requirements implied by the OSMI benchmark.]
The immediate requirements we gather from such a complex experiment workflow are (a) the interplay between large computational components executed on GPUs that are interwoven with the overall execution in an iterative simulation executed on multiple servers (b) the scheduling of thousands of independent calculations executed on the GPUs while hyperparameters and data sets need to be feed to the executing GPUs, and (c) the gathering of the results in a mathematically sound dataset, and (d) the execution of such a workflow on different architectures to showcase the wide variety of performance differences while configuring the workflow specifically for the various target machines.
\end{BOX}

\subsection{Conditional and branching workflows}

Other types of workflows involving loosely coupled workflow components are becoming more popular in scientific applications. They involve branches or criteria-based loops, thus violating the fundamental assumptions of a DAG. Parameter estimation is an example of such a workflow because the number of iterations is not generally known a priori. SmartSim's ability to aggregate output from ensembles and generate new ensembles was applied to an OpenFOAM case \citep{Maric2024OpenFOAM}. In that example, an optimizer at every iteration generates candidate parameter sets which are then used to launch new cases. The output from those cases is then ingested by the optimizer for the next iteration. As is typical of optimization problems, this cycle ends when either the loss function converges, stalls, or reaches a certain number of optimizations.

Reinforcement learning is another type of workflow that involves non-cyclic and potentially branching workflow execution. In \citep{Font2024}, an ML model is used to control the behavior of a turbulent flow surrounding a rising bubble. The ML model is able to modify the flow by controlling actuators.  The RL model deploys multiple agents in various environments to explore and refine the optimal actions. Similarly, \citep{kurz2023deep} train a surrogate model of turbulence using an RL framework. The agent predicts an eddy viscosity where the RL model is incentivized to match the energy spectra in a turbulence-resolving model. As in \citep{Font2024}, the scientific simulation is used as an environment to evaluate the agents' strategies. In both these cases, the need to continue to iterate and test requires dynamic configuration and execution where the number of cycles is not known a priori.

In all of these examples, the workflow is not easily representable by a DAG, but rather requires the evaluation of logic as a fundamental component of the workflow. Additionally, aspects of these cases map onto a producer-consumer paradigm for exchanging data, facillitated by a central datastore to store intermediate results.

\subsection{Additional Cloudmesh Application Usecases}

Besides exploring the usage of EE for OSMI, we have tested CC while running various applications including MNIST, Multilayer Perceptron, LSTM (Long short-term memory), Auto-Encoder, Convolutional, and Recurrent Neural Networks, Distributed Training, and PyTorch training.  A much larger application using earthquake prediction has also been used. Results of using it outside of the earthquake code are available in \cite{las-2023-escience}.

%
%
%
%
%
%

\section{Related Work}
\label{sec:related}

In recent years, significant progress has been made in the development and standardization of community-driven workflow benchmarks for High Performance Computing (HPC). This section reviews the key related work in this area, structured into several key topics. First, we explore the evolution of workflow management systems (WMS), which form the backbone of benchmark execution and automation in HPC environments. Next, we categorize and discuss various types of workflows, ranging from traditional workflows to newer paradigms that are still emerging, including HPC-specific workflows and those designed for hybrid HPC/AI applications, such as large language models (LLMs). Finally, we examine workflow benchmarks themselves, highlighting efforts to establish standardized metrics and methodologies for performance evaluation across different systems. By reviewing these areas, we aim to provide a comprehensive understanding of the current landscape and identify gaps for future development in community benchmark workflows for HPC.

\subsection{Workflow Management Systems}

A workflow management system (WMS) is an essential tool for automating and orchestrating complex computational processes, particularly in HPC environments. There are more than 360 known workflow management systems, each tailored to meet specific needs within diverse application domains, and the list (see \citep{workflow-list}) is growing. These systems vary widely in design and functionality, reflecting the unique requirements of the workflows they support. The choice of WMS is heavily influenced by the characteristics of the workflows themselves, such as scale, complexity, computational resources, and the type of tasks being executed.

In HPC, workflows can range from traditional batch processing jobs to sophisticated, data-intensive simulations and AI-driven applications. As workflows differ significantly across domains, so too do their management systems, which are designed with varying levels of parallelism, fault tolerance, scheduling algorithms, and scalability to address these needs. While some WMS are highly specialized for certain kinds of scientific computing or data analysis, others are built for more general-purpose or hybrid computing environments, reflecting the diversity of computational tasks encountered in modern HPC research. Thus, rather than adhering to a single standardized architecture, WMS design is deeply influenced by the specific demands and constraints of the workflows they support.

\subsection{Workflows}

Despite the critical role they play in organizing, automating, and optimizing complex scientific computations across various domains, computational workflows suffer from the lack of community consensus regarding what specifically defines a {\em\bf workflow} \citep{ferreira_da_silva2022, wilkinson2025}. Workflows can be designed to handle a broad range of tasks, from traditional batch jobs to emerging data-driven and AI-centric processes. This section discusses three categories of workflows: traditional workflows, {\em\bf emerging} workflows, and HPC/AI workflows, each of which has unique characteristics and requirements.

\subsubsection{Traditional Workflows}

Traditional computational workflows are well-established, often utilizing systems such as Pegasus \citep{www-pegasus} and Kepler \citep{www-kepler}. These systems have been widely used in scientific computing and high-throughput environments, where workflows typically involve a series of interdependent tasks that must be executed in a specified order. These workflows often consist of batch processing jobs that execute simulations, analyses, or data processing pipelines, where the main challenge is ensuring the reliability, scalability, and efficient scheduling of tasks across distributed computing resources.


\subsubsection{Emerging Workflows}

Emerging workflows are characterized by their adaptability to modern, dynamic environments, where workflows are not merely static task lists but instead are flexible and configurable based on system or resource availability. Tools like Nextflow \citep{di_tommaso_nextflow_2017}, Parsl 
 \citep{babuji2019}, Globus Compute (formerly known as funcX \citep{chard2020}), and ExaWorks \citep{alsaadi2024} represent the evolution of workflow management in response to new computational paradigms, such as cloud computing, distributed systems, and high-performance heterogeneous environments.

Nextflow, for example, provides a platform for building and running scalable data-driven workflows, integrating seamlessly with cloud and high-performance computing infrastructures. It supports a variety of computational platforms, including Kubernetes and SLURM, and facilitates FAIR workflows \citep{wilkinson2025, wilkinson2022} through integration with WorkflowHub \citep{gustafsson2024} and reproducibility through version-controlled, containerized environments. Parsl, similarly, is designed for parallel and distributed computing, using Python to define workflows and enabling dynamic task scheduling based on resource availability. Globus Compute is a serverless function execution framework that abstracts away many of the complexities of managing compute resources, allowing users to focus on defining tasks rather than infrastructure. ExaWorks is a more recent entrant that emphasizes flexible workflows capable of scaling to exascale HPC systems, handling both data-intensive and compute-intensive tasks efficiently, by combining the strengths of several WMSs into one software development kit.

These emerging systems are typically designed with the flexibility to work in hybrid, heterogeneous environments where workflows must adapt to changing computational resources, from cloud instances to high-performance clusters and supercomputers.

\subsubsection{HPC-AI workflows and LLMs}




The growing convergence of HPC and artificial intelligence (AI) has led to the emergence of specialized workflows that combine traditional computational tasks with modern AI-driven approaches \citep{mcclure2020}. A notable subset of these workflows involves large language models (LLMs), which require highly specialized computational infrastructure and advanced workflow management techniques. HPC/AI workflows often involve stages such as pre-processing large datasets, training machine learning models, and fine-tuning LLMs, all of which demand substantial computational resources, parallelism, and advanced orchestration.

HPC/AI workflows are typically defined in environments such as TensorFlow, PyTorch, and other deep learning frameworks, where the workflow management must ensure efficient data pipeline handling, distributed training, and scalable execution. LLMs, such as GPT \citep{gpt-report} and BERT \cite{bert-report}, are often incorporated into these workflows at various stages, from model pretraining on massive datasets to fine-tuning for specific tasks in natural language processing. These workflows require not only the ability to scale across multiple nodes but also sophisticated task scheduling and resource management to accommodate the substantial computational and memory demands of training and inference with LLMs.

LLMs are computationally expensive methods that need to be trained primarily on large amounts of data. Training LLMs often requires hundreds to thousands of graphics processing units (GPUs)~\citep{jiang2024megascalescalinglargelanguage}. These GPUs must have sufficient video random access memory (VRAM) such that they can retain model parameters, often in the magnitude of billions or trillions, during training. The input data required to train the LLM, such as text corpora, demands large disk requirements; for example, Common Crawl, a repository of web data, uses hundreds of terabytes~\citep{8311752}. Running inference on LLMs is much more feasible, even for at-home computing environments, where resource-friendly LLMs such as Gemma \citep{gemma} or Vicuna-7b can fit their 14GB VRAM requirement within a high-end consumer GPU~\citep{xu2024surveyresourceefficientllmmultimodal}.
HPC workflows offer high-performance computing capabilities that allow LLMs to quickly and efficiently process massive amounts of data. This is especially relevant as LLMs are driving compute requirements towards Zettaflop levels, which HPC systems are well-tuned to address \citep{ferreira_da_silva2024}.

In the context of large-scale AI applications, workflow systems often integrate with specialized hardware, such as GPUs and Tensor Processing Units (TPUs), and must ensure optimal utilization of these resources across a distributed network. Workflow management tools in this domain must also handle the intricacies of model versioning, data pipelines, and monitoring across potentially heterogeneous architectures. This makes HPC/AI workflows, especially those involving LLMs, some of the most complex and resource-demanding workflows in modern computational research.

\subsection{Workflow Benchmarks}

The growing complexity and heterogeneity of high-performance computing (HPC) workflows have led to an increasing need for standardized benchmarking approaches \citep{badia2024integrating}. Several recent efforts have focused on developing frameworks and methodologies to evaluate workflow performance systematically. Among these, WfCommons \citep{coleman2022-1}, WfBench \citep{coleman2022-2}, and OpenEBench \citep{Capella-Gutierrez2017} provide significant contributions to the field.

WfCommons \citep{coleman2022-1} is a comprehensive framework that provides tools for the modeling, generation, and benchmarking of scientific workflows. It enables users to extract workflow characteristics from real-world executions, generate synthetic yet realistic workflow instances, and evaluate workflow execution on different computing environments. By facilitating workflow reproducibility and comparison, WfCommons serves as an essential tool for both workflow designers and HPC researchers.

WfBench \citep{coleman2022-2} is a dedicated benchmarking framework designed to evaluate workflow management systems (WMS) in terms of their execution efficiency, scalability, and resource utilization. It provides a set of predefined benchmarking workloads that mimic real scientific applications, allowing researchers to assess how different WMS implementations perform under varying computational loads. By systematically comparing workflow execution across platforms, WfBench aids in identifying performance bottlenecks and optimizing resource allocation.

OpenEBench \citep{Capella-Gutierrez2017}, developed as part of the European Open Science Cloud (EOSC) initiative, focuses on benchmarking workflows within life sciences. It provides a collaborative platform for evaluating workflow robustness, scalability, and reproducibility by leveraging community-driven benchmarking datasets. OpenEBench facilitates comparative analyses of bioinformatics workflows, ensuring that computational pipelines meet the demands of scientific reproducibility and efficiency.

These efforts collectively contribute to advancing workflow benchmarking methodologies, enabling the HPC community to develop more efficient and scalable workflows. By standardizing benchmarking practices, these frameworks help researchers optimize performance and resource utilization in increasingly complex computational environments.

\section{Conclusion}
\label{sec:conclusion}

We have seen from our discussion that, throughout the years, a number of aspects regarding scientific workflows have influenced our research work. The goal in all of these efforts is to strive to simplify the task of managing large-scale scientific experiments by the users. We identified abstractions, programming APIs, runtime support libraries, resource management, standardization, evaluation, and applications that are essential for a comprehensive strategy addressing many of the complex subtasks.

We have focused on scientific workflows running on large-scale HPC resources while focusing on deep learning and AI algorithms. As part of such applications, we identified that experiments iterating over a static or dynamically managed set of parameters is of utmost importance. We identified that we not only need to deal with hyperparameters but also integrate parameters into the batch jobs that constitute such experiments. This is based on the fact that we also need to iterate over potential resource-defining parameters such as GPUs, the use of specific file systems, or even the use of particular libraries that need to be provisioned or used as part of an experiment.

Most importantly we discovered that these requirements have been fulfilled by two completely independently developed efforts. One being SmartSim and the other is the combination of the Experiment Executor and the Compute Coordinator which are both plugins to the Cloudmesh toolkit. In working together, we have found surprising similarities in the taxonomy, design, and applications of these two packages. Due to this similarity, we believe that the approaches we have taken and the solutions we have come up independently from each other map onto fundamental requirements that emerging HPC and AI workflows will require.

While we have distilled the requirements from our own experiences in scientific workflows, we by no means claim that these form a complete set. In particular, the emerging applications of converged AI and HPC \citep{brewer2024ai} represent a new source of requirements and design constraints that have yet to be discovered. Despite this uncertainty, the requirements discussed in this paper likely represent a necessary subset. As workflow solutions continue to develop concurrently with these emerging applications, the pooled experience from domain scientists and workflow toolkit developers will be crucial to standardizing concepts and characteristics for the next generation of scientific applications. 
\clearpage

\section{Nomenclature}

\subsection{Resource Identification Initiative}

{\bf Organization:} \verb|RRID:SCR_011743|

\section*{Conflict of Interest Statement}

The authors declare that the research was conducted in the absence of any commercial or financial relationships that could be construed as a potential conflict of interest.

\section*{Author Contributions}

{\em GvL} is the author of the Experiment Executor and many other components that are distributed as bag of plugins to Cloudmesh.  He has modified modifications to how the OSMI benchmark operates while leveraging some of the elementary features contained in the Cloudmesh experiment management. He has decades worth of HPC dating back to 1984. 

{\em WB} is the author of the OSMI code and benchmark, and contributed related research on surrogate model and digital twin workflows. His experience from using DOE machines is integrated into this paper.

{\em SRW} contributed to the discussions of the FAIR Principles, Open Science, and workflows. He serves in the GO FAIR US Office, and he is the co-chair of the Workflows Community Initiative (WCI) FAIR Computational Workflows working group.

{\em AS} is a lead developer of SmartSim, which was independently designed and implemented from Cloudmesh. Through various collaborations, he has tested and gathered requirements across multiple applications that shaped this paper. He has experience with open development of scientific software and the dissemination of large datasets through his contributions to large-scale climate modeling efforts in the United States and Canada. 

{\em JPF} developed the Cloudmesh-vpn plugin for integrating split VPNs as well as independently tested the workflow code for multiple scientific applications such as earthquake and cloudmask. He also maintained the workflow Compute Coordinator and job generator libraries.

{\em HP} developed the plugin to Cloudmesh for the HPC clusters in the cloud. 

{\em CK} contributed to the discussion of the FAIR Principles, Open Science, and research data management concepts. She is head of GO FAIR US, a US-based consortium focused on FAIR implementation, a lead on the National Science Data Fabric project, and principal investigator of the NSF-funded Research Coordination Network FARR: FAIR in Machine Learning, AI Readiness, AI Reproducibility. She is the Secretary General of the International Science Council's Committee on Data (CODATA).

{\em GCF} is the author of the earthquake code and facilitates the interactions with the MLCommons Science Working group as a group leader of that effort. 

\section*{Funding}

Work was in part funded by the NSF CyberTraining: CIC: CyberTraining for Students and Technologies from Generation Z with the award numbers 1829704 and 2200409 and NIST 60NANB21D151T.  The work was also funded by the Department of Energy under the grant Award No. DE-SC0023452. The work was conducted at the Biocomplexity Institute and Initiative at the University of Virginia.

\section*{Acknowledgments}

This research was sponsored in part by and used resources of the Oak Ridge Leadership Computing Facility (OLCF), which is a DOE Office of Science User Facility at the Oak Ridge National Laboratory (ORNL) supported by the U.S. Department of Energy under Contract No. DE-AC05-00OR22725. The US government retains and the publisher, by accepting the article for publication, acknowledges that the US government retains a nonexclusive, paid-up, irrevocable, worldwide license to publish or reproduce the published form of this manuscript, or allow others to do so, for US government purposes. DOE will provide public access to these results of federally sponsored research in accordance with the DOE Public Access Plan (\url{https://www.energy.gov/doe-public-access-plan}). Kirkpatrick's work was made possible through the National Science Foundation awards \#2226453 and 2138811. Shao was supported by internal funding from Hewlett Packard Enterprise. He additionally gratefully acknowledges the contributions of SmartSim developers, specifically Alyssa Cote for providing a figure used in this paper.

\section*{Data Availability Statement}

The code is all in the public domain and available on GitHub at the following locations:

\begin{itemize}

\item {\bf cloudmesh-cc} -- Is a code to control workflows to be executed on
  remote computing
  resources. \url{https://github.com/cloudmesh/cloudmesh-cc}

\item {\bf cloudmesh-ee} -- Is a code to generate batch scripts for
  hyperparameter studies high-performance computers so they can be
  executed on different supercomputers by multiple
  accounts. \url{https://github.com/cloudmesh/cloudmesh-ee}

\item {\bf cloudmesh-vpn} -- Is a plugin that allows to use a VPN client as part of the client-focused workflow supported by the Cloudmesh command and shell. Recently we added support for split VPN allowing access to multiple resources controlled by multiple VPNs.
\url{https://github.com/cloudmesh/cloudmesh-vpn}

\item {\bf cloudmesh} -- Cloudmesh is a large collection of repositories for
  accessing cloud and HPC
  resources. \url{https://github.com/orgs/cloudmesh/repositories}

\item {\bf OSMI} -- Is a surrogate-model inference benchmark. \url{https://github.com/laszewsk/osmi-bench-new}



\end{itemize}

\newpage
\appendix

\section{Supplementary: Selected Related Research from Co-authors}
\label{sec:related}

It is impossible for a single paper to summarize all related research in this area. We refer to the other papers in this collection of papers published as part of this special issue. Therefore, we restrict our summary of related and selected research to activities conducted by the authors.

{\bf von Laszewski} has worked in the area of scientific workflows for about 30 years. This includes the introduction of a novel metacomputing framework \citep{las-99-loosely,las-94-ecwmf,las-96-ecwmf} that was used to schedule jobs in a distributed fashion on multiple supercomputers and also access supercomputers of significant architectural design. 
This was continued by the usage of workflows in problem-solving environments \citep{las-01-pse}. This was followed by integrating many of the conceptual designs into the Globus Toolkit with the first application using workflows as part of Grids \citep{las-00-sbc}. The lesson from this motivated us to focus on developing the Java Commodity Grid Kit (Java CoG Kit) \citep{las-06-workcoordination,
las-06-workflow-book,
las-06-exp-a,
las-05-workflowrepo,
las-05-workflow-jgc,
las-05-exp,
las-04-abstraction-j,
las-03-gridcomputing,
las-02-javacog,
las-00-grande,
las-01-cog-concurency}. 
During the peak of Grid Computing over 100 demonstrations on the Supercomputing exhibition floor used the Java CoG kit. As part of these activities, he pioneered a remote execution service InfoGramm 
\citep{las-02-infogram}
that in addition to serving as a service returning information about remote resources also allowed the execution of programs and scripts executed remotely as part of Grids allowing workflows to utilize multiple Grid resources at the same time. Early systems such as GridAnt \citep{las-04-gridant} 
did provide the ability to formulate Grid Workflows into frameworks familiar to Java developers. A much-enhanced workflow framework system was introduced into the  Java CoG Kit Karajan \citep{las-06-workflow-book} that in addition to using DAGs also allowed the specification of iterations into the workflow to help in the analysis of advanced photon source images and other applications. It also includes the introduction of futures \cite{friedman-futures}. Prior work to Karajan includes \citep{las-04-gridant,las-01-cog-concurency,las-96-ecwmf}. The availability of the loops allowed superior performance as the application-specific demands could be integrated. Workflows could be specified through an API, but also through the integration of XML specification. The workflows could be dynamic and changed based on runtime conditions. Some of the ideas from this work were continued into the Swift framework while leveraging the futures from Karajan in support of fast, reliable, loosely coupled parallel computation \citep{las--7-swift}. 
As part of the CoG Kit, von Laszewski and his colleagues also invented the first service controllable file transfer service with GUI to coordinate multiple file transfers. While this work was focused mostly on implementations done in Java, a new development using mostly Python was started with the Cyberaide toolkit \citep{las-09-ccgrid} that later on was renamed to Cloudmesh. As the name indicates the emphasis here was the integration of cloud resources rather than the focus of utilizing and enhancing the Globus Toolkit services. However, it also included initially the integration with Globus that focused on file transfer 
\citep{las-04-ftp-journal}
\citep{las-03-ftp}.
This tool could support many different cloud services from which some no longer exist such as Eucalyptus \cite{eucalyptus} and OpenCirrus \cite{opencirrus}. The services supported included execution services on AWS, Google, Azure, and OpenStack (KIT, and Chameleon Cloud). It also included data transfer services. The workflows emphasized here were not server-to-server services, but client-to-server services. One of the goals was to create workflows that let a scientific user develop workflows that can be controlled from their laptop in such a fashion that the workflows can be started and monitored from the laptop, allowing also the shutdown of the laptop and restart and discovering its state from ongoing workflow executions. 
The Cloudmesh toolkit \citep{las-17-cloudmesh} philosophy includes the distribution of a number of plugins into an extensible command line and command shell framework. While separating them into different packages extensions and different client needs can be fulfilled more easily because the user can select the needed plugins so that Cloudmesh offers a highly customizable solution for the different users. Early plugins include compute and file transfer resource services for AWS, Azure, Google, and OpenStack. However, most recently we have focused on experiment management which we describe in more detail within this paper due to the advent of large-scale HPC centers with the use of GPUs to increase computational capabilities. 
Additionally, von Laszewski participated in the efforts of Cylon for data engineering that simplifies data engineering workflow tasks \citep{cylon,cylon-radical}. 

Although we also worked on infrastructure provisioning for scientific
workflows that include image management
\citep{las-12-imagemanagement}, management of cloud infrastructures
including \citep{las-20-10gce,las-14-bigdata,las-12-fg-bookchapter}
\citep{las-17-futuregrid} and creation of virtual clusters
\citep{las-16-virtcluster,las-19-harc-comet}, as well as federation
\citep{las-08-federated-cloud}, we will not discuss them here in more
detail and refer to the provided references as they also provide
valuable lessons in regard to integration of provisioning into
workflows.

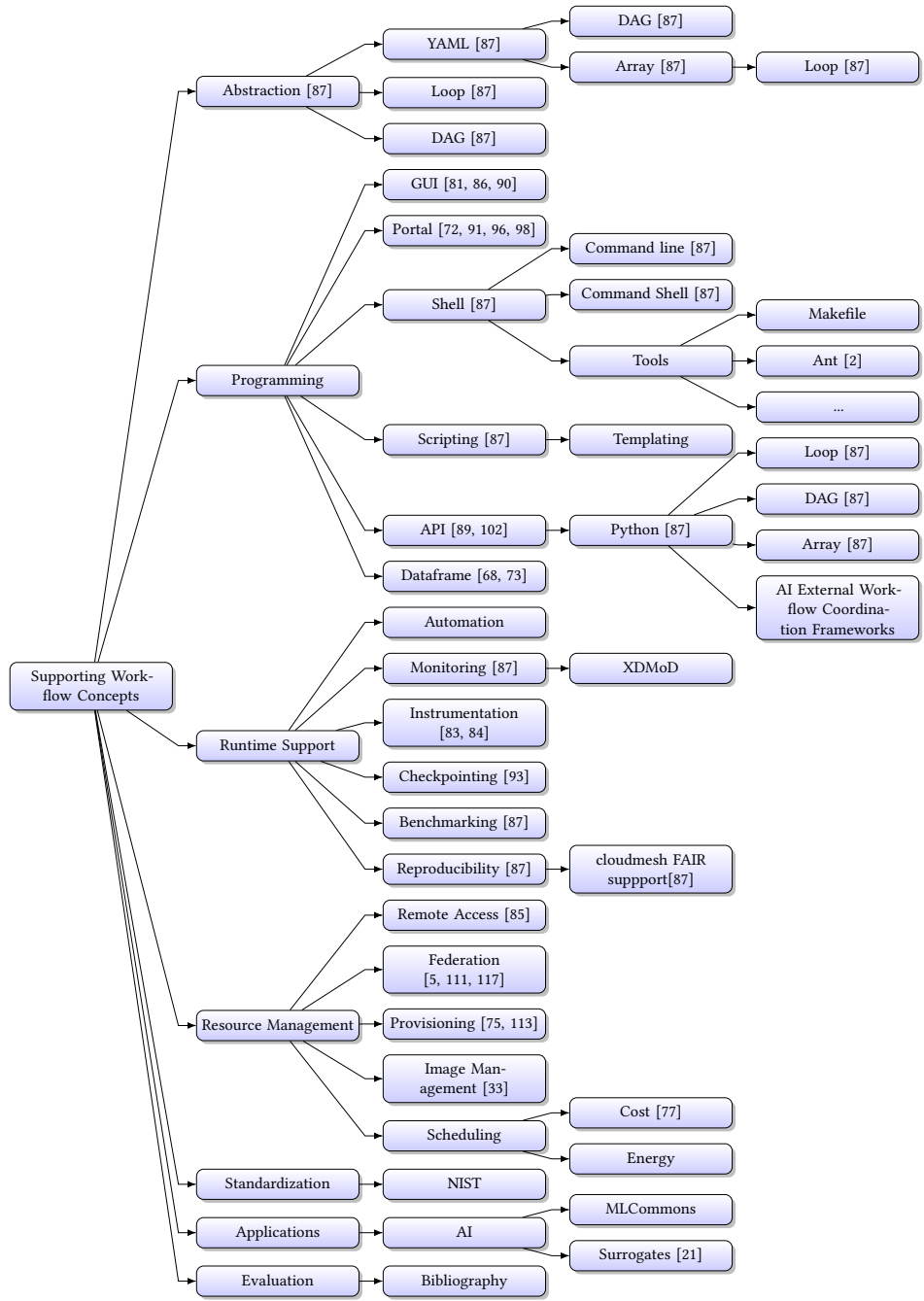
\begin{figure*}[p!]
  \centering
  \resizebox{12.5cm}{!}{%
\begin{forest}
skan tree
   [Supporting Workflow Concepts
        [Abstraction \citep{cloudmesh-ee}
          [YAML \citep{cloudmesh-ee}
            [DAG \citep{cloudmesh-ee}]
            [Array \citep{cloudmesh-ee}
              [Loop \citep{cloudmesh-ee}] 
            ]
          ]
          [Loop \citep{cloudmesh-ee}]
          [DAG \citep{cloudmesh-ee}]
        ]
        [Programming
           [GUI \citep{cloudmesh-cc,las-03-ftp,las-05-workflow-jgc}]
           [Portal \citep{las-99-rostock,las-06-guss-j,las-01-pse,las-02-cactus-j}]
           [Shell \citep{cloudmesh-ee}
             [Command line \citep{cloudmesh-ee}]
             [Command Shell \citep{cloudmesh-ee}]
             [Tools
                [Makefile]
                [Ant \citep{las-04-gridant}]
                [ ... ]
             ]
           ]
           [Scripting \citep{cloudmesh-ee}
              [Templating]
           ]
           [API \citep{las-02-javacog,las-17-cloudmesh}
             [Python \citep{cloudmesh-ee}
             [Loop \citep{cloudmesh-ee}]
             [DAG \citep{cloudmesh-ee}]
             [Array \citep{cloudmesh-ee}]
             [AI External Workflow Coordination Frameworks]
             ]
           ]
           [Dataframe \citep{cylon,cylon-radical}]
        ]
        [Runtime Support
           [Automation]
           [Monitoring \citep{cloudmesh-ee}
              [XDMoD]
           ]
           [Instrumentation \citep{cloudmesh-stopwatch,cloudmesh-gpu}]
           [Checkpointing \citep{las-2023-mlcommons-edu-eq}]
           [Benchmarking \citep{cloudmesh-ee}]
           [Reproducibility \citep{cloudmesh-ee}
               [cloudmesh FAIR suppport\citep{cloudmesh-ee}]
           ]
        ]
        [Resource Management
          [Remote Access \citep{cloudmesh-vpn}]
          [Federation \cite{las-08-federated-cloud,las-12-fedcloud-proc,antypas2021}]
          [Provisioning \citep{las-19-harc-comet,las-17-comet}]
          [Image Management \citep{las-12-imagemanagement}]
          [Scheduling
            [Cost \citep{las-01-greed}]
            [Energy ]
          ]
        ]
        [Standardization 
          [NIST]
        ]
        [Applications
            [AI
               [MLCommons]
               [Surrogates \cite{brewer2021production}]
            ] 
        ]
        [Evaluation 
           [Bibliography]
        ]
   ]
]
\end{forest}
}
   \caption{Scheduling challenges applied to all levels. We added not
     all but selected publications that we worked on as part of these
     workflow challenges.} 
  \label{F:graph-challanges}
\end{figure*}

\begin{figure}
    \centering
    \includegraphics[width=0.8\linewidth]{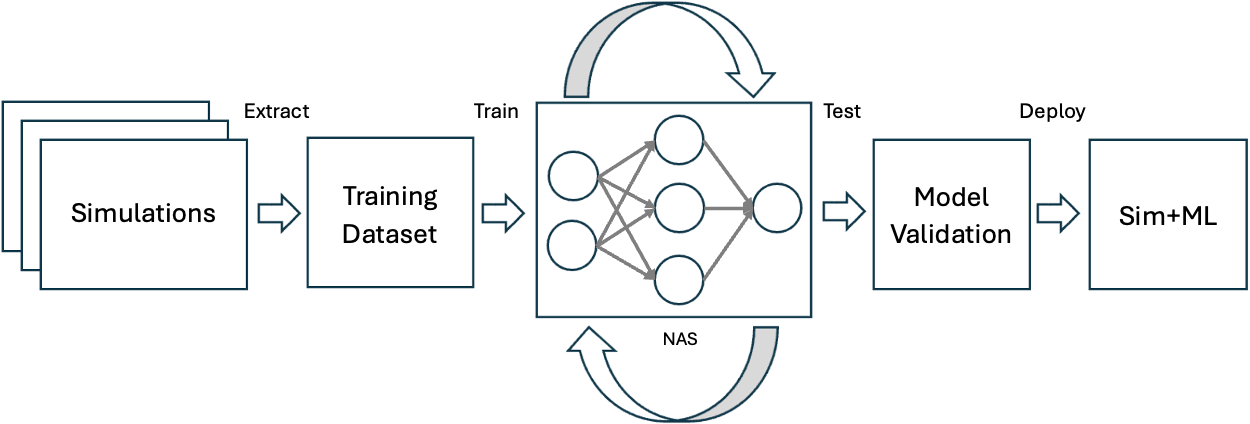}
    \caption{Machine-learned surrogate model training and deployment workflow \cite{brewer2023entropy}.}
    \label{fig:surrogate}
\end{figure}

{\bf Brewer} has most recently focused on  {\bf Surrogate Model Workflows}. Figure \ref{fig:surrogate} provides a schematic of a typical machine-learned \textit{surrogate model} training and deployment workflow. Simulations are run on HPC using a variety of input parameters, from which data is extracted to curate a training dataset. Intelligent subsampling techniques, such as the principal of maximum entropy \cite{brewer2023entropy}, are used to curate an optimal training dataset. Hyperparameter optimization, such as DeepHyper \cite{balaprakash2018deephyper} or DLEO \cite{martinez2018deep}, is used to perform neural architecture search (NAS) in order to design an optimal architecture. Model validation techniques, such as using PI3NN \cite{liu2021pi3nn} use prediction intervals to assess proper coverage of the training data (in-distribution vs. out-of-distribution) via uncertainty quantification. Finally, optimal deployment studies are performed to determine the optimal deployment parameters, such as concurrency, batch size, and precision \cite{brewer2021production}. The surrogate model may be deployed as a means of replacing a computationally expensive portion of the simulation, for example, the machine-learned turbulence model \cite{bhushan2021development}, or replace the entire simulation, e.g., FourCastNet climate model \cite{pathak2022fourcastnet}.

A digital twin is a virtual replica of a physical asset, that mimics the behavior of the asset, and communicates in a bi-directional manner with its physical counterpart \cite{nas2023foundational}.   Brewer et al. \cite{brewer2024digital} recently developed a digital twin framework for HPC, called ExaDigiT, which integrates five different levels of information: (1) 3D asset modeling and visualization using augmented reality (AR), (2) telemetry/sensor data streaming from the physical asset, (3) machine learned (ML) models to mimic behavior in a data-driven manner, (4) modeling and simulation to mimic behavior based on first principles, and (5) reinforcement learning. Telemetry data is used for training AI/ML models and validating modeling and simulation. Modeling and simulation are used as a training environment for training a reinforcement learning agent, which provides autonomous control and optimization in the form of a feedback agent to the physical asset. This framework has been used to develop a digital twin of the Frontier supercomputer, the first Exascale supercomputer, which can dynamically schedule system workloads, predicts power at any level of granularity (from chip-level to total system) and cooling throughout the system and its supporting central energy plant, as well as dynamically predicts its power usage effectiveness (PUE). Such a twin can be used for performing what-if scenarios (e.g., what-if a pump fails), system optimizations (e.g., power and cooling), and virtual prototyping of future systems. Several different instantiations of data center digital twins are reviewed in \cite{athavale2024digital}. A benchmark has yet to be developed for such a complex workflow, but we plan to work on this in the future.

\begin{figure}
    \centering
    \includegraphics[width=0.8\linewidth]{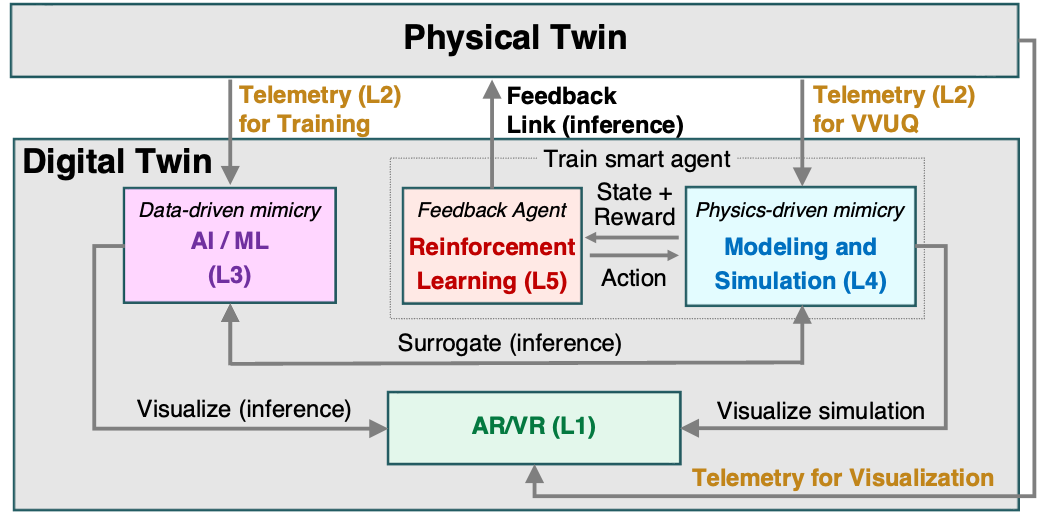}
    \caption{Digital twin workflow \cite{brewer2024digital}.}
    \label{fig:dt}
\end{figure}

{\bf Wilkinson} has published recently on numerous topics in the area of scientific workflows \citep{ferreira_da_silva2024, ferreira_da_silva2022, badia2024integrating}, including work that incorporates benchmarks \citep{coleman2022-2}, cross-facility resources \citep{antypas2021}, provenance \citep{souza2023}, HPC \citep{wilkinson2022-2}, and quantum computing \citep{bieberich2023} for domains such as high-energy physics \citep{ananthraj2018}, bioinformatics \citep{lee2021,wilson2021}, and geology \citep{mcclure2020}. The main focus of his current work is on the application of the FAIR principles to computational ecosystems generally and computational workflows specifically \citep{wilkinson2022, caw2021-report}, the latter for which he co-chairs the Workflows Community Initiative's FAIR Computational Workflows working group, which recently published the results from its first two years \citep{wilkinson2025}. He also contributes to and co-administers WorkflowHub \citep{gustafsson2024}, and he serves in the GO FAIR US Office.

{\bf Shao} approaches workflow management from both the point of view of a domain scientist (with a particular emphasis on climate modeling) and a computer scientist investigating the emerging workflow paradigms and philosophies needed to combine AI/ML techniques with HPC-scale scientific simulations. In particular, most traditional numerical modeling operates as a pipeline with each stage focusing on the execution of a single application and the file system used to exchange data between stages. Ensemble-based modeling (often used in climate/weather) represents a horizontally scaled pipeline. Each individual member ensemble may differ by the values of their tunable parameters and/or the initial/boundary conditions, but run independently of each other. HPC simulation and AI applications often require a more asynchronous computational paradigm with data exchange that occurs across loosely coupled components (i.e. processing elements transfer data through an intermediary). The SmartSim library provides well-maintained, open-source tooling that enables domain scientists to describe and execute their own complex workflows.

While originally designed for in-the-loop inference, increasingly the library has been used by users whose workflows apply AI techniques to ensembles of simulation including reinforcement learning. In general, these are characterized by a more complex set of outcomes/artifacts than traditional scientific modeling. Instead of just simulation data, workflow artifacts may include trained AI models or control schemes to be used in conjunction with digital/physical twins. 

{\bf Kirkpatrick} collaborates with workflow experts at several NSF and DOE-funded labs. Activities have included an invited keynote at a recent international workflows workshop, activities through GO FAIR US to promote the extension of the FAIR Principles for Workflows, and participation in other workshops \cite{kirkpatrick2023}. Her most recent scholarship includes a co-authored a section from a Dagstuhl seminar proceeding, ``Integrating HPC, AI, and Workflows for Scientific Data Analysis'' on sustainability in HPC and AI-driven scientific workflows \citep{badia2024integrating}.

{\bf Pitkar} has more than 20 years of experience in the field of Information Technology. He is currently working as an IT Engineering Leader at Cummins Inc. in the Engineering and Automation department. He is passionate about automation and leads the Kubernetes platform team. His areas of focus are Platform Engineering, Cloud computing and automation.

{\bf Fleischer} works with open-source computer-vision applications towards traffic management and pedestrian safety analysis. The completion of an accurate object-detection model requires an integrated workflow process, from data annotation to configuring learning parameters using the Darknet/YOLO (You Only Look Once) suite~\citep{charette_2022, bochkovskiy2020yolov4optimalspeedaccuracy}. As standardized benchmarking is only feasible within containerized environments, he uses platforms such as Docker and Apptainer to facilitate portable training in HPC environments. Under the supervision of von Laszewski, he has conducted similar model training experiments on applications such as earthquake and meteorological forecasting.

To provide a better view of the various aspects of workflows we have organized them into a graph as shown in Figure \ref{F:graph-challanges}.


\clearpage


\bibliographystyle{ACM-Reference-Format}

\bibliography{%
vonLaszewski-2025-workflow}


\newpage

\bibliographystyle{Frontiers-Vancouver} 

\bibliography{vonLaszewski-2025-workflow}


\begin{thebibliography}{128}


\ifx \showCODEN    \undefined \def \showCODEN     #1{\unskip}     \fi
\ifx \showDOI      \undefined \def \showDOI       #1{#1}\fi
\ifx \showISBNx    \undefined \def \showISBNx     #1{\unskip}     \fi
\ifx \showISBNxiii \undefined \def \showISBNxiii  #1{\unskip}     \fi
\ifx \showISSN     \undefined \def \showISSN      #1{\unskip}     \fi
\ifx \showLCCN     \undefined \def \showLCCN      #1{\unskip}     \fi
\ifx \shownote     \undefined \def \shownote      #1{#1}          \fi
\ifx \showarticletitle \undefined \def \showarticletitle #1{#1}   \fi
\ifx \showURL      \undefined \def \showURL       {\relax}        \fi
\providecommand\bibfield[2]{#2}
\providecommand\bibinfo[2]{#2}
\providecommand\natexlab[1]{#1}
\providecommand\showeprint[2][]{arXiv:#2}

\bibitem[Alsaadi et~al\mbox{.}(2024)]%
        {alsaadi2024}
\bibfield{author}{\bibinfo{person}{Aymen Alsaadi}, \bibinfo{person}{Mihael
  Hategan-Marandiuc}, \bibinfo{person}{Ketan Maheshwari},
  \bibinfo{person}{Andre Merzky}, \bibinfo{person}{Mikhail Titov},
  \bibinfo{person}{Matteo Turilli}, \bibinfo{person}{Andreas Wilke},
  \bibinfo{person}{Justin~M. Wozniak}, \bibinfo{person}{Kyle Chard},
  \bibinfo{person}{Rafael~Ferreira da Silva}, \bibinfo{person}{Shantenu Jha},
  {and} \bibinfo{person}{Daniel Laney}.} \bibinfo{year}{2024}\natexlab{}.
\newblock \showarticletitle{Exascale Workflow Applications and Middleware: An
  ExaWorks Retrospective}.
\newblock  (\bibinfo{year}{2024}).
\newblock
\urldef\tempurl%
\url{https://doi.org/10.48550/arXiv.2411.10637}
\showDOI{\tempurl}
\showeprint[arxiv]{2411.10637}~[cs.SE]


\bibitem[Amin et~al\mbox{.}(2004)]%
        {las-04-gridant}
\bibfield{author}{\bibinfo{person}{Kaizar Amin}, \bibinfo{person}{Mihael
  Hategan}, \bibinfo{person}{Gregor von Laszewski}, \bibinfo{person}{Nestor~J.
  Zaluzec}, \bibinfo{person}{Shawn Hampton}, {and} \bibinfo{person}{Albert
  Rossi}.} \bibinfo{year}{2004}\natexlab{}.
\newblock \showarticletitle{{GridAnt: A Client-Controllable Grid Workflow
  System}}. In \bibinfo{booktitle}{\emph{37th Hawai'i International Conference
  on System Science}}, Vol.~\bibinfo{volume}{7}. \bibinfo{publisher}{IEEE
  Computer Society, Los Alamitos, CA, USA}, \bibinfo{address}{Island of Hawaii,
  Big Island}.
\newblock
\showISSN{1530-1605}
\urldef\tempurl%
\url{https://doi.org/10.1109/HICSS.2004.1265491}
\showDOI{\tempurl}
\newblock
\shownote{{The original paper is: von Laszewski, Gregor, Kaizar Amin, Shawn
  Hampton, and Sandeep Nijsure. Technical report, Argonne National Laboratory,
  31 July 2002. https://laszewski.github.io/papers/vonLaszewski-gridant.pdf.}}.


\bibitem[Amin et~al\mbox{.}(2006)]%
        {las-04-abstraction-j}
\bibfield{author}{\bibinfo{person}{Kaizar Amin}, \bibinfo{person}{Gregor von
  Laszewski}, \bibinfo{person}{Rashid~Al Ali}, \bibinfo{person}{Omer Rana},
  {and} \bibinfo{person}{David Walker}.} \bibinfo{year}{2006}\natexlab{}.
\newblock \showarticletitle{{An Abstraction Model for a Grid Execution
  Framework}}.
\newblock \bibinfo{journal}{\emph{Euromicro Journal of Systems Architecture}}
  \bibinfo{volume}{52}, \bibinfo{number}{2} (\bibinfo{year}{2006}),
  \bibinfo{pages}{73--87}.
\newblock
\showISSN{1383-7621}
\urldef\tempurl%
\url{https://doi.org/10.1016/j.sysarc.2004.10.007}
\showDOI{\tempurl}


\bibitem[Ananthraj et~al\mbox{.}(2018)]%
        {ananthraj2018}
\bibfield{author}{\bibinfo{person}{V Ananthraj}, \bibinfo{person}{K De},
  \bibinfo{person}{S Jha}, \bibinfo{person}{A Klimentov}, \bibinfo{person}{D
  Oleynik}, \bibinfo{person}{S Oral}, \bibinfo{person}{A Merzky},
  \bibinfo{person}{R Mashinistov}, \bibinfo{person}{S Panitkin},
  \bibinfo{person}{P Svirin}, \bibinfo{person}{M Turilli}, \bibinfo{person}{J
  Wells}, {and} \bibinfo{person}{S Wilkinson}.}
  \bibinfo{year}{2018}\natexlab{}.
\newblock \showarticletitle{Towards Exascale Computing for High Energy Physics:
  The ATLAS Experience at ORNL}. In \bibinfo{booktitle}{\emph{2018 IEEE 14th
  International Conference on e-Science (e-Science)}}.
  \bibinfo{pages}{341--342}.
\newblock
\urldef\tempurl%
\url{https://doi.org/10.1109/eScience.2018.00086}
\showDOI{\tempurl}


\bibitem[Antypas et~al\mbox{.}(2021)]%
        {antypas2021}
\bibfield{author}{\bibinfo{person}{K.~B. Antypas}, \bibinfo{person}{D.~J.
  Bard}, \bibinfo{person}{J.~P. Blaschke}, \bibinfo{person}{R. Shane~Canon},
  \bibinfo{person}{Bjoern Enders}, \bibinfo{person}{Mallikarjun~Arjun Shankar},
  \bibinfo{person}{Suhas Somnath}, \bibinfo{person}{Dale Stansberry},
  \bibinfo{person}{Thomas~D. Uram}, {and} \bibinfo{person}{Sean~R. Wilkinson}.}
  \bibinfo{year}{2021}\natexlab{}.
\newblock \showarticletitle{Enabling discovery data science through
  cross-facility workflows}. In \bibinfo{booktitle}{\emph{2021 IEEE
  International Conference on Big Data (Big Data)}}.
  \bibinfo{pages}{3671--3680}.
\newblock
\urldef\tempurl%
\url{https://doi.org/10.1109/BigData52589.2021.9671421}
\showDOI{\tempurl}


\bibitem[Athavale et~al\mbox{.}(2024)]%
        {athavale2024digital}
\bibfield{author}{\bibinfo{person}{Jyotika Athavale}, \bibinfo{person}{Cullen
  Bash}, \bibinfo{person}{Wesley Brewer}, \bibinfo{person}{Matthias Maiterth},
  \bibinfo{person}{Dejan Milojicic}, \bibinfo{person}{Harry Petty}, {and}
  \bibinfo{person}{Soumyendu Sarkar}.} \bibinfo{year}{2024}\natexlab{}.
\newblock \showarticletitle{Digital Twins for Data Centers}.
\newblock \bibinfo{journal}{\emph{Computer}} \bibinfo{volume}{57},
  \bibinfo{number}{10} (\bibinfo{year}{2024}), \bibinfo{pages}{151--158}.
\newblock


\bibitem[Avetisyan et~al\mbox{.}(2010)]%
        {opencirrus}
\bibfield{author}{\bibinfo{person}{Arutyun~I. Avetisyan}, \bibinfo{person}{Roy
  Campbell}, \bibinfo{person}{Indranil Gupta}, \bibinfo{person}{Michael~T.
  Heath}, \bibinfo{person}{Steven~Y. Ko}, \bibinfo{person}{Gregory~R. Ganger},
  \bibinfo{person}{Michael~A. Kozuch}, \bibinfo{person}{David O'Hallaron},
  \bibinfo{person}{Marcel Kunze}, \bibinfo{person}{Thomas~T. Kwan},
  \bibinfo{person}{Kevin Lai}, \bibinfo{person}{Martha Lyons},
  \bibinfo{person}{Dejan~S. Milojicic}, \bibinfo{person}{Hing~Yan Lee},
  \bibinfo{person}{Yeng~Chai Soh}, \bibinfo{person}{Ng~Kwang Ming},
  \bibinfo{person}{Jing-Yuan Luke}, {and} \bibinfo{person}{Han Namgoong}.}
  \bibinfo{year}{2010}\natexlab{}.
\newblock \showarticletitle{Open Cirrus: A Global Cloud Computing Testbed}.
\newblock \bibinfo{journal}{\emph{Computer}} \bibinfo{volume}{43},
  \bibinfo{number}{4} (\bibinfo{year}{2010}), \bibinfo{pages}{35--43}.
\newblock
\urldef\tempurl%
\url{https://doi.org/10.1109/MC.2010.111}
\showDOI{\tempurl}


\bibitem[{AWS}(2024)]%
        {awspcs:online}
\bibfield{author}{\bibinfo{person}{{AWS}}.} \bibinfo{year}{2024}\natexlab{}.
\newblock \showarticletitle{What is AWS ParallelCluster}.
\newblock \bibinfo{journal}{\emph{NA}} (\bibinfo{date}{Oct.}
  \bibinfo{year}{2024}).
\newblock
\urldef\tempurl%
\url{https://docs.aws.amazon.com/parallelcluster/latest/ug/what-is-aws-parallelcluster.html}
\showURL{%
\tempurl}
\newblock
\shownote{[Online; Accessed on 03/01/2025]}.


\bibitem[{AWS}(2025)]%
        {www-aws-pricing}
\bibfield{author}{\bibinfo{person}{{AWS}}.} \bibinfo{year}{2025}\natexlab{}.
\newblock \bibinfo{title}{{HPC Workload Service – AWS Parallel Computing
  Service Pricing}}.
\newblock \bibinfo{howpublished}{Web Page}.
\newblock
\urldef\tempurl%
\url{https://aws.amazon.com/pcs/pricing/}
\showURL{%
\tempurl}
\newblock
\shownote{[Online; accessed on 03/08/2025]}.


\bibitem[Babuji et~al\mbox{.}(2019)]%
        {babuji2019}
\bibfield{author}{\bibinfo{person}{Yadu Babuji}, \bibinfo{person}{Anna
  Woodard}, \bibinfo{person}{Zhuozhao Li}, \bibinfo{person}{Daniel~S. Katz},
  \bibinfo{person}{Ben Clifford}, \bibinfo{person}{Rohan Kumar},
  \bibinfo{person}{Lukasz Lacinski}, \bibinfo{person}{Ryan Chard},
  \bibinfo{person}{Justin~M. Wozniak}, \bibinfo{person}{Ian Foster},
  \bibinfo{person}{Michael Wilde}, {and} \bibinfo{person}{Kyle Chard}.}
  \bibinfo{year}{2019}\natexlab{}.
\newblock \showarticletitle{Parsl: Pervasive Parallel Programming in Python}.
  In \bibinfo{booktitle}{\emph{Proceedings of the 28th International Symposium
  on High-Performance Parallel and Distributed Computing}} (Phoenix, AZ, USA)
  \emph{(\bibinfo{series}{HPDC '19})}. \bibinfo{publisher}{Association for
  Computing Machinery}, \bibinfo{address}{New York, NY, USA},
  \bibinfo{pages}{25–36}.
\newblock
\showISBNx{9781450366700}
\urldef\tempurl%
\url{https://doi.org/10.1145/3307681.3325400}
\showDOI{\tempurl}


\bibitem[Badia et~al\mbox{.}(2024)]%
        {badia2024integrating}
\bibfield{author}{\bibinfo{person}{Rosa~M. Badia}, \bibinfo{person}{Laure
  Berti-Equille}, \bibinfo{person}{Rafael Ferreira Da~Silva}, {and}
  \bibinfo{person}{Ulf Leser}.} \bibinfo{year}{2024}\natexlab{}.
\newblock \bibinfo{booktitle}{\emph{Integrating HPC, AI, and Workflows for
  Scientific Data Analysis: Report from Dagstuhl Seminar 23352}}.
\newblock \bibinfo{type}{{T}echnical {R}eport}. \bibinfo{institution}{Oak Ridge
  National Laboratory (ORNL), Oak Ridge, TN (United States)}.
\newblock
\urldef\tempurl%
\url{https://doi.org/10.2172/2341398}
\showDOI{\tempurl}


\bibitem[Balaprakash et~al\mbox{.}(2018)]%
        {balaprakash2018deephyper}
\bibfield{author}{\bibinfo{person}{Prasanna Balaprakash},
  \bibinfo{person}{Michael Salim}, \bibinfo{person}{Thomas~D Uram},
  \bibinfo{person}{Venkat Vishwanath}, {and} \bibinfo{person}{Stefan~M Wild}.}
  \bibinfo{year}{2018}\natexlab{}.
\newblock \showarticletitle{{DeepHyper}: Asynchronous hyperparameter search for
  deep neural networks}. In \bibinfo{booktitle}{\emph{2018 IEEE 25th
  international conference on high performance computing (HiPC)}}. IEEE,
  \bibinfo{pages}{42--51}.
\newblock
\urldef\tempurl%
\url{https://doi.org/10.1109/HiPC.2018.00014}
\showDOI{\tempurl}


\bibitem[Bhushan et~al\mbox{.}(2021)]%
        {bhushan2021development}
\bibfield{author}{\bibinfo{person}{Shanti Bhushan}, \bibinfo{person}{Greg~W
  Burgreen}, \bibinfo{person}{Wesley Brewer}, {and} \bibinfo{person}{Ian~D
  Dettwiller}.} \bibinfo{year}{2021}\natexlab{}.
\newblock \showarticletitle{Development and validation of a machine learned
  turbulence model}.
\newblock \bibinfo{journal}{\emph{Energies}} \bibinfo{volume}{14},
  \bibinfo{number}{5} (\bibinfo{year}{2021}), \bibinfo{pages}{1465}.
\newblock
\urldef\tempurl%
\url{https://doi.org/10.3390/en14051465}
\showDOI{\tempurl}


\bibitem[Bhushan et~al\mbox{.}(2023)]%
        {bhushan2023assessment}
\bibfield{author}{\bibinfo{person}{Shanti Bhushan}, \bibinfo{person}{Greg~W
  Burgreen}, \bibinfo{person}{Wesley Brewer}, {and} \bibinfo{person}{Ian~D
  Dettwiller}.} \bibinfo{year}{2023}\natexlab{}.
\newblock \showarticletitle{Assessment of neural network augmented {Reynolds
  averaged Navier Stokes} turbulence model in extrapolation modes}.
\newblock \bibinfo{journal}{\emph{Physics of Fluids}} \bibinfo{volume}{35},
  \bibinfo{number}{5} (\bibinfo{year}{2023}).
\newblock
\urldef\tempurl%
\url{https://doi.org/10.1063/5.0146456}
\showDOI{\tempurl}


\bibitem[Bieberich et~al\mbox{.}(2023)]%
        {bieberich2023}
\bibfield{author}{\bibinfo{person}{Samuel~T. Bieberich},
  \bibinfo{person}{Ketan~C. Maheshwari}, \bibinfo{person}{Sean~R. Wilkinson},
  \bibinfo{person}{Prasanna Date}, \bibinfo{person}{In-Saeng Suh}, {and}
  \bibinfo{person}{Rafael~Ferreira da Silva}.} \bibinfo{year}{2023}\natexlab{}.
\newblock \showarticletitle{Bridging HPC and Quantum Systems using Scientific
  Workflows}.
\newblock  (\bibinfo{year}{2023}).
\newblock
\urldef\tempurl%
\url{https://doi.org/10.48550/arXiv.2310.03286}
\showDOI{\tempurl}
\showeprint[arxiv]{2310.03286}~[cs.ET]


\bibitem[Bochkovskiy et~al\mbox{.}(2020)]%
        {bochkovskiy2020yolov4optimalspeedaccuracy}
\bibfield{author}{\bibinfo{person}{Alexey Bochkovskiy},
  \bibinfo{person}{Chien-Yao Wang}, {and} \bibinfo{person}{Hong-Yuan~Mark
  Liao}.} \bibinfo{year}{2020}\natexlab{}.
\newblock \showarticletitle{YOLOv4: Optimal Speed and Accuracy of Object
  Detection}.
\newblock \bibinfo{journal}{\emph{arXiv}} (\bibinfo{year}{2020}).
\newblock
\urldef\tempurl%
\url{https://doi.org/10.48550/arXiv.2004.10934}
\showDOI{\tempurl}
\showeprint{2004.10934}
\newblock
\shownote{cs.CV}.


\bibitem[Boyer et~al\mbox{.}(2022)]%
        {boyer2022scalable}
\bibfield{author}{\bibinfo{person}{Mathew Boyer}, \bibinfo{person}{Wesley
  Brewer}, \bibinfo{person}{Dylan Jude}, {and} \bibinfo{person}{Ian
  Dettwiller}.} \bibinfo{year}{2022}\natexlab{}.
\newblock \showarticletitle{Scalable Integration of Computational Physics
  Simulations with Machine Learning}. In \bibinfo{booktitle}{\emph{2022
  IEEE/ACM International Workshop on Artificial Intelligence and Machine
  Learning for Scientific Applications (AI4S)}}. IEEE, \bibinfo{pages}{44--49}.
\newblock
\urldef\tempurl%
\url{https://doi.org/10.1109/AI4S56813.2022.00013}
\showDOI{\tempurl}


\bibitem[Brewer et~al\mbox{.}(2020)]%
        {brewer2020inference}
\bibfield{author}{\bibinfo{person}{Wesley Brewer}, \bibinfo{person}{Greg Behm},
  \bibinfo{person}{Alan Scheinine}, \bibinfo{person}{Ben Parsons},
  \bibinfo{person}{Wesley Emeneker}, {and} \bibinfo{person}{Robert~P Trevino}.}
  \bibinfo{year}{2020}\natexlab{}.
\newblock \showarticletitle{Inference benchmarking on {HPC} systems}. In
  \bibinfo{booktitle}{\emph{2020 IEEE High Performance Extreme Computing
  Conference (HPEC)}}. IEEE, \bibinfo{pages}{1--9}.
\newblock
\urldef\tempurl%
\url{https://doi.org/10.1109/HPEC43674.2020.9286138}
\showDOI{\tempurl}


\bibitem[Brewer et~al\mbox{.}(2024a)]%
        {brewer2024ai}
\bibfield{author}{\bibinfo{person}{Wes Brewer}, \bibinfo{person}{Ana Gainaru},
  \bibinfo{person}{Fr{\'e}d{\'e}ric Suter}, \bibinfo{person}{Feiyi Wang},
  \bibinfo{person}{Murali Emani}, {and} \bibinfo{person}{Shantenu Jha}.}
  \bibinfo{year}{2024}\natexlab{a}.
\newblock \showarticletitle{{AI-coupled HPC} Workflow Applications, Middleware
  and Performance}.
\newblock \bibinfo{journal}{\emph{arXiv preprint arXiv:2406.14315}}
  (\bibinfo{year}{2024}).
\newblock
\urldef\tempurl%
\url{https://doi.org/10.48550/arXiv.2406.14315}
\showDOI{\tempurl}


\bibitem[Brewer et~al\mbox{.}(2024b)]%
        {brewer2024digital}
\bibfield{author}{\bibinfo{person}{Wesley Brewer}, \bibinfo{person}{Matthias
  Maiterth}, \bibinfo{person}{Vineet Kumar}, \bibinfo{person}{Rafal Wojda},
  \bibinfo{person}{Sedrick Bouknight}, \bibinfo{person}{Jesse Hines},
  \bibinfo{person}{Woong Shin}, \bibinfo{person}{Scott Greenwood},
  \bibinfo{person}{David Grant}, \bibinfo{person}{Wesley Williams}, {and}
  \bibinfo{person}{Feiyi Wang}.} \bibinfo{year}{2024}\natexlab{b}.
\newblock \showarticletitle{A digital twin framework for liquid-cooled
  supercomputers as demonstrated at exascale}. In
  \bibinfo{booktitle}{\emph{Proceedings of the International Conference for
  High Performance Computing, Networking, Storage and Analysis ({SC})}}.
\newblock


\bibitem[Brewer et~al\mbox{.}(2021)]%
        {brewer2021production}
\bibfield{author}{\bibinfo{person}{Wesley Brewer}, \bibinfo{person}{Daniel
  Martinez}, \bibinfo{person}{Mathew Boyer}, \bibinfo{person}{Dylan Jude},
  \bibinfo{person}{Andy Wissink}, \bibinfo{person}{Ben Parsons},
  \bibinfo{person}{Junqi Yin}, {and} \bibinfo{person}{Valentine Anantharaj}.}
  \bibinfo{year}{2021}\natexlab{}.
\newblock \showarticletitle{Production deployment of machine-learned rotorcraft
  surrogate models on {HPC}}. In \bibinfo{booktitle}{\emph{2021 IEEE/ACM
  Workshop on Machine Learning in High Performance Computing Environments
  (MLHPC)}}. IEEE, \bibinfo{pages}{21--32}.
\newblock
\urldef\tempurl%
\url{https://doi.org/10.1109/MLHPC54614.2021.00008}
\showDOI{\tempurl}


\bibitem[Brewer et~al\mbox{.}(2023)]%
        {brewer2023entropy}
\bibfield{author}{\bibinfo{person}{Wesley Brewer}, \bibinfo{person}{Daniel
  Martinez}, \bibinfo{person}{Muralikrishnan Gopalakrishnan~Meena},
  \bibinfo{person}{Aditya Kashi}, \bibinfo{person}{Katarzyna Borowiec},
  \bibinfo{person}{Siyan Liu}, \bibinfo{person}{Christopher Pilmaier},
  \bibinfo{person}{Greg Burgreen}, {and} \bibinfo{person}{Shanti Bhushan}.}
  \bibinfo{year}{2023}\natexlab{}.
\newblock \showarticletitle{Entropy-driven Optimal Sub-sampling of Fluid
  Dynamics for Developing Machine-learned Surrogates}. In
  \bibinfo{booktitle}{\emph{Proceedings of the SC'23 Workshops of The
  International Conference on High Performance Computing, Network, Storage, and
  Analysis}}. \bibinfo{pages}{73--80}.
\newblock
\urldef\tempurl%
\url{https://doi.org/10.1145/3624062.3626084}
\showDOI{\tempurl}


\bibitem[Capella-Gutierrez et~al\mbox{.}(2017)]%
        {Capella-Gutierrez2017}
\bibfield{author}{\bibinfo{person}{Salvador Capella-Gutierrez},
  \bibinfo{person}{Diana de~la Iglesia}, \bibinfo{person}{Juergen Haas},
  \bibinfo{person}{Analia Lourenco}, \bibinfo{person}{Jos{\'e}~Mar{\'\i}a
  Fern{\'a}ndez}, \bibinfo{person}{Dmitry Repchevsky},
  \bibinfo{person}{Christophe Dessimoz}, \bibinfo{person}{Torsten Schwede},
  \bibinfo{person}{Cedric Notredame}, \bibinfo{person}{Josep~Ll Gelpi}, {and}
  \bibinfo{person}{Alfonso Valencia}.} \bibinfo{year}{2017}\natexlab{}.
\newblock \showarticletitle{Lessons Learned: Recommendations for Establishing
  Critical Periodic Scientific Benchmarking}.
\newblock \bibinfo{journal}{\emph{bioRxiv}} (\bibinfo{year}{2017}).
\newblock
\urldef\tempurl%
\url{https://doi.org/10.1101/181677}
\showDOI{\tempurl}
\showeprint{https://www.biorxiv.org/content/early/2017/08/31/181677.full.pdf}


\bibitem[{Carpentries Introduction}(2025)]%
        {def-workflow}
\bibfield{author}{\bibinfo{person}{{Carpentries Introduction}}.}
  \bibinfo{year}{2025}\natexlab{}.
\newblock \bibinfo{title}{Introduction to Workflows with Common Workflow
  Language}.
\newblock
\newblock
\urldef\tempurl%
\url{https://carpentries-incubator.github.io/cwl-novice-tutorial/01-introduction/index.html}
\showURL{%
\tempurl}
\newblock
\shownote{[Online; accessed 03/01/2025]}.


\bibitem[Chard et~al\mbox{.}(2020)]%
        {chard2020}
\bibfield{author}{\bibinfo{person}{Ryan Chard}, \bibinfo{person}{Yadu Babuji},
  \bibinfo{person}{Zhuozhao Li}, \bibinfo{person}{Tyler Skluzacek},
  \bibinfo{person}{Anna Woodard}, \bibinfo{person}{Ben Blaiszik},
  \bibinfo{person}{Ian Foster}, {and} \bibinfo{person}{Kyle Chard}.}
  \bibinfo{year}{2020}\natexlab{}.
\newblock \showarticletitle{funcX: A Federated Function Serving Fabric for
  Science}. In \bibinfo{booktitle}{\emph{Proceedings of the 29th International
  Symposium on High-Performance Parallel and Distributed Computing}}
  (Stockholm, Sweden) \emph{(\bibinfo{series}{HPDC '20})}.
  \bibinfo{publisher}{Association for Computing Machinery},
  \bibinfo{address}{New York, NY, USA}, \bibinfo{pages}{65–76}.
\newblock
\showISBNx{9781450370523}
\urldef\tempurl%
\url{https://doi.org/10.1145/3369583.3392683}
\showDOI{\tempurl}


\bibitem[Charette(2022)]%
        {charette_2022}
\bibfield{author}{\bibinfo{person}{Stéphane Charette}.}
  \bibinfo{year}{2022}\natexlab{}.
\newblock \bibinfo{journal}{\emph{Stéphane's Darknet FAQ}}
  (\bibinfo{date}{Apr} \bibinfo{year}{2022}).
\newblock
\urldef\tempurl%
\url{https://www.ccoderun.ca/programming/darknet_faq/}
\showURL{%
\tempurl}


\bibitem[Coleman et~al\mbox{.}(2022a)]%
        {coleman2022-2}
\bibfield{author}{\bibinfo{person}{Tain\~a Coleman}, \bibinfo{person}{Henri
  Casanova}, \bibinfo{person}{Ketan Maheshwari}, \bibinfo{person}{Lo\"ic
  Pottier}, \bibinfo{person}{Sean~R. Wilkinson}, \bibinfo{person}{Justin
  Wozniak}, \bibinfo{person}{Frédéric Suter}, \bibinfo{person}{Mallikarjun
  Shankar}, {and} \bibinfo{person}{Rafael~Ferreira Da~Silva}.}
  \bibinfo{year}{2022}\natexlab{a}.
\newblock \showarticletitle{WfBench: Automated Generation of Scientific
  Workflow Benchmarks}. In \bibinfo{booktitle}{\emph{2022 IEEE/ACM
  International Workshop on Performance Modeling, Benchmarking and Simulation
  of High Performance Computer Systems (PMBS)}}. \bibinfo{pages}{100--111}.
\newblock
\urldef\tempurl%
\url{https://doi.org/10.1109/PMBS56514.2022.00014}
\showDOI{\tempurl}


\bibitem[Coleman et~al\mbox{.}(2022b)]%
        {coleman2022-1}
\bibfield{author}{\bibinfo{person}{Tain\~a Coleman}, \bibinfo{person}{Henri
  Casanova}, \bibinfo{person}{Lo\"ic Pottier}, \bibinfo{person}{Manav Kaushik},
  \bibinfo{person}{Ewa Deelman}, {and} \bibinfo{person}{Rafael Ferreira~da
  Silva}.} \bibinfo{year}{2022}\natexlab{b}.
\newblock \showarticletitle{WfCommons: A Framework for Enabling Scientific
  Workflow Research and Development}.
\newblock \bibinfo{journal}{\emph{Future Generation Computer Systems}}
  \bibinfo{volume}{128} (\bibinfo{year}{2022}), \bibinfo{pages}{16--27}.
\newblock
\urldef\tempurl%
\url{https://doi.org/10.1016/j.future.2021.09.043}
\showDOI{\tempurl}


\bibitem[da~Silva et~al\mbox{.}(2023)]%
        {ferreira_da_silva2022}
\bibfield{author}{\bibinfo{person}{Rafael~Ferreira da Silva},
  \bibinfo{person}{Rosa~M. Badia}, \bibinfo{person}{Venkat Bala},
  \bibinfo{person}{Debbie Bard}, \bibinfo{person}{Peer-Timo Bremer},
  \bibinfo{person}{Ian Buckley}, \bibinfo{person}{Silvina Caino-Lores},
  \bibinfo{person}{Kyle Chard}, \bibinfo{person}{Carole Goble},
  \bibinfo{person}{Shantenu Jha}, \bibinfo{person}{Daniel~S. Katz},
  \bibinfo{person}{Daniel Laney}, \bibinfo{person}{Manish Parashar},
  \bibinfo{person}{Frederic Suter}, \bibinfo{person}{Nick Tyler},
  \bibinfo{person}{Thomas Uram}, \bibinfo{person}{Ilkay Altintas},
  \bibinfo{person}{Stefan Andersson}, \bibinfo{person}{William Arndt},
  \bibinfo{person}{Juan Aznar}, \bibinfo{person}{Jonathan Bader},
  \bibinfo{person}{Bartosz Balis}, \bibinfo{person}{Chris Blanton},
  \bibinfo{person}{Kelly~Rosa Braghetto}, \bibinfo{person}{Aharon Brodutch},
  \bibinfo{person}{Paul Brunk}, \bibinfo{person}{Henri Casanova},
  \bibinfo{person}{Alba~Cervera Lierta}, \bibinfo{person}{Justin Chigu},
  \bibinfo{person}{Taina Coleman}, \bibinfo{person}{Nick Collier},
  \bibinfo{person}{Iacopo Colonnelli}, \bibinfo{person}{Frederik Coppens},
  \bibinfo{person}{Michael Crusoe}, \bibinfo{person}{Will Cunningham},
  \bibinfo{person}{Bruno de Paula~Kinoshita}, \bibinfo{person}{Paolo~Di
  Tommaso}, \bibinfo{person}{Charles Doutriaux}, \bibinfo{person}{Matthew
  Downton}, \bibinfo{person}{Wael Elwasif}, \bibinfo{person}{Bjoern Enders},
  \bibinfo{person}{Chris Erdmann}, \bibinfo{person}{Thomas Fahringer},
  \bibinfo{person}{Ludmilla Figueiredo}, \bibinfo{person}{Rosa Filgueira},
  \bibinfo{person}{Martin Foltin}, \bibinfo{person}{Anne Fouilloux},
  \bibinfo{person}{Luiz Gadelha}, \bibinfo{person}{Andy Gallo},
  \bibinfo{person}{Artur~Garcia Saez}, \bibinfo{person}{Daniel Garijo},
  \bibinfo{person}{Roman Gerlach}, \bibinfo{person}{Ryan Grant},
  \bibinfo{person}{Samuel Grayson}, \bibinfo{person}{Patricia Grubel},
  \bibinfo{person}{Johan Gustafsson}, \bibinfo{person}{Valerie Hayot-Sasson},
  \bibinfo{person}{Oscar Hernandez}, \bibinfo{person}{Marcus Hilbrich},
  \bibinfo{person}{AnnMary Justine}, \bibinfo{person}{Ian Laflotte},
  \bibinfo{person}{Fabian Lehmann}, \bibinfo{person}{Andre Luckow},
  \bibinfo{person}{Jakob Luettgau}, \bibinfo{person}{Ketan Maheshwari},
  \bibinfo{person}{Motohiko Matsuda}, \bibinfo{person}{Doriana Medic},
  \bibinfo{person}{Pete Mendygral}, \bibinfo{person}{Marek Michalewicz},
  \bibinfo{person}{Jorji Nonaka}, \bibinfo{person}{Maciej Pawlik},
  \bibinfo{person}{Loic Pottier}, \bibinfo{person}{Line Pouchard},
  \bibinfo{person}{Mathias Putz}, \bibinfo{person}{Santosh~Kumar Radha},
  \bibinfo{person}{Lavanya Ramakrishnan}, \bibinfo{person}{Sashko Ristov},
  \bibinfo{person}{Paul Romano}, \bibinfo{person}{Daniel Rosendo},
  \bibinfo{person}{Martin Ruefenacht}, \bibinfo{person}{Katarzyna Rycerz},
  \bibinfo{person}{Nishant Saurabh}, \bibinfo{person}{Volodymyr Savchenko},
  \bibinfo{person}{Martin Schulz}, \bibinfo{person}{Christine Simpson},
  \bibinfo{person}{Raul Sirvent}, \bibinfo{person}{Tyler Skluzacek},
  \bibinfo{person}{Stian Soiland-Reyes}, \bibinfo{person}{Renan Souza},
  \bibinfo{person}{Sreenivas~Rangan Sukumar}, \bibinfo{person}{Ziheng Sun},
  \bibinfo{person}{Alan Sussman}, \bibinfo{person}{Douglas Thain},
  \bibinfo{person}{Mikhail Titov}, \bibinfo{person}{Benjamin Tovar},
  \bibinfo{person}{Aalap Tripathy}, \bibinfo{person}{Matteo Turilli},
  \bibinfo{person}{Bartosz Tuznik}, \bibinfo{person}{Hubertus van Dam},
  \bibinfo{person}{Aurelio Vivas}, \bibinfo{person}{Logan Ward},
  \bibinfo{person}{Patrick Widener}, \bibinfo{person}{Sean Wilkinson},
  \bibinfo{person}{Justyna Zawalska}, {and} \bibinfo{person}{Mahnoor
  Zulfiqar}.} \bibinfo{year}{2023}\natexlab{}.
\newblock \showarticletitle{Workflows Community Summit 2022: A Roadmap
  Revolution}.
\newblock  (\bibinfo{date}{March} \bibinfo{year}{2023}).
\newblock
\urldef\tempurl%
\url{https://doi.org/10.5281/zenodo.7750670}
\showDOI{\tempurl}


\bibitem[DeLeon et~al\mbox{.}(2015)]%
        {las-15-tas}
\bibfield{author}{\bibinfo{person}{Robert~L. DeLeon},
  \bibinfo{person}{Thomas~R. Furlani}, \bibinfo{person}{Steven~M. Gallo},
  \bibinfo{person}{Joseph~P. White}, \bibinfo{person}{Matthew~D. Jones},
  \bibinfo{person}{Abani Patra}, \bibinfo{person}{Martins Innus},
  \bibinfo{person}{Thomas Yearke}, \bibinfo{person}{Jeffrey~T. Palmer},
  \bibinfo{person}{Jeanette~M. Sperhac}, \bibinfo{person}{Ryan Rathsam},
  \bibinfo{person}{Nikolay Simakov}, \bibinfo{person}{Gregor von Laszewski},
  {and} \bibinfo{person}{Fugang Wang}.} \bibinfo{year}{2015}\natexlab{}.
\newblock \showarticletitle{TAS View of XSEDE Users and Usage}. In
  \bibinfo{booktitle}{\emph{Proceedings of the 2015 XSEDE Conference:
  Scientific Advancements Enabled by Enhanced Cyberinfrastructure}} (St. Louis,
  Missouri) \emph{(\bibinfo{series}{XSEDE '15})}. \bibinfo{publisher}{ACM},
  \bibinfo{address}{New York, NY, USA}, Article \bibinfo{articleno}{21},
  \bibinfo{numpages}{8}~pages.
\newblock
\showISBNx{978-1-4503-3720-5}
\urldef\tempurl%
\url{https://doi.org/10.1145/2792745.2792766}
\showDOI{\tempurl}


\bibitem[Devlin et~al\mbox{.}(2019)]%
        {bert-report}
\bibfield{author}{\bibinfo{person}{Jacob Devlin}, \bibinfo{person}{Ming-Wei
  Chang}, \bibinfo{person}{Kenton Lee}, {and} \bibinfo{person}{Kristina
  Toutanova}.} \bibinfo{year}{2019}\natexlab{}.
\newblock \showarticletitle{BERT: Pre-training of Deep Bidirectional
  Transformers for Language Understanding}.
\newblock  (\bibinfo{year}{2019}).
\newblock
\urldef\tempurl%
\url{https://doi.org/10.48550/arXiv.1810.04805}
\showDOI{\tempurl}
\showeprint[arxiv]{1810.04805}~[cs.CL]


\bibitem[Di~Tommaso et~al\mbox{.}(2017)]%
        {di_tommaso_nextflow_2017}
\bibfield{author}{\bibinfo{person}{Paolo Di~Tommaso}, \bibinfo{person}{Maria
  Chatzou}, \bibinfo{person}{Evan~W Floden}, \bibinfo{person}{Pablo~Prieto
  Barja}, \bibinfo{person}{Emilio Palumbo}, {and} \bibinfo{person}{Cedric
  Notredame}.} \bibinfo{year}{2017}\natexlab{}.
\newblock \showarticletitle{Nextflow enables reproducible computational
  workflows}.
\newblock \bibinfo{journal}{\emph{Nature Biotechnology}} \bibinfo{volume}{35},
  \bibinfo{number}{4} (\bibinfo{date}{April} \bibinfo{year}{2017}),
  \bibinfo{pages}{316--319}.
\newblock
\showISSN{1087-0156, 1546-1696}
\urldef\tempurl%
\url{https://doi.org/10.1038/nbt.3820}
\showDOI{\tempurl}


\bibitem[Diaz et~al\mbox{.}(2012)]%
        {las-12-imagemanagement}
\bibfield{author}{\bibinfo{person}{Javier Diaz}, \bibinfo{person}{Gregor von
  Laszewski}, \bibinfo{person}{Fugang Wang}, {and} \bibinfo{person}{Geoffrey~C.
  Fox}.} \bibinfo{year}{2012}\natexlab{}.
\newblock \showarticletitle{{Abstract Image Management and Universal Image
  Registration for Cloud and HPC Infrastructures}}. In
  \bibinfo{booktitle}{\emph{IEEE Cloud 2012}}. \bibinfo{address}{Honolulu}.
\newblock
\urldef\tempurl%
\url{https://doi.org/10.1109/CLOUD.2012.94}
\showDOI{\tempurl}


\bibitem[Feng and Cameron(2007)]%
        {green500}
\bibfield{author}{\bibinfo{person}{Wu-chun Feng} {and} \bibinfo{person}{Kirk
  Cameron}.} \bibinfo{year}{2007}\natexlab{}.
\newblock \showarticletitle{{The Green500 List: Encouraging Sustainable
  Supercomputing}}.
\newblock \bibinfo{journal}{\emph{Computer}} \bibinfo{volume}{40},
  \bibinfo{number}{12} (\bibinfo{date}{dec} \bibinfo{year}{2007}),
  \bibinfo{pages}{50–55}.
\newblock
\showISSN{0018-9162}
\urldef\tempurl%
\url{https://doi.org/10.1109/MC.2007.445}
\showDOI{\tempurl}


\bibitem[Ferreira~da Silva et~al\mbox{.}(2024)]%
        {ferreira_da_silva2024}
\bibfield{author}{\bibinfo{person}{Rafael Ferreira~da Silva},
  \bibinfo{person}{Deborah Bard}, \bibinfo{person}{Kyle Chard},
  \bibinfo{person}{de~Witt Shaun}, \bibinfo{person}{Ian~T. Foster},
  \bibinfo{person}{Tom Gibbs}, \bibinfo{person}{Carole Goble},
  \bibinfo{person}{William Godoy}, \bibinfo{person}{Johan Gustafsson},
  \bibinfo{person}{Utz-Uwe Haus}, \bibinfo{person}{Stephen Hudson},
  \bibinfo{person}{Shantenu Jha}, \bibinfo{person}{Laila Los},
  \bibinfo{person}{Drew Paine}, \bibinfo{person}{Frédéric Suter},
  \bibinfo{person}{Logan Ward}, \bibinfo{person}{Sean Wilkinson},
  \bibinfo{person}{Marcos Amaris}, \bibinfo{person}{Yadu Babuji},
  \bibinfo{person}{Jonathan Bader}, \bibinfo{person}{Riccardo Balin},
  \bibinfo{person}{Daniel Balouek}, \bibinfo{person}{Sarah Beecroft},
  \bibinfo{person}{Khalid Belhajjame}, \bibinfo{person}{Rajat Bhattarai},
  \bibinfo{person}{Wes Brewer}, \bibinfo{person}{Paul Brunk},
  \bibinfo{person}{Silvina Caino-Lores}, \bibinfo{person}{Henri Casanova},
  \bibinfo{person}{Daniela Cassol}, \bibinfo{person}{Jared Coleman},
  \bibinfo{person}{Taina Coleman}, \bibinfo{person}{Iacopo Colonnelli},
  \bibinfo{person}{Anderson~Andrei Da~Silva}, \bibinfo{person}{Daniel de
  Oliveira}, \bibinfo{person}{Pascal Elahi}, \bibinfo{person}{Nour Elfaramawy},
  \bibinfo{person}{Wael Elwasif}, \bibinfo{person}{Brian Etz},
  \bibinfo{person}{Thomas Fahringer}, \bibinfo{person}{Wesley Ferreira},
  \bibinfo{person}{Rosa Filgueira}, \bibinfo{person}{Jacob Fosso~Tande},
  \bibinfo{person}{Luiz Gadelha}, \bibinfo{person}{Andy Gallo},
  \bibinfo{person}{Daniel Garijo}, \bibinfo{person}{Yiannis Georgiou},
  \bibinfo{person}{Philipp Gritsch}, \bibinfo{person}{Patricia Grubel},
  \bibinfo{person}{Amal Gueroudji}, \bibinfo{person}{Quentin Guilloteau},
  \bibinfo{person}{Carlo Hamalainen}, \bibinfo{person}{Rolando Hong~Enriquez},
  \bibinfo{person}{Lauren Huet}, \bibinfo{person}{Kevin Hunter~Kesling},
  \bibinfo{person}{Paula Iborra}, \bibinfo{person}{Shiva Jahangiri},
  \bibinfo{person}{Jan Janssen}, \bibinfo{person}{Joe Jordan},
  \bibinfo{person}{Sehrish Kanwal}, \bibinfo{person}{Liliane Kunstmann},
  \bibinfo{person}{Fabian Lehmann}, \bibinfo{person}{Ulf Leser},
  \bibinfo{person}{Chen Li}, \bibinfo{person}{Peini Liu},
  \bibinfo{person}{Jakob Luettgau}, \bibinfo{person}{Richard Lupat},
  \bibinfo{person}{Jose M.~Fernandez}, \bibinfo{person}{Ketan Maheshwari},
  \bibinfo{person}{Tanu Malik}, \bibinfo{person}{Jack Marquez},
  \bibinfo{person}{Motohiko Matsuda}, \bibinfo{person}{Doriana Medic},
  \bibinfo{person}{Somayeh Mohammadi}, \bibinfo{person}{Alberto Mulone},
  \bibinfo{person}{John-Luke Navarro}, \bibinfo{person}{Kin~Wai Ng},
  \bibinfo{person}{Klaus Noelp}, \bibinfo{person}{Bruno P.~Kinoshita},
  \bibinfo{person}{Ryan Prout}, \bibinfo{person}{Michael R.~Crusoe},
  \bibinfo{person}{Sashko Ristov}, \bibinfo{person}{Stefan Robila},
  \bibinfo{person}{Daniel Rosendo}, \bibinfo{person}{Billy Rowell},
  \bibinfo{person}{Jedrzej Rybicki}, \bibinfo{person}{Hector Sanchez},
  \bibinfo{person}{Nishant Saurabh}, \bibinfo{person}{Sumit~Kumar Saurav},
  \bibinfo{person}{Tom Scogland}, \bibinfo{person}{Dinindu Senanayake},
  \bibinfo{person}{Woong Shin}, \bibinfo{person}{Raul Sirvent},
  \bibinfo{person}{Tyler Skluzacek}, \bibinfo{person}{Barry Sly-Delgado},
  \bibinfo{person}{Stian Soiland-Reyes}, \bibinfo{person}{Abel Souza},
  \bibinfo{person}{Renan Souza}, \bibinfo{person}{Domenico Talia},
  \bibinfo{person}{Nathan Tallent}, \bibinfo{person}{Lauritz Thamsen},
  \bibinfo{person}{Mikhail Titov}, \bibinfo{person}{Benjamin Tovar},
  \bibinfo{person}{Karan Vahi}, \bibinfo{person}{Eric Vardar-Irrgang},
  \bibinfo{person}{Edite Vartina}, \bibinfo{person}{Yuandou Wang},
  \bibinfo{person}{Merridee Wouters}, \bibinfo{person}{Qi Yu},
  \bibinfo{person}{Ziad Al~Bkhetan}, {and} \bibinfo{person}{Mahnoor Zulfiqar}.}
  \bibinfo{year}{2024}\natexlab{}.
\newblock \bibinfo{booktitle}{\emph{Workflows {Community} {Summit} 2024:
  {Future} {Trends} and {Challenges} in {Scientific} {Workflows}}}.
\newblock \bibinfo{type}{{T}echnical {R}eport}. \bibinfo{institution}{Zenodo}.
\newblock
\urldef\tempurl%
\url{https://doi.org/10.5281/ZENODO.13844758}
\showDOI{\tempurl}


\bibitem[Font et~al\mbox{.}(2024)]%
        {Font2024}
\bibfield{author}{\bibinfo{person}{Bernat Font}, \bibinfo{person}{Francisco
  Alcántara-Ávila}, \bibinfo{person}{Jean Rabault}, \bibinfo{person}{Ricardo
  Vinuesa}, {and} \bibinfo{person}{Oriol Lehmkuhl}.}
  \bibinfo{year}{2024}\natexlab{}.
\newblock \showarticletitle{Active flow control of a turbulent separation
  bubble through deep reinforcement learning}.
\newblock \bibinfo{journal}{\emph{Journal of Physics: Conference Series}}
  \bibinfo{volume}{2753}, \bibinfo{number}{1} (\bibinfo{date}{apr}
  \bibinfo{year}{2024}), \bibinfo{pages}{012022}.
\newblock
\urldef\tempurl%
\url{https://doi.org/10.1088/1742-6596/2753/1/012022}
\showDOI{\tempurl}


\bibitem[Fox et~al\mbox{.}(2012)]%
        {las-12-fg-bookchapter}
\bibfield{author}{\bibinfo{person}{Geoffrey~C. Fox}, \bibinfo{person}{Gregor
  von Laszewski}, \bibinfo{person}{Javier Diaz}, \bibinfo{person}{Kate Keahey},
  \bibinfo{person}{Jose Fortes}, \bibinfo{person}{Renato Figueiredo},
  \bibinfo{person}{Shava Smallen}, \bibinfo{person}{Warren Smith}, {and}
  \bibinfo{person}{Andrew Grimshaw}.} \bibinfo{year}{2012}\natexlab{}.
\newblock \showarticletitle{{FutureGrid - a reconfigurable testbed for Cloud,
  HPC and Grid Computing}}.
\newblock In \bibinfo{booktitle}{\emph{{Contemporary HPC Architectures}}
  (\bibinfo{edition}{draft} ed.)}.
\newblock
\urldef\tempurl%
\url{https://laszewski.github.io/papers/vonLaszewski-12-fg-bookchapter.pdf}
\showURL{%
\tempurl}


\bibitem[Fox et~al\mbox{.}(2017)]%
        {las-17-futuregrid}
\bibfield{author}{\bibinfo{person}{Geoffrey~C Fox}, \bibinfo{person}{Gregor von
  Laszewski}, \bibinfo{person}{Javier Diaz}, \bibinfo{person}{Kate Keahey},
  \bibinfo{person}{Jose Fortes}, \bibinfo{person}{Renato Figueiredo},
  \bibinfo{person}{Shava Smallen}, \bibinfo{person}{Warren Smith}, {and}
  \bibinfo{person}{Andrew Grimshaw}.} \bibinfo{year}{2017}\natexlab{}.
\newblock \showarticletitle{Futuregrid: a Reconfigurable Testbed for Cloud,
  Hpc, and Grid Computing}.
\newblock In \bibinfo{booktitle}{\emph{Contemporary High Performance
  Computing}}. \bibinfo{publisher}{Chapman and Hall/CRC},
  \bibinfo{pages}{603--635}.
\newblock


\bibitem[Friedman and Wise(1978)]%
        {friedman-futures}
\bibfield{author}{\bibinfo{person}{Friedman} {and} \bibinfo{person}{Wise}.}
  \bibinfo{year}{1978}\natexlab{}.
\newblock \showarticletitle{Aspects of Applicative Programming for Parallel
  Processing}.
\newblock \bibinfo{journal}{\emph{IEEE Trans. Comput.}} \bibinfo{volume}{C-27},
  \bibinfo{number}{4} (\bibinfo{year}{1978}), \bibinfo{pages}{289--296}.
\newblock
\urldef\tempurl%
\url{https://doi.org/10.1109/TC.1978.1675100}
\showDOI{\tempurl}


\bibitem[{Gemma Team} et~al\mbox{.}(2024)]%
        {gemma}
\bibfield{author}{\bibinfo{person}{Thomas~Mesnard {Gemma Team}},
  \bibinfo{person}{Cassidy Hardin}, \bibinfo{person}{Robert Dadashi},
  \bibinfo{person}{Surya Bhupatiraju}, \bibinfo{person}{Shreya Pathak},
  \bibinfo{person}{Laurent Sifre}, \bibinfo{person}{Morgane Rivière},
  \bibinfo{person}{Mihir~Sanjay Kale}, \bibinfo{person}{Juliette Love},
  \bibinfo{person}{Pouya Tafti}, \bibinfo{person}{Léonard Hussenot},
  \bibinfo{person}{Pier~Giuseppe Sessa}, \bibinfo{person}{Aakanksha Chowdhery},
  \bibinfo{person}{Adam Roberts}, \bibinfo{person}{Aditya Barua},
  \bibinfo{person}{Alex Botev}, \bibinfo{person}{Alex Castro-Ros},
  \bibinfo{person}{Ambrose Slone}, \bibinfo{person}{Amélie Héliou},
  \bibinfo{person}{Andrea Tacchetti}, \bibinfo{person}{Anna Bulanova},
  \bibinfo{person}{Antonia Paterson}, \bibinfo{person}{Beth Tsai},
  \bibinfo{person}{Bobak Shahriari}, \bibinfo{person}{Charline~Le Lan},
  \bibinfo{person}{Christopher~A. Choquette-Choo}, \bibinfo{person}{Clément
  Crepy}, \bibinfo{person}{Daniel Cer}, \bibinfo{person}{Daphne Ippolito},
  \bibinfo{person}{David Reid}, \bibinfo{person}{Elena Buchatskaya},
  \bibinfo{person}{Eric Ni}, \bibinfo{person}{Eric Noland},
  \bibinfo{person}{Geng Yan}, \bibinfo{person}{George Tucker},
  \bibinfo{person}{George-Christian Muraru}, \bibinfo{person}{Grigory
  Rozhdestvenskiy}, \bibinfo{person}{Henryk Michalewski}, \bibinfo{person}{Ian
  Tenney}, \bibinfo{person}{Ivan Grishchenko}, \bibinfo{person}{Jacob Austin},
  \bibinfo{person}{James Keeling}, \bibinfo{person}{Jane Labanowski},
  \bibinfo{person}{Jean-Baptiste Lespiau}, \bibinfo{person}{Jeff Stanway},
  \bibinfo{person}{Jenny Brennan}, \bibinfo{person}{Jeremy Chen},
  \bibinfo{person}{Johan Ferret}, \bibinfo{person}{Justin Chiu},
  \bibinfo{person}{Justin Mao-Jones}, \bibinfo{person}{Katherine Lee},
  \bibinfo{person}{Kathy Yu}, \bibinfo{person}{Katie Millican},
  \bibinfo{person}{Lars~Lowe Sjoesund}, \bibinfo{person}{Lisa Lee},
  \bibinfo{person}{Lucas Dixon}, \bibinfo{person}{Machel Reid},
  \bibinfo{person}{Maciej Mikuła}, \bibinfo{person}{Mateo Wirth},
  \bibinfo{person}{Michael Sharman}, \bibinfo{person}{Nikolai Chinaev},
  \bibinfo{person}{Nithum Thain}, \bibinfo{person}{Olivier Bachem},
  \bibinfo{person}{Oscar Chang}, \bibinfo{person}{Oscar Wahltinez},
  \bibinfo{person}{Paige Bailey}, \bibinfo{person}{Paul Michel},
  \bibinfo{person}{Petko Yotov}, \bibinfo{person}{Rahma Chaabouni},
  \bibinfo{person}{Ramona Comanescu}, \bibinfo{person}{Reena Jana},
  \bibinfo{person}{Rohan Anil}, \bibinfo{person}{Ross McIlroy},
  \bibinfo{person}{Ruibo Liu}, \bibinfo{person}{Ryan Mullins},
  \bibinfo{person}{Samuel~L Smith}, \bibinfo{person}{Sebastian Borgeaud},
  \bibinfo{person}{Sertan Girgin}, \bibinfo{person}{Sholto Douglas},
  \bibinfo{person}{Shree Pandya}, \bibinfo{person}{Siamak Shakeri},
  \bibinfo{person}{Soham De}, \bibinfo{person}{Ted Klimenko},
  \bibinfo{person}{Tom Hennigan}, \bibinfo{person}{Vlad Feinberg},
  \bibinfo{person}{Wojciech Stokowiec}, \bibinfo{person}{Yu hui Chen},
  \bibinfo{person}{Zafarali Ahmed}, \bibinfo{person}{Zhitao Gong},
  \bibinfo{person}{Tris Warkentin}, \bibinfo{person}{Ludovic Peran},
  \bibinfo{person}{Minh Giang}, \bibinfo{person}{Clément Farabet},
  \bibinfo{person}{Oriol Vinyals}, \bibinfo{person}{Jeff Dean},
  \bibinfo{person}{Koray Kavukcuoglu}, \bibinfo{person}{Demis Hassabis},
  \bibinfo{person}{Zoubin Ghahramani}, \bibinfo{person}{Douglas Eck},
  \bibinfo{person}{Joelle Barral}, \bibinfo{person}{Fernando Pereira},
  \bibinfo{person}{Eli Collins}, \bibinfo{person}{Armand Joulin},
  \bibinfo{person}{Noah Fiedel}, \bibinfo{person}{Evan Senter},
  \bibinfo{person}{Alek Andreev}, {and} \bibinfo{person}{Kathleen Kenealy}.}
  \bibinfo{year}{2024}\natexlab{}.
\newblock \showarticletitle{Gemma: Open Models Based on Gemini Research and
  Technology}.
\newblock \bibinfo{journal}{\emph{arXiv}} (\bibinfo{date}{April}
  \bibinfo{year}{2024}).
\newblock
\urldef\tempurl%
\url{https://doi.org/10.48550/arXiv.2403.08295}
\showDOI{\tempurl}
\showeprint{2403.08295}
\newblock
\shownote{cs.CL}.


\bibitem[Gustafsson et~al\mbox{.}(2024)]%
        {gustafsson2024}
\bibfield{author}{\bibinfo{person}{Ove Johan~Ragnar Gustafsson},
  \bibinfo{person}{Sean~R. Wilkinson}, \bibinfo{person}{Finn Bacall},
  \bibinfo{person}{Luca Pireddu}, \bibinfo{person}{Stian Soiland-Reyes},
  \bibinfo{person}{Simone Leo}, \bibinfo{person}{Stuart Owen},
  \bibinfo{person}{Nick Juty}, \bibinfo{person}{José~M. Fernández},
  \bibinfo{person}{Björn Grüning}, \bibinfo{person}{Tom Brown},
  \bibinfo{person}{Hervé Ménager}, \bibinfo{person}{Salvador
  Capella-Gutierrez}, \bibinfo{person}{Frederik Coppens}, {and}
  \bibinfo{person}{Carole Goble}.} \bibinfo{year}{2024}\natexlab{}.
\newblock \showarticletitle{{WorkflowHub}: a registry for computational
  workflows}.
\newblock  (\bibinfo{year}{2024}).
\newblock
\urldef\tempurl%
\url{https://doi.org/10.48550/arXiv.2410.06941}
\showDOI{\tempurl}
\showeprint[arxiv]{2410.06941}~[cs.DL]


\bibitem[{Hewlett Packard Enterprise}(2025)]%
        {smartsim-repo}
\bibfield{author}{\bibinfo{person}{{Hewlett Packard Enterprise}}.}
  \bibinfo{year}{2025}\natexlab{}.
\newblock \showarticletitle{SmartSim Source Code and Documentation}.
\newblock \bibinfo{journal}{\emph{GitHub}} (\bibinfo{date}{March}
  \bibinfo{year}{2025}).
\newblock
\urldef\tempurl%
\url{https://github.com/CrayLabs/SmartSim/}
\showURL{%
\tempurl}


\bibitem[{InCommon}(2024)]%
        {incommon}
\bibfield{author}{\bibinfo{person}{{InCommon}}.}
  \bibinfo{year}{2024}\natexlab{}.
\newblock \showarticletitle{{Home Page}}.
\newblock \bibinfo{howpublished}{Web Page}.
\newblock  (\bibinfo{date}{Oct.} \bibinfo{year}{2024}).
\newblock
\urldef\tempurl%
\url{https://incommon.org/}
\showURL{%
\tempurl}
\newblock
\shownote{[Online; accessed: 03/01/2025]}.


\bibitem[Jacobsen et~al\mbox{.}(2020)]%
        {jacobsen2020fair}
\bibfield{author}{\bibinfo{person}{Annika Jacobsen}, \bibinfo{person}{Ricardo
  de Miranda~Azevedo}, \bibinfo{person}{Nick Juty}, \bibinfo{person}{Dominique
  Batista}, \bibinfo{person}{Simon Coles}, \bibinfo{person}{Ronald Cornet},
  \bibinfo{person}{M{\'e}lanie Courtot}, \bibinfo{person}{Merc{\`e} Crosas},
  \bibinfo{person}{Michel Dumontier}, \bibinfo{person}{Chris~T Evelo},
  {et~al\mbox{.}}} \bibinfo{year}{2020}\natexlab{}.
\newblock \showarticletitle{FAIR principles: interpretations and implementation
  considerations}.
\newblock \bibinfo{journal}{\emph{Data intelligence}} \bibinfo{volume}{2},
  \bibinfo{number}{1-2} (\bibinfo{year}{2020}), \bibinfo{pages}{10--29}.
\newblock
\urldef\tempurl%
\url{https://doi.org/10.1162/dint_r_00024}
\showDOI{\tempurl}


\bibitem[Jiang et~al\mbox{.}(2024)]%
        {jiang2024megascalescalinglargelanguage}
\bibfield{author}{\bibinfo{person}{Ziheng Jiang}, \bibinfo{person}{Haibin Lin},
  \bibinfo{person}{Yinmin Zhong}, \bibinfo{person}{Qi Huang},
  \bibinfo{person}{Yangrui Chen}, \bibinfo{person}{Zhi Zhang},
  \bibinfo{person}{Yanghua Peng}, \bibinfo{person}{Xiang Li},
  \bibinfo{person}{Cong Xie}, \bibinfo{person}{Shibiao Nong},
  \bibinfo{person}{Yulu Jia}, \bibinfo{person}{Sun He},
  \bibinfo{person}{Hongmin Chen}, \bibinfo{person}{Zhihao Bai},
  \bibinfo{person}{Qi Hou}, \bibinfo{person}{Shipeng Yan},
  \bibinfo{person}{Ding Zhou}, \bibinfo{person}{Yiyao Sheng},
  \bibinfo{person}{Zhuo Jiang}, \bibinfo{person}{Haohan Xu},
  \bibinfo{person}{Haoran Wei}, \bibinfo{person}{Zhang Zhang},
  \bibinfo{person}{Pengfei Nie}, \bibinfo{person}{Leqi Zou},
  \bibinfo{person}{Sida Zhao}, \bibinfo{person}{Liang Xiang},
  \bibinfo{person}{Zherui Liu}, \bibinfo{person}{Zhe Li},
  \bibinfo{person}{Xiaoying Jia}, \bibinfo{person}{Jianxi Ye},
  \bibinfo{person}{Xin Jin}, {and} \bibinfo{person}{Xin Liu}.}
  \bibinfo{year}{2024}\natexlab{}.
\newblock \showarticletitle{MegaScale: Scaling Large Language Model Training to
  More Than 10,000 GPUs}.
\newblock  (\bibinfo{year}{2024}).
\newblock
\urldef\tempurl%
\url{https://doi.org/10.48550/arXiv.2402.15627}
\showDOI{\tempurl}
\showeprint{2402.15627}
\newblock
\shownote{cs.LG}.


\bibitem[{Kepler}(2011)]%
        {www-kepler}
{Kepler} \bibinfo{year}{2011}\natexlab{}.
\newblock \bibinfo{title}{The Kepler Project}.
\newblock
\newblock
\urldef\tempurl%
\url{https://kepler-project.org/}
\showURL{%
\tempurl}
\newblock
\shownote{[Online; accessed 03/01/2025]}.


\bibitem[Kherroubi~Garcia et~al\mbox{.}(2025)]%
        {garcia2025}
\bibfield{author}{\bibinfo{person}{Ismael Kherroubi~Garcia},
  \bibinfo{person}{Christopher Erdmann}, \bibinfo{person}{Sandra Gesing},
  \bibinfo{person}{Michael Barton}, \bibinfo{person}{Lauren Cadwallader},
  \bibinfo{person}{Geerten Hengeveld}, \bibinfo{person}{Christine~R.
  Kirkpatrick}, \bibinfo{person}{Kathryn Knight}, \bibinfo{person}{Carsten
  Lemmen}, \bibinfo{person}{Rebecca Ringuette}, \bibinfo{person}{Qing Zhan},
  \bibinfo{person}{Melissa Harrison}, \bibinfo{person}{Feilim Mac~Gabhann},
  \bibinfo{person}{Natalie Meyers}, \bibinfo{person}{Cailean Osborne},
  \bibinfo{person}{Charlotte Till}, \bibinfo{person}{Paul Brenner},
  \bibinfo{person}{Matt Buys}, \bibinfo{person}{Min Chen},
  \bibinfo{person}{Allen Lee}, \bibinfo{person}{Jason Papin}, {and}
  \bibinfo{person}{Yuhan Rao}.} \bibinfo{year}{2025}\natexlab{}.
\newblock \showarticletitle{Ten simple rules for good model-sharing practices}.
\newblock \bibinfo{journal}{\emph{PLOS Computational Biology}}
  \bibinfo{volume}{21}, \bibinfo{number}{1} (\bibinfo{date}{01}
  \bibinfo{year}{2025}), \bibinfo{pages}{1--21}.
\newblock
\urldef\tempurl%
\url{https://doi.org/10.1371/journal.pcbi.1012702}
\showDOI{\tempurl}


\bibitem[Kirkpatrick(2023)]%
        {kirkpatrick2023}
\bibfield{author}{\bibinfo{person}{Christine Kirkpatrick}.}
  \bibinfo{year}{2023}\natexlab{}.
\newblock \showarticletitle{FAIRIST of them all: Meeting researchers where they
  are with just-in-time, FAIR implementation advice}.
\newblock \bibinfo{howpublished}{Invited talk at the WORKS workshop at the
  Supercomputing Conference 2023, Denver}.
\newblock  (\bibinfo{year}{2023}).
\newblock


\bibitem[Kirkpatrick et~al\mbox{.}(2025)]%
        {kirkpatrick2025}
\bibfield{author}{\bibinfo{person}{Christine Kirkpatrick},
  \bibinfo{person}{Gregor~von Laszewski}, \bibinfo{person}{Gregg Barrett},
  \bibinfo{person}{Wesley Brewer}, \bibinfo{person}{Julie Christopher},
  \bibinfo{person}{Inês Dutra}, \bibinfo{person}{Murali Emani},
  \bibinfo{person}{Piotr Luszczek}, \bibinfo{person}{Mallikarjun~(Arjun)
  Shankar}, \bibinfo{person}{Juri Papay}, \bibinfo{person}{Jeyan
  Thiyagalingam}, {and} \bibinfo{person}{Geoffrey Fox}.}
  \bibinfo{year}{2025}\natexlab{}.
\newblock \showarticletitle{Optimizing Machine Learning Benchmarking: A FAIR
  Approach to Energy Efficiency and Data Transparency}.
\newblock  (\bibinfo{year}{2025}).
\newblock
\newblock
\shownote{[unpublished draft]}.


\bibitem[Kurz et~al\mbox{.}(2023)]%
        {kurz2023deep}
\bibfield{author}{\bibinfo{person}{Marius Kurz}, \bibinfo{person}{Philipp
  Offenh{\"a}user}, {and} \bibinfo{person}{Andrea Beck}.}
  \bibinfo{year}{2023}\natexlab{}.
\newblock \showarticletitle{Deep reinforcement learning for turbulence modeling
  in large eddy simulations}.
\newblock \bibinfo{journal}{\emph{International journal of heat and fluid
  flow}}  \bibinfo{volume}{99} (\bibinfo{year}{2023}), \bibinfo{pages}{109094}.
\newblock
\urldef\tempurl%
\url{https://doi.org/10.1016/j.ijheatfluidflow.2022.109094}
\showDOI{\tempurl}


\bibitem[Lee et~al\mbox{.}(2021)]%
        {lee2021}
\bibfield{author}{\bibinfo{person}{Hyungro Lee}, \bibinfo{person}{Andre
  Merzky}, \bibinfo{person}{Li Tan}, \bibinfo{person}{Mikhail Titov},
  \bibinfo{person}{Matteo Turilli}, \bibinfo{person}{Dario Alfe},
  \bibinfo{person}{Agastya Bhati}, \bibinfo{person}{Alex Brace},
  \bibinfo{person}{Austin Clyde}, \bibinfo{person}{Peter Coveney},
  \bibinfo{person}{Heng Ma}, \bibinfo{person}{Arvind Ramanathan},
  \bibinfo{person}{Rick Stevens}, \bibinfo{person}{Anda Trifan},
  \bibinfo{person}{Hubertus Van~Dam}, \bibinfo{person}{Shunzhou Wan},
  \bibinfo{person}{Sean Wilkinson}, {and} \bibinfo{person}{Shantenu Jha}.}
  \bibinfo{year}{2021}\natexlab{}.
\newblock \showarticletitle{Scalable HPC \& AI infrastructure for COVID-19
  therapeutics}. In \bibinfo{booktitle}{\emph{Proceedings of the Platform for
  Advanced Scientific Computing Conference}} (Geneva, Switzerland)
  \emph{(\bibinfo{series}{PASC '21})}. \bibinfo{publisher}{Association for
  Computing Machinery}, \bibinfo{address}{New York, NY, USA}, Article
  \bibinfo{articleno}{2}, \bibinfo{numpages}{13}~pages.
\newblock
\showISBNx{9781450385633}
\urldef\tempurl%
\url{https://doi.org/10.1145/3468267.3470573}
\showDOI{\tempurl}


\bibitem[Liu et~al\mbox{.}(2021)]%
        {liu2021pi3nn}
\bibfield{author}{\bibinfo{person}{Siyan Liu}, \bibinfo{person}{Pei Zhang},
  \bibinfo{person}{Dan Lu}, {and} \bibinfo{person}{Guannan Zhang}.}
  \bibinfo{year}{2021}\natexlab{}.
\newblock \showarticletitle{PI3NN: Out-of-distribution-aware prediction
  intervals from three neural networks}.
\newblock \bibinfo{journal}{\emph{arXiv preprint arXiv:2108.02327}}
  (\bibinfo{year}{2021}).
\newblock


\bibitem[Maric et~al\mbox{.}(2024)]%
        {Maric2024OpenFOAM}
\bibfield{author}{\bibinfo{person}{Tomislav Maric},
  \bibinfo{person}{Mohammed~Elwardi Fadeli}, \bibinfo{person}{Alessandro
  Rigazzi}, \bibinfo{person}{Andrew Shao}, {and} \bibinfo{person}{Andre
  Weiner}.} \bibinfo{year}{2024}\natexlab{}.
\newblock \showarticletitle{Combining machine learning with computational fluid
  dynamics using OpenFOAM and SmartSim}.
\newblock \bibinfo{journal}{\emph{Meccanica}} (\bibinfo{date}{20 Apr}
  \bibinfo{year}{2024}).
\newblock
\showISSN{1572-9648}
\urldef\tempurl%
\url{https://doi.org/10.1007/s11012-024-01797-z}
\showDOI{\tempurl}


\bibitem[Martinez et~al\mbox{.}(2018)]%
        {martinez2018deep}
\bibfield{author}{\bibinfo{person}{Daniel Martinez}, \bibinfo{person}{Wesley
  Brewer}, \bibinfo{person}{Gregory Behm}, \bibinfo{person}{Andrew Strelzoff},
  \bibinfo{person}{Andrew Wilson}, {and} \bibinfo{person}{Daniel Wade}.}
  \bibinfo{year}{2018}\natexlab{}.
\newblock \showarticletitle{Deep learning evolutionary optimization for
  regression of rotorcraft vibrational spectra}. In
  \bibinfo{booktitle}{\emph{2018 IEEE/ACM Machine Learning in HPC Environments
  (MLHPC)}}. IEEE, \bibinfo{pages}{57--66}.
\newblock
\urldef\tempurl%
\url{https://doi.org/10.1109/MLHPC.2018.8638645}
\showDOI{\tempurl}


\bibitem[Martinez-Gonzalez et~al\mbox{.}(2022)]%
        {martinez2022roam}
\bibfield{author}{\bibinfo{person}{Daniel~A Martinez-Gonzalez},
  \bibinfo{person}{Dylan Jude}, \bibinfo{person}{Jayanarayanan Sitaraman},
  \bibinfo{person}{Wesley Brewer}, {and} \bibinfo{person}{Andrew~M Wissink}.}
  \bibinfo{year}{2022}\natexlab{}.
\newblock \showarticletitle{{ROAM-ML}: A reduced order aerodynamic module
  augmented with neural network digital surrogates}. In
  \bibinfo{booktitle}{\emph{AIAA SCITECH 2022 Forum}}. \bibinfo{pages}{1248}.
\newblock
\urldef\tempurl%
\url{https://doi.org/10.2514/6.2022-1248}
\showDOI{\tempurl}


\bibitem[McClure et~al\mbox{.}(2020)]%
        {mcclure2020}
\bibfield{author}{\bibinfo{person}{James~E. McClure}, \bibinfo{person}{Junqi
  Yin}, \bibinfo{person}{Ryan~T. Armstrong}, \bibinfo{person}{Ketan~C.
  Maheshwari}, \bibinfo{person}{Sean Wilkinson}, \bibinfo{person}{Lucas Vlcek},
  \bibinfo{person}{Ying Da~Wang}, \bibinfo{person}{Mark~A. Berrill}, {and}
  \bibinfo{person}{Mark Rivers}.} \bibinfo{year}{2020}\natexlab{}.
\newblock \showarticletitle{Toward Real-Time Analysis of Synchrotron
  Micro-Tomography Data: Accelerating Experimental Workflows with AI and HPC}.
  In \bibinfo{booktitle}{\emph{Driving Scientific and Engineering Discoveries
  Through the Convergence of HPC, Big Data and AI}},
  \bibfield{editor}{\bibinfo{person}{Jeffrey Nichols}, \bibinfo{person}{Becky
  Verastegui}, \bibinfo{person}{Arthur~`Barney' Maccabe},
  \bibinfo{person}{Oscar Hernandez}, \bibinfo{person}{Suzanne Parete-Koon},
  {and} \bibinfo{person}{Theresa Ahearn}} (Eds.). \bibinfo{publisher}{Springer
  International Publishing}, \bibinfo{address}{Cham},
  \bibinfo{pages}{226--239}.
\newblock
\showISBNx{978-3-030-63393-6}
\urldef\tempurl%
\url{https://doi.org/10.1007/978-3-030-63393-6_15}
\showDOI{\tempurl}


\bibitem[Mehmood et~al\mbox{.}(2017)]%
        {8311752}
\bibfield{author}{\bibinfo{person}{Muhammad~Amir Mehmood},
  \bibinfo{person}{Hafiz~Muhammad Shafiq}, {and} \bibinfo{person}{Abdul
  Waheed}.} \bibinfo{year}{2017}\natexlab{}.
\newblock \showarticletitle{Understanding regional context of World Wide Web
  using common crawl corpus}. In \bibinfo{booktitle}{\emph{2017 IEEE 13th
  Malaysia International Conference on Communications (MICC)}}.
  \bibinfo{pages}{164--169}.
\newblock
\urldef\tempurl%
\url{https://doi.org/10.1109/MICC.2017.8311752}
\showDOI{\tempurl}


\bibitem[{MLcommons}(2024)]%
        {mlcommonsBenchmarkMLPerf}
\bibfield{author}{\bibinfo{person}{{MLcommons}}.}
  \bibinfo{year}{2024}\natexlab{}.
\newblock \bibinfo{title}{{B}enchmark {MLP}erf {T}raining: {HPC} V2.0 Results}.
\newblock
\newblock
\urldef\tempurl%
\url{https://mlcommons.org/benchmarks/training-hpc/}
\showURL{%
\tempurl}
\newblock
\shownote{[Online; accessed 03/01/2025]}.


\bibitem[MLPerf(2023)]%
        {mlperftrainingpolicies}
\bibfield{author}{\bibinfo{person}{MLCommons MLPerf}.}
  \bibinfo{year}{2023}\natexlab{}.
\newblock \bibinfo{title}{{MLCommons MLPerf}}.
\newblock
\newblock
\urldef\tempurl%
\url{https://github.com/mlcommons/training_policies/blob/master/hpc_training_rules.adoc#8-benchmark-results}
\showURL{%
\tempurl}


\bibitem[Munhoz et~al\mbox{.}(2023)]%
        {munhoz_performance_2023}
\bibfield{author}{\bibinfo{person}{Vanderlei Munhoz}, \bibinfo{person}{Antoine
  Bonfils}, \bibinfo{person}{Márcio Castro}, {and} \bibinfo{person}{Odorico
  Mendizabal}.} \bibinfo{year}{2023}\natexlab{}.
\newblock \showarticletitle{A {Performance} {Comparison} of {HPC} {Workloads}
  on {Traditional} and {Cloud}-{Based} {HPC} {Clusters}}. In
  \bibinfo{booktitle}{\emph{2023 {International} {Symposium} on {Computer}
  {Architecture} and {High} {Performance} {Computing} {Workshops}
  ({SBAC}-{PADW})}}. \bibinfo{publisher}{IEEE}, \bibinfo{address}{Porto Alegre,
  Brazil}, \bibinfo{pages}{108--114}.
\newblock
\showISBNx{979-8-3503-8160-3}
\urldef\tempurl%
\url{https://doi.org/10.1109/SBAC-PADW60351.2023.00026}
\showDOI{\tempurl}


\bibitem[{National Academies of Sciences, Engineering, and Medicine}(2023)]%
        {nas2023foundational}
\bibfield{author}{\bibinfo{person}{{National Academies of Sciences,
  Engineering, and Medicine}}.} \bibinfo{year}{2023}\natexlab{}.
\newblock \bibinfo{booktitle}{\emph{Foundational Research Gaps and Future
  Directions for Digital Twins}}.
\newblock \bibinfo{publisher}{The National Academies Press},
  \bibinfo{address}{Washington, DC}.
\newblock
\urldef\tempurl%
\url{https://doi.org/10.17226/26894}
\showDOI{\tempurl}


\bibitem[Nurmi et~al\mbox{.}(2009)]%
        {eucalyptus}
\bibfield{author}{\bibinfo{person}{Daniel Nurmi}, \bibinfo{person}{Rich
  Wolski}, \bibinfo{person}{Chris Grzegorczyk}, \bibinfo{person}{Graziano
  Obertelli}, \bibinfo{person}{Sunil Soman}, \bibinfo{person}{Lamia Youseff},
  {and} \bibinfo{person}{Dmitrii Zagorodnov}.} \bibinfo{year}{2009}\natexlab{}.
\newblock \showarticletitle{The Eucalyptus Open-Source Cloud-Computing System}.
  In \bibinfo{booktitle}{\emph{2009 9th IEEE/ACM International Symposium on
  Cluster Computing and the Grid}}. \bibinfo{pages}{124--131}.
\newblock
\urldef\tempurl%
\url{https://doi.org/10.1109/CCGRID.2009.93}
\showDOI{\tempurl}


\bibitem[{NVIDIA}(2025)]%
        {mlperf-nvidia-benchmark}
\bibfield{author}{\bibinfo{person}{{NVIDIA}}.} \bibinfo{year}{2025}\natexlab{}.
\newblock \showarticletitle{{MLPerf AI Benchmarks}}.
\newblock  (\bibinfo{date}{March} \bibinfo{year}{2025}).
\newblock
\urldef\tempurl%
\url{https://www.nvidia.com/en-us/data-center/resources/mlperf-benchmarks/}
\showURL{%
\tempurl}


\bibitem[OpenAI et~al\mbox{.}(2024)]%
        {gpt-report}
\bibfield{author}{\bibinfo{person}{OpenAI}, \bibinfo{person}{Josh Achiam},
  \bibinfo{person}{Steven Adler}, \bibinfo{person}{Sandhini Agarwal},
  \bibinfo{person}{Lama Ahmad}, \bibinfo{person}{Ilge Akkaya},
  \bibinfo{person}{Florencia~Leoni Aleman}, \bibinfo{person}{Diogo Almeida},
  \bibinfo{person}{Janko Altenschmidt}, \bibinfo{person}{Sam Altman},
  \bibinfo{person}{Shyamal Anadkat}, \bibinfo{person}{Red Avila},
  \bibinfo{person}{Igor Babuschkin}, \bibinfo{person}{Suchir Balaji},
  \bibinfo{person}{Valerie Balcom}, \bibinfo{person}{Paul Baltescu},
  \bibinfo{person}{Haiming Bao}, \bibinfo{person}{Mohammad Bavarian},
  \bibinfo{person}{Jeff Belgum}, \bibinfo{person}{Irwan Bello},
  \bibinfo{person}{Jake Berdine}, \bibinfo{person}{Gabriel Bernadett-Shapiro},
  \bibinfo{person}{Christopher Berner}, \bibinfo{person}{Lenny Bogdonoff},
  \bibinfo{person}{Oleg Boiko}, \bibinfo{person}{Madelaine Boyd},
  \bibinfo{person}{Anna-Luisa Brakman}, \bibinfo{person}{Greg Brockman},
  \bibinfo{person}{Tim Brooks}, \bibinfo{person}{Miles Brundage},
  \bibinfo{person}{Kevin Button}, \bibinfo{person}{Trevor Cai},
  \bibinfo{person}{Rosie Campbell}, \bibinfo{person}{Andrew Cann},
  \bibinfo{person}{Brittany Carey}, \bibinfo{person}{Chelsea Carlson},
  \bibinfo{person}{Rory Carmichael}, \bibinfo{person}{Brooke Chan},
  \bibinfo{person}{Che Chang}, \bibinfo{person}{Fotis Chantzis},
  \bibinfo{person}{Derek Chen}, \bibinfo{person}{Sully Chen},
  \bibinfo{person}{Ruby Chen}, \bibinfo{person}{Jason Chen},
  \bibinfo{person}{Mark Chen}, \bibinfo{person}{Ben Chess},
  \bibinfo{person}{Chester Cho}, \bibinfo{person}{Casey Chu},
  \bibinfo{person}{Hyung~Won Chung}, \bibinfo{person}{Dave Cummings},
  \bibinfo{person}{Jeremiah Currier}, \bibinfo{person}{Yunxing Dai},
  \bibinfo{person}{Cory Decareaux}, \bibinfo{person}{Thomas Degry},
  \bibinfo{person}{Noah Deutsch}, \bibinfo{person}{Damien Deville},
  \bibinfo{person}{Arka Dhar}, \bibinfo{person}{David Dohan},
  \bibinfo{person}{Steve Dowling}, \bibinfo{person}{Sheila Dunning},
  \bibinfo{person}{Adrien Ecoffet}, \bibinfo{person}{Atty Eleti},
  \bibinfo{person}{Tyna Eloundou}, \bibinfo{person}{David Farhi},
  \bibinfo{person}{Liam Fedus}, \bibinfo{person}{Niko Felix},
  \bibinfo{person}{Simón~Posada Fishman}, \bibinfo{person}{Juston Forte},
  \bibinfo{person}{Isabella Fulford}, \bibinfo{person}{Leo Gao},
  \bibinfo{person}{Elie Georges}, \bibinfo{person}{Christian Gibson},
  \bibinfo{person}{Vik Goel}, \bibinfo{person}{Tarun Gogineni},
  \bibinfo{person}{Gabriel Goh}, \bibinfo{person}{Rapha Gontijo-Lopes},
  \bibinfo{person}{Jonathan Gordon}, \bibinfo{person}{Morgan Grafstein},
  \bibinfo{person}{Scott Gray}, \bibinfo{person}{Ryan Greene},
  \bibinfo{person}{Joshua Gross}, \bibinfo{person}{Shixiang~Shane Gu},
  \bibinfo{person}{Yufei Guo}, \bibinfo{person}{Chris Hallacy},
  \bibinfo{person}{Jesse Han}, \bibinfo{person}{Jeff Harris},
  \bibinfo{person}{Yuchen He}, \bibinfo{person}{Mike Heaton},
  \bibinfo{person}{Johannes Heidecke}, \bibinfo{person}{Chris Hesse},
  \bibinfo{person}{Alan Hickey}, \bibinfo{person}{Wade Hickey},
  \bibinfo{person}{Peter Hoeschele}, \bibinfo{person}{Brandon Houghton},
  \bibinfo{person}{Kenny Hsu}, \bibinfo{person}{Shengli Hu},
  \bibinfo{person}{Xin Hu}, \bibinfo{person}{Joost Huizinga},
  \bibinfo{person}{Shantanu Jain}, \bibinfo{person}{Shawn Jain},
  \bibinfo{person}{Joanne Jang}, \bibinfo{person}{Angela Jiang},
  \bibinfo{person}{Roger Jiang}, \bibinfo{person}{Haozhun Jin},
  \bibinfo{person}{Denny Jin}, \bibinfo{person}{Shino Jomoto},
  \bibinfo{person}{Billie Jonn}, \bibinfo{person}{Heewoo Jun},
  \bibinfo{person}{Tomer Kaftan}, \bibinfo{person}{Łukasz Kaiser},
  \bibinfo{person}{Ali Kamali}, \bibinfo{person}{Ingmar Kanitscheider},
  \bibinfo{person}{Nitish~Shirish Keskar}, \bibinfo{person}{Tabarak Khan},
  \bibinfo{person}{Logan Kilpatrick}, \bibinfo{person}{Jong~Wook Kim},
  \bibinfo{person}{Christina Kim}, \bibinfo{person}{Yongjik Kim},
  \bibinfo{person}{Jan~Hendrik Kirchner}, \bibinfo{person}{Jamie Kiros},
  \bibinfo{person}{Matt Knight}, \bibinfo{person}{Daniel Kokotajlo},
  \bibinfo{person}{Łukasz Kondraciuk}, \bibinfo{person}{Andrew Kondrich},
  \bibinfo{person}{Aris Konstantinidis}, \bibinfo{person}{Kyle Kosic},
  \bibinfo{person}{Gretchen Krueger}, \bibinfo{person}{Vishal Kuo},
  \bibinfo{person}{Michael Lampe}, \bibinfo{person}{Ikai Lan},
  \bibinfo{person}{Teddy Lee}, \bibinfo{person}{Jan Leike},
  \bibinfo{person}{Jade Leung}, \bibinfo{person}{Daniel Levy},
  \bibinfo{person}{Chak~Ming Li}, \bibinfo{person}{Rachel Lim},
  \bibinfo{person}{Molly Lin}, \bibinfo{person}{Stephanie Lin},
  \bibinfo{person}{Mateusz Litwin}, \bibinfo{person}{Theresa Lopez},
  \bibinfo{person}{Ryan Lowe}, \bibinfo{person}{Patricia Lue},
  \bibinfo{person}{Anna Makanju}, \bibinfo{person}{Kim Malfacini},
  \bibinfo{person}{Sam Manning}, \bibinfo{person}{Todor Markov},
  \bibinfo{person}{Yaniv Markovski}, \bibinfo{person}{Bianca Martin},
  \bibinfo{person}{Katie Mayer}, \bibinfo{person}{Andrew Mayne},
  \bibinfo{person}{Bob McGrew}, \bibinfo{person}{Scott~Mayer McKinney},
  \bibinfo{person}{Christine McLeavey}, \bibinfo{person}{Paul McMillan},
  \bibinfo{person}{Jake McNeil}, \bibinfo{person}{David Medina},
  \bibinfo{person}{Aalok Mehta}, \bibinfo{person}{Jacob Menick},
  \bibinfo{person}{Luke Metz}, \bibinfo{person}{Andrey Mishchenko},
  \bibinfo{person}{Pamela Mishkin}, \bibinfo{person}{Vinnie Monaco},
  \bibinfo{person}{Evan Morikawa}, \bibinfo{person}{Daniel Mossing},
  \bibinfo{person}{Tong Mu}, \bibinfo{person}{Mira Murati},
  \bibinfo{person}{Oleg Murk}, \bibinfo{person}{David Mély},
  \bibinfo{person}{Ashvin Nair}, \bibinfo{person}{Reiichiro Nakano},
  \bibinfo{person}{Rajeev Nayak}, \bibinfo{person}{Arvind Neelakantan},
  \bibinfo{person}{Richard Ngo}, \bibinfo{person}{Hyeonwoo Noh},
  \bibinfo{person}{Long Ouyang}, \bibinfo{person}{Cullen O'Keefe},
  \bibinfo{person}{Jakub Pachocki}, \bibinfo{person}{Alex Paino},
  \bibinfo{person}{Joe Palermo}, \bibinfo{person}{Ashley Pantuliano},
  \bibinfo{person}{Giambattista Parascandolo}, \bibinfo{person}{Joel Parish},
  \bibinfo{person}{Emy Parparita}, \bibinfo{person}{Alex Passos},
  \bibinfo{person}{Mikhail Pavlov}, \bibinfo{person}{Andrew Peng},
  \bibinfo{person}{Adam Perelman}, \bibinfo{person}{Filipe de Avila
  Belbute~Peres}, \bibinfo{person}{Michael Petrov},
  \bibinfo{person}{Henrique~Ponde de Oliveira~Pinto},
  \bibinfo{person}{Michael}, \bibinfo{person}{Pokorny},
  \bibinfo{person}{Michelle Pokrass}, \bibinfo{person}{Vitchyr~H. Pong},
  \bibinfo{person}{Tolly Powell}, \bibinfo{person}{Alethea Power},
  \bibinfo{person}{Boris Power}, \bibinfo{person}{Elizabeth Proehl},
  \bibinfo{person}{Raul Puri}, \bibinfo{person}{Alec Radford},
  \bibinfo{person}{Jack Rae}, \bibinfo{person}{Aditya Ramesh},
  \bibinfo{person}{Cameron Raymond}, \bibinfo{person}{Francis Real},
  \bibinfo{person}{Kendra Rimbach}, \bibinfo{person}{Carl Ross},
  \bibinfo{person}{Bob Rotsted}, \bibinfo{person}{Henri Roussez},
  \bibinfo{person}{Nick Ryder}, \bibinfo{person}{Mario Saltarelli},
  \bibinfo{person}{Ted Sanders}, \bibinfo{person}{Shibani Santurkar},
  \bibinfo{person}{Girish Sastry}, \bibinfo{person}{Heather Schmidt},
  \bibinfo{person}{David Schnurr}, \bibinfo{person}{John Schulman},
  \bibinfo{person}{Daniel Selsam}, \bibinfo{person}{Kyla Sheppard},
  \bibinfo{person}{Toki Sherbakov}, \bibinfo{person}{Jessica Shieh},
  \bibinfo{person}{Sarah Shoker}, \bibinfo{person}{Pranav Shyam},
  \bibinfo{person}{Szymon Sidor}, \bibinfo{person}{Eric Sigler},
  \bibinfo{person}{Maddie Simens}, \bibinfo{person}{Jordan Sitkin},
  \bibinfo{person}{Katarina Slama}, \bibinfo{person}{Ian Sohl},
  \bibinfo{person}{Benjamin Sokolowsky}, \bibinfo{person}{Yang Song},
  \bibinfo{person}{Natalie Staudacher}, \bibinfo{person}{Felipe~Petroski Such},
  \bibinfo{person}{Natalie Summers}, \bibinfo{person}{Ilya Sutskever},
  \bibinfo{person}{Jie Tang}, \bibinfo{person}{Nikolas Tezak},
  \bibinfo{person}{Madeleine~B. Thompson}, \bibinfo{person}{Phil Tillet},
  \bibinfo{person}{Amin Tootoonchian}, \bibinfo{person}{Elizabeth Tseng},
  \bibinfo{person}{Preston Tuggle}, \bibinfo{person}{Nick Turley},
  \bibinfo{person}{Jerry Tworek}, \bibinfo{person}{Juan Felipe~Cerón Uribe},
  \bibinfo{person}{Andrea Vallone}, \bibinfo{person}{Arun Vijayvergiya},
  \bibinfo{person}{Chelsea Voss}, \bibinfo{person}{Carroll Wainwright},
  \bibinfo{person}{Justin~Jay Wang}, \bibinfo{person}{Alvin Wang},
  \bibinfo{person}{Ben Wang}, \bibinfo{person}{Jonathan Ward},
  \bibinfo{person}{Jason Wei}, \bibinfo{person}{CJ Weinmann},
  \bibinfo{person}{Akila Welihinda}, \bibinfo{person}{Peter Welinder},
  \bibinfo{person}{Jiayi Weng}, \bibinfo{person}{Lilian Weng},
  \bibinfo{person}{Matt Wiethoff}, \bibinfo{person}{Dave Willner},
  \bibinfo{person}{Clemens Winter}, \bibinfo{person}{Samuel Wolrich},
  \bibinfo{person}{Hannah Wong}, \bibinfo{person}{Lauren Workman},
  \bibinfo{person}{Sherwin Wu}, \bibinfo{person}{Jeff Wu},
  \bibinfo{person}{Michael Wu}, \bibinfo{person}{Kai Xiao},
  \bibinfo{person}{Tao Xu}, \bibinfo{person}{Sarah Yoo}, \bibinfo{person}{Kevin
  Yu}, \bibinfo{person}{Qiming Yuan}, \bibinfo{person}{Wojciech Zaremba},
  \bibinfo{person}{Rowan Zellers}, \bibinfo{person}{Chong Zhang},
  \bibinfo{person}{Marvin Zhang}, \bibinfo{person}{Shengjia Zhao},
  \bibinfo{person}{Tianhao Zheng}, \bibinfo{person}{Juntang Zhuang},
  \bibinfo{person}{William Zhuk}, {and} \bibinfo{person}{Barret Zoph}.}
  \bibinfo{year}{2024}\natexlab{}.
\newblock \showarticletitle{GPT-4 Technical Report}.
\newblock  (\bibinfo{year}{2024}).
\newblock
\urldef\tempurl%
\url{https://doi.org/10.48550/arXiv.2303.08774}
\showDOI{\tempurl}
\showeprint[arxiv]{2303.08774}~[cs.CL]


\bibitem[Partee et~al\mbox{.}(2022)]%
        {partee2022using}
\bibfield{author}{\bibinfo{person}{Sam Partee}, \bibinfo{person}{Matthew
  Ellis}, \bibinfo{person}{Alessandro Rigazzi}, \bibinfo{person}{Andrew~E.
  Shao}, \bibinfo{person}{Scott Bachman}, \bibinfo{person}{Gustavo Marques},
  {and} \bibinfo{person}{Benjamin Robbins}.} \bibinfo{year}{2022}\natexlab{}.
\newblock \showarticletitle{Using Machine Learning at scale in numerical
  simulations with SmartSim: An application to ocean climate modeling}.
\newblock \bibinfo{journal}{\emph{Journal of Computational Science}}
  \bibinfo{volume}{62} (\bibinfo{year}{2022}), \bibinfo{pages}{101707}.
\newblock
\showISSN{1877-7503}
\urldef\tempurl%
\url{https://doi.org/10.1016/j.jocs.2022.101707}
\showDOI{\tempurl}


\bibitem[Pathak et~al\mbox{.}(2022)]%
        {pathak2022fourcastnet}
\bibfield{author}{\bibinfo{person}{Jaideep Pathak}, \bibinfo{person}{Shashank
  Subramanian}, \bibinfo{person}{Peter Harrington}, \bibinfo{person}{Sanjeev
  Raja}, \bibinfo{person}{Ashesh Chattopadhyay}, \bibinfo{person}{Morteza
  Mardani}, \bibinfo{person}{Thorsten Kurth}, \bibinfo{person}{David Hall},
  \bibinfo{person}{Zongyi Li}, \bibinfo{person}{Kamyar Azizzadenesheli},
  {et~al\mbox{.}}} \bibinfo{year}{2022}\natexlab{}.
\newblock \showarticletitle{{FourCastNet}: A global data-driven high-resolution
  weather model using adaptive {Fourier} neural operators}.
\newblock \bibinfo{journal}{\emph{arXiv preprint arXiv:2202.11214}}
  (\bibinfo{year}{2022}).
\newblock


\bibitem[{Pegasus}(2025)]%
        {www-pegasus}
{Pegasus} \bibinfo{year}{2025}\natexlab{}.
\newblock \bibinfo{title}{Pegasus WMS – Automate, recover, and debug
  scientific computations}.
\newblock
\newblock
\urldef\tempurl%
\url{https://pegasus.isi.edu/}
\showURL{%
\tempurl}
\newblock
\shownote{[Online; accessed 03/01/2025]}.


\bibitem[Perera et~al\mbox{.}(2023)]%
        {cylon}
\bibfield{author}{\bibinfo{person}{Niranda Perera}, \bibinfo{person}{Arup~Kumar
  Sarker}, \bibinfo{person}{Mills Staylor}, \bibinfo{person}{Gregor {von
  Laszewski}}, \bibinfo{person}{Kaiying Shan}, \bibinfo{person}{Supun
  Kamburugamuve}, \bibinfo{person}{Chathura Widanage},
  \bibinfo{person}{Vibhatha Abeykoon}, \bibinfo{person}{Thejaka~Amila
  Kanewela}, {and} \bibinfo{person}{Geoffrey Fox}.}
  \bibinfo{year}{2023}\natexlab{}.
\newblock \showarticletitle{In-depth analysis on parallel processing patterns
  for high-performance Dataframes}.
\newblock \bibinfo{journal}{\emph{Future Generation Computer Systems}}
  \bibinfo{volume}{149} (\bibinfo{year}{2023}), \bibinfo{pages}{250--264}.
\newblock
\showISSN{0167-739X}
\urldef\tempurl%
\url{https://doi.org/10.1016/j.future.2023.07.007}
\showDOI{\tempurl}


\bibitem[{PRACE}(2024a)]%
        {prace-fact}
\bibfield{author}{\bibinfo{person}{{PRACE}}.} \bibinfo{year}{2024}\natexlab{a}.
\newblock \bibinfo{title}{Fact Sheet PRACE Access}.
\newblock \bibinfo{howpublished}{Web Page}.
\newblock
\urldef\tempurl%
\url{https://prace-ri.eu/wp-content/uploads/Fact-Sheet-PRACE-Access.pdf}
\showURL{%
\tempurl}
\newblock
\shownote{[Online; accessed 03/01/2025]}.


\bibitem[{PRACE}(2024b)]%
        {www-prace}
\bibfield{author}{\bibinfo{person}{{PRACE}}.} \bibinfo{year}{2024}\natexlab{b}.
\newblock \bibinfo{title}{{HPC Infrastructure}}.
\newblock \bibinfo{howpublished}{Web Page}.
\newblock
\urldef\tempurl%
\url{https://prace-ri.eu/prace-archive/infrastructure-support/prace-hpc-infrastructure/}
\showURL{%
\tempurl}
\newblock
\shownote{[Online: accessed 03/01/2025]}.


\bibitem[Reddi et~al\mbox{.}(2020)]%
        {reddi2020mlperf}
\bibfield{author}{\bibinfo{person}{Vijay~Janapa Reddi},
  \bibinfo{person}{Christine Cheng}, \bibinfo{person}{David Kanter},
  \bibinfo{person}{Peter Mattson}, \bibinfo{person}{Guenther Schmuelling},
  \bibinfo{person}{Carole-Jean Wu}, \bibinfo{person}{Brian Anderson},
  \bibinfo{person}{Maximilien Breughe}, \bibinfo{person}{Mark Charlebois},
  \bibinfo{person}{William Chou}, {et~al\mbox{.}}}
  \bibinfo{year}{2020}\natexlab{}.
\newblock \showarticletitle{{MLPerf} inference benchmark}. In
  \bibinfo{booktitle}{\emph{2020 ACM/IEEE 47th Annual International Symposium
  on Computer Architecture (ISCA)}}. IEEE, \bibinfo{pages}{446--459}.
\newblock
\urldef\tempurl%
\url{https://doi.org/10.1109/ISCA45697.2020.00045}
\showDOI{\tempurl}


\bibitem[Russell et~al\mbox{.}(2002)]%
        {las-02-cactus-j}
\bibfield{author}{\bibinfo{person}{Michael Russell}, \bibinfo{person}{Gabrielle
  Allen}, \bibinfo{person}{Ian Foster}, \bibinfo{person}{Ed Seidel},
  \bibinfo{person}{Jason Novotny}, \bibinfo{person}{John Shalf},
  \bibinfo{person}{Gregor von Laszewski}, {and} \bibinfo{person}{Greg Daues}.}
  \bibinfo{year}{2002}\natexlab{}.
\newblock \showarticletitle{{The Astrophysics Simulation Collaboratory: A
  Science Portal Enabling Community Software Development}}.
\newblock \bibinfo{journal}{\emph{Journal on Cluster Computing}}
  \bibinfo{volume}{5}, \bibinfo{number}{3} (\bibinfo{date}{July}
  \bibinfo{year}{2002}), \bibinfo{pages}{297--304}.
\newblock
\showISSN{1386-7857}
\urldef\tempurl%
\url{https://doi.org/10.1023/A:1015629422149}
\showDOI{\tempurl}


\bibitem[Sarker et~al\mbox{.}(2024)]%
        {cylon-radical}
\bibfield{author}{\bibinfo{person}{Arup~Kumar Sarker}, \bibinfo{person}{Aymen
  Alsaadi}, \bibinfo{person}{Niranda Perera}, \bibinfo{person}{Mills Staylor},
  \bibinfo{person}{Gregor von Laszewski}, \bibinfo{person}{Matteo Turilli},
  \bibinfo{person}{Ozgur~Ozan Kilic}, \bibinfo{person}{Mikhail Titov},
  \bibinfo{person}{Andre Merzky}, \bibinfo{person}{Shantenu Jha}, {and}
  \bibinfo{person}{Geoffrey Fox}.} \bibinfo{year}{2024}\natexlab{}.
\newblock \showarticletitle{Design and Implementation of an Analysis Pipeline
  for Heterogeneous Data}.
\newblock  (\bibinfo{year}{2024}).
\newblock
\showeprint[arxiv]{2403.15721}~[cs.DC]
\urldef\tempurl%
\url{https://arxiv.org/abs/2403.15721}
\showURL{%
\tempurl}


\bibitem[Souza et~al\mbox{.}(2023)]%
        {souza2023}
\bibfield{author}{\bibinfo{person}{Renan Souza}, \bibinfo{person}{Tyler~J.
  Skluzacek}, \bibinfo{person}{Sean~R. Wilkinson}, \bibinfo{person}{Maxim
  Ziatdinov}, {and} \bibinfo{person}{Rafael~Ferreira da Silva}.}
  \bibinfo{year}{2023}\natexlab{}.
\newblock \showarticletitle{Towards Lightweight Data Integration Using
  Multi-Workflow Provenance and Data Observability}. In
  \bibinfo{booktitle}{\emph{2023 IEEE 19th International Conference on
  e-Science (e-Science)}}. \bibinfo{pages}{1--10}.
\newblock
\urldef\tempurl%
\url{https://doi.org/10.1109/e-Science58273.2023.10254822}
\showDOI{\tempurl}


\bibitem[Strande et~al\mbox{.}(2017)]%
        {las-17-comet}
\bibfield{author}{\bibinfo{person}{Shawn~M Strande}, \bibinfo{person}{Haisong
  Cai}, \bibinfo{person}{Trevor Cooper}, \bibinfo{person}{Karen Flammer},
  \bibinfo{person}{Christopher Irving}, \bibinfo{person}{Gregor von Laszewski},
  \bibinfo{person}{Amit Majumdar}, \bibinfo{person}{Dmistry Mishin},
  \bibinfo{person}{Philip Papadopoulos}, \bibinfo{person}{Wayne Pfeiffer},
  {et~al\mbox{.}}} \bibinfo{year}{2017}\natexlab{}.
\newblock \showarticletitle{Comet: Tales from the Long Tail: Two Years in and
  10,000 users later}. In \bibinfo{booktitle}{\emph{Proceedings of the Practice
  and Experience in Advanced Research Computing 2017 on Sustainability, Success
  and Impact}}. ACM, \bibinfo{pages}{38}.
\newblock


\bibitem[Top500(2025)]%
        {www-top500}
Top500 \bibinfo{year}{2025}\natexlab{}.
\newblock \bibinfo{title}{Homepage}.
\newblock
\newblock
\urldef\tempurl%
\url{https://www.top500.org/}
\showURL{%
\tempurl}
\newblock
\shownote{[Online; accessed 03/01/2025]}.


\bibitem[Vazhkudai and von Laszewski(2001)]%
        {las-01-greed}
\bibfield{author}{\bibinfo{person}{Sudharshan Vazhkudai} {and}
  \bibinfo{person}{Gregor von Laszewski}.} \bibinfo{year}{2001}\natexlab{}.
\newblock \showarticletitle{{A Greedy Grid - The Grid Economic Engine
  Directive}}. In \bibinfo{booktitle}{\emph{Proceedings of the 15th
  International Parallel and Distributed Processing Symposium, International
  Workshop on Internet Computing and E-Commerce (ICEC'01)}}
  \emph{(\bibinfo{series}{IPDPS '01})}. \bibinfo{publisher}{IEEE Computer
  Society, Washington, DC, USA}, \bibinfo{address}{San Francisco, California,
  USA}, \bibinfo{pages}{173--}.
\newblock
\showISBNx{0-7695-0990-8}
\urldef\tempurl%
\url{http://dl.acm.org/citation.cfm?id=645609.662793}
\showURL{%
\tempurl}


\bibitem[von Laszewski(1996)]%
        {las-96-ecwmf}
\bibfield{author}{\bibinfo{person}{Gregor von Laszewski}.}
  \bibinfo{year}{1996}\natexlab{}.
\newblock \showarticletitle{{An Interactive Parallel Programming Environment
  Applied in Atmospheric Science}}. In \bibinfo{booktitle}{\emph{{Making Its
  Mark, Proceedings of the 6th Workshop on the Use of Parallel Processors in
  Meteorology}}}, \bibfield{editor}{\bibinfo{person}{G.-R. Hoffman} {and}
  \bibinfo{person}{N.~Kreitz}} (Eds.). European Centre for Medium Weather
  Forecast, \bibinfo{publisher}{World Scientific}, \bibinfo{address}{Reading,
  UK}, \bibinfo{pages}{311--325}.
\newblock
\showISBNx{978-9810233501}
\urldef\tempurl%
\url{https://laszewski.github.io/papers/vonLaszewski-ecwmf-interactive.pdf}
\showURL{%
\tempurl}


\bibitem[von Laszewski(1999)]%
        {las-99-loosely}
\bibfield{author}{\bibinfo{person}{Gregor von Laszewski}.}
  \bibinfo{year}{1999}\natexlab{}.
\newblock \showarticletitle{{A Loosely Coupled Metacomputer: Cooperating Job
  Submissions Across Multiple Supercomputing Sites}}.
\newblock \bibinfo{journal}{\emph{Concurrency: Practice and Experience}}
  \bibinfo{volume}{11}, \bibinfo{number}{15} (\bibinfo{date}{Dec.}
  \bibinfo{year}{1999}), \bibinfo{pages}{933--948}.
\newblock
\showISSN{1096-9128}
\urldef\tempurl%
\url{https://doi.org/10.1002/(SICI)1096-9128(19991225)11:15<933::AID-CPE461>3.0.CO;2-J}
\showDOI{\tempurl}
\newblock
\shownote{The initial version of this paper was available in 1996}.


\bibitem[von Laszewski(2005a)]%
        {las-05-exp}
\bibfield{author}{\bibinfo{person}{Gregor von Laszewski}.}
  \bibinfo{year}{2005}\natexlab{a}.
\newblock \bibinfo{booktitle}{\emph{{The Java CoG Kit Experiment Manager}}}.
\newblock \bibinfo{type}{{T}echnical {R}eport} P1259.
  \bibinfo{institution}{Argonne National Laboratory}.
\newblock
\urldef\tempurl%
\url{https://laszewski.github.io/papers/vonLaszewski-exp.pdf}
\showURL{%
\tempurl}


\bibitem[von Laszewski(2005b)]%
        {las-05-workflow-jgc}
\bibfield{author}{\bibinfo{person}{Gregor von Laszewski}.}
  \bibinfo{year}{2005}\natexlab{b}.
\newblock \showarticletitle{{Workflow Concepts of the Java CoG Kit}}.
\newblock \bibinfo{journal}{\emph{Journal of Grid Computing}}
  \bibinfo{volume}{3} (\bibinfo{date}{Jan.} \bibinfo{year}{2005}),
  \bibinfo{pages}{239--258}.
\newblock
Issue 3-4.
\showISSN{1570-7873}
\urldef\tempurl%
\url{https://doi.org/10.1007/s10723-005-9013-5}
\showDOI{\tempurl}


\bibitem[von Laszewski(2020)]%
        {yamldb}
\bibfield{author}{\bibinfo{person}{Gregor von Laszewski}.}
  \bibinfo{year}{2020}\natexlab{}.
\newblock \bibinfo{title}{cloudmesh/yamldb}.
\newblock
\newblock
\urldef\tempurl%
\url{https://github.com/cloudmesh/yamldb}
\showURL{%
\tempurl}
\newblock
\shownote{[Online; accessed 03/01/2025]}.


\bibitem[von Laszewski(2022a)]%
        {cloudmesh-stopwatch}
\bibfield{author}{\bibinfo{person}{Gregor von Laszewski}.}
  \bibinfo{year}{2022}\natexlab{a}.
\newblock \bibinfo{title}{{Cloudmesh Common StopWatch}}.
\newblock
\newblock
\urldef\tempurl%
\url{https://github.com/cloudmesh/cloudmesh-common/blob/main/
  cloudmesh/common/StopWatch.py}
\showURL{%
\tempurl}


\bibitem[von Laszewski(2022b)]%
        {cloudmesh-gpu}
\bibfield{author}{\bibinfo{person}{Gregor von Laszewski}.}
  \bibinfo{year}{2022}\natexlab{b}.
\newblock \bibinfo{title}{{Cloudmesh GPU Monitor}}.
\newblock
\newblock
\urldef\tempurl%
\url{https://github.com/cloudmesh/cloudmesh-gpu}
\showURL{%
\tempurl}
\newblock
\shownote{[Accessed 03/01/2025]}.


\bibitem[von Laszewski(2022c)]%
        {cloudmesh-vpn}
\bibfield{author}{\bibinfo{person}{Gregor von Laszewski}.}
  \bibinfo{year}{2022}\natexlab{c}.
\newblock \bibinfo{title}{{Cloudmesh VPN}}.
\newblock
\newblock
\urldef\tempurl%
\url{https://github.com/cloudmesh/cloudmesh-vpn}
\showURL{%
\tempurl}
\newblock
\shownote{[Accessed 03/01/2025]}.


\bibitem[von Laszewski(2023a)]%
        {cloudmesh-cc}
\bibfield{author}{\bibinfo{person}{Gregor von Laszewski}.}
  \bibinfo{year}{2023}\natexlab{a}.
\newblock \bibinfo{title}{{Cloudmesh Compute Coordinator}}.
\newblock
\newblock
\urldef\tempurl%
\url{https://github.com/cloudmesh/cloudmesh-cc}
\showURL{%
\tempurl}
\newblock
\shownote{[Online; accessed 03/01/2025]}.


\bibitem[von Laszewski(2023b)]%
        {cloudmesh-ee}
\bibfield{author}{\bibinfo{person}{Gregor von Laszewski}.}
  \bibinfo{year}{2023}\natexlab{b}.
\newblock \bibinfo{title}{{Cloudmesh Experiment Executor}}.
\newblock \bibinfo{howpublished}{GitHub}.
\newblock
\urldef\tempurl%
\url{https://github.com/cloudmesh/cloudmesh-ee}
\showURL{%
\tempurl}
\newblock
\shownote{[Online; accessed 03/14/2025]}.


\bibitem[von Laszewski(2024)]%
        {www-cloudmesh-org}
\bibfield{author}{\bibinfo{person}{Gregor von Laszewski}.}
  \bibinfo{year}{2024}\natexlab{}.
\newblock \bibinfo{title}{Cloudmesh Version 4}.
\newblock
\newblock
\urldef\tempurl%
\url{https://cloudmesh.github.io/cloudmesh-manual/index.html}
\showURL{%
\tempurl}
\newblock
\shownote{[Online; accessed 03/01/2025]}.


\bibitem[von Laszewski et~al\mbox{.}(2017)]%
        {las-17-cloudmesh}
\bibfield{author}{\bibinfo{person}{Gregor von Laszewski}, \bibinfo{person}{Badi
  Abdul-Wahid}, \bibinfo{person}{Fugang Wang}, \bibinfo{person}{Hyungro Lee},
  \bibinfo{person}{Geoffrey~C Fox}, {and} \bibinfo{person}{Wo Chang}.}
  \bibinfo{year}{2017}\natexlab{}.
\newblock \bibinfo{booktitle}{\emph{Cloudmesh in support of the NIST Big Data
  Architecture Framework}}.
\newblock \bibinfo{type}{{T}echnical {R}eport}. \bibinfo{institution}{Technical
  report, Indiana University, Bloomingtion IN 47408, USA}.
\newblock


\bibitem[von Laszewski et~al\mbox{.}(2003a)]%
        {las-03-ftp}
\bibfield{author}{\bibinfo{person}{Gregor von Laszewski},
  \bibinfo{person}{Beulah Alunkal}, \bibinfo{person}{Jarek Gawor},
  \bibinfo{person}{Ravi Madhuri}, \bibinfo{person}{Pawel Plaszczak}, {and}
  \bibinfo{person}{Xian-He Sun}.} \bibinfo{year}{2003}\natexlab{a}.
\newblock \showarticletitle{{A File Transfer Component for Grids}}. In
  \bibinfo{booktitle}{\emph{Proceedings of the International Conferenece on
  Parallel and Distributed Processing Techniques and Applications}},
  \bibfield{editor}{\bibinfo{person}{H.R. Arabnia} {and}
  \bibinfo{person}{Youngson Mun}} (Eds.), Vol.~\bibinfo{volume}{1}.
  \bibinfo{publisher}{CSREA Press}, \bibinfo{address}{Las Vegas},
  \bibinfo{pages}{24--30}.
\newblock
\urldef\tempurl%
\url{https://laszewski.github.io/papers/vonLaszewski-gridftp.pdf}
\showURL{%
\tempurl}


\bibitem[von Laszewski et~al\mbox{.}(2007a)]%
        {las-06-guss-j}
\bibfield{author}{\bibinfo{person}{Gregor von Laszewski},
  \bibinfo{person}{Jonathan DiCarlo}, {and} \bibinfo{person}{Bill Allcock}.}
  \bibinfo{year}{2007}\natexlab{a}.
\newblock \showarticletitle{{A Portal for Visualizing Grid Usage}}.
\newblock \bibinfo{journal}{\emph{Concurrency and Computation: Practice and
  Experience}} \bibinfo{volume}{19}, \bibinfo{number}{12} (\bibinfo{date}{Aug.}
  \bibinfo{year}{2007}), \bibinfo{pages}{1683--1692}.
\newblock
\showISSN{1532-0626}
\urldef\tempurl%
\url{https://doi.org/10.1002/cpe.v19:12}
\showDOI{\tempurl}


\bibitem[von Laszewski et~al\mbox{.}(2023a)]%
        {las-2022-templated}
\bibfield{author}{\bibinfo{person}{Gregor von Laszewski}, \bibinfo{person}{J.P.
  Fleischer}, \bibinfo{person}{Geoffrey~C. Fox}, \bibinfo{person}{Juri Papay},
  \bibinfo{person}{Sam Jackson}, {and} \bibinfo{person}{Jeyan Thiyagalingam}.}
  \bibinfo{year}{2023}\natexlab{a}.
\newblock \showarticletitle{Templated Hybrid Reusable Computational Analytics
  Workflow Management with Cloudmesh, Applied to the Deep Learning MLCommons
  Cloudmask Application}. In \bibinfo{booktitle}{\emph{2023 IEEE 19th
  International Conference on e-Science (e-Science)}}. \bibinfo{pages}{1--6}.
\newblock
\urldef\tempurl%
\url{https://doi.org/10.1109/e-Science58273.2023.10254942}
\showDOI{\tempurl}


\bibitem[von Laszewski et~al\mbox{.}(2023b)]%
        {las-2023-mlcommons-edu-eq}
\bibfield{author}{\bibinfo{person}{Gregor von Laszewski}, \bibinfo{person}{J.P.
  Fleischer}, \bibinfo{person}{R. Knuuti}, \bibinfo{person}{G.C. Fox},
  \bibinfo{person}{J. Kolessar}, \bibinfo{person}{T.S. Butler}, {and}
  \bibinfo{person}{J. Fox}.} \bibinfo{year}{2023}\natexlab{b}.
\newblock \showarticletitle{Opportunities for enhancing MLCommons efforts while
  leveraging insights from educational MLCommons earthquake benchmarks
  efforts}.
\newblock \bibinfo{journal}{\emph{Frontiers in High Performance Computing,}}
  \bibinfo{volume}{1}, \bibinfo{number}{1233877} (\bibinfo{date}{October}
  \bibinfo{year}{2023}), \bibinfo{pages}{31}.
\newblock
\urldef\tempurl%
\url{https://doi.org/10.3389/fhpcp.2023.1233877}
\showURL{%
\tempurl}


\bibitem[von Laszewski et~al\mbox{.}(2022)]%
        {las-2022-hybrid}
\bibfield{author}{\bibinfo{person}{Gregor von Laszewski},
  \bibinfo{person}{J.~P. Fleischer}, {and} \bibinfo{person}{Geoffrey~C. Fox}.}
  \bibinfo{year}{2022}\natexlab{}.
\newblock \showarticletitle{Hybrid Reusable Computational Analytics Workflow
  Management with Cloudmesh}.
\newblock  (\bibinfo{year}{2022}).
\newblock
\showeprint[arxiv]{2210.16941}~[cs.DC]
\urldef\tempurl%
\url{https://arxiv.org/abs/2210.16941}
\showURL{%
\tempurl}


\bibitem[von Laszewski et~al\mbox{.}(2023c)]%
        {las-frontiers-edu}
\bibfield{author}{\bibinfo{person}{Gregor von Laszewski},
  \bibinfo{person}{J.~P. Fleischer}, \bibinfo{person}{Robert Knuuti},
  \bibinfo{person}{Geoffrey~C. Fox}, \bibinfo{person}{Jake Kolessar},
  \bibinfo{person}{Thomas~S. Butler}, {and} \bibinfo{person}{Judy Fox}.}
  \bibinfo{year}{2023}\natexlab{c}.
\newblock \showarticletitle{Opportunities for enhancing MLCommons efforts while
  leveraging insights from educational MLCommons earthquake benchmarks
  efforts}.
\newblock \bibinfo{journal}{\emph{Frontiers in High Performance Computing}}
  \bibinfo{volume}{1} (\bibinfo{year}{2023}).
\newblock
\showISSN{2813-7337}
\urldef\tempurl%
\url{https://doi.org/10.3389/fhpcp.2023.1233877}
\showDOI{\tempurl}


\bibitem[von Laszewski and Foster(1999)]%
        {las-99-rostock}
\bibfield{author}{\bibinfo{person}{Gregor von Laszewski} {and}
  \bibinfo{person}{Ian Foster}.} \bibinfo{year}{1999}\natexlab{}.
\newblock \showarticletitle{{Grid Infrastructure to Support Science Portals for
  Large Scale Instruments}}. In \bibinfo{booktitle}{\emph{{Proceedings of the
  Workshop Distributed Computing on the Web (DCW)}}}.
  \bibinfo{publisher}{University of Rostock, Germany}, \bibinfo{pages}{1--16}.
\newblock
\newblock
\shownote{{\em (Invited Talk)}}.


\bibitem[von Laszewski et~al\mbox{.}(2001a)]%
        {las-01-cog-concurency}
\bibfield{author}{\bibinfo{person}{Gregor von Laszewski}, \bibinfo{person}{Ian
  Foster}, \bibinfo{person}{Jarek Gawor}, {and} \bibinfo{person}{Peter Lane}.}
  \bibinfo{year}{2001}\natexlab{a}.
\newblock \showarticletitle{{A Java Commodity Grid Kit}}.
\newblock \bibinfo{journal}{\emph{Concurrency and Computation: Practice and
  Experience}} \bibinfo{volume}{13}, \bibinfo{number}{8-9}
  (\bibinfo{year}{2001}), \bibinfo{pages}{645--662}.
\newblock
\showISSN{1532-0634}
\urldef\tempurl%
\url{https://doi.org/10.1002/cpe.572}
\showDOI{\tempurl}


\bibitem[von Laszewski et~al\mbox{.}(2001b)]%
        {las-01-pse}
\bibfield{author}{\bibinfo{person}{Gregor von Laszewski}, \bibinfo{person}{Ian
  Foster}, \bibinfo{person}{Jarek Gawor}, \bibinfo{person}{Peter Lane},
  \bibinfo{person}{Nell Rehn}, {and} \bibinfo{person}{Mike Russell}.}
  \bibinfo{year}{2001}\natexlab{b}.
\newblock \showarticletitle{{Designing Grid-based Problem Solving Environments
  and Portals}}. In \bibinfo{booktitle}{\emph{Proceedings of the 34th Annual
  Hawaii International Conference on System Sciences (HICSS-34)}}
  \emph{(\bibinfo{series}{HICSS '01}, Vol.~\bibinfo{volume}{9})}.
  \bibinfo{publisher}{IEEE Computer Society, Washington, DC, USA},
  \bibinfo{address}{Maui, Hawaii}, \bibinfo{pages}{9028--}.
\newblock
\showISBNx{0-7695-0981-9}
\urldef\tempurl%
\url{https://laszewski.github.io/papers/vonLaszewski-cog-pse-final.pdf}
\showURL{%
\tempurl}


\bibitem[von Laszewski et~al\mbox{.}(2000a)]%
        {las-00-grande}
\bibfield{author}{\bibinfo{person}{Gregor von Laszewski}, \bibinfo{person}{Ian
  Foster}, \bibinfo{person}{Jarek Gawor}, \bibinfo{person}{Warren Smith}, {and}
  \bibinfo{person}{Steve Tuecke}.} \bibinfo{year}{2000}\natexlab{a}.
\newblock \showarticletitle{{CoG Kits: A Bridge between Commodity Distributed
  Computing and High-Performance Grids}}. In
  \bibinfo{booktitle}{\emph{Proceedings of the ACM 2000 conference on Java
  Grande}} (San Francisco, California, United States)
  \emph{(\bibinfo{series}{JAVA'00})}. \bibinfo{publisher}{ACM, New York, NY,
  USA}, \bibinfo{address}{San Francisco, CA}, \bibinfo{pages}{97--106}.
\newblock
\showISBNx{1-58113-288-3}
\urldef\tempurl%
\url{https://doi.org/10.1145/337449.337491}
\showDOI{\tempurl}


\bibitem[von Laszewski et~al\mbox{.}(2010)]%
        {las-20-10gce}
\bibfield{author}{\bibinfo{person}{Gregor von Laszewski},
  \bibinfo{person}{Geoffrey~C. Fox}, \bibinfo{person}{Fugang Wang},
  \bibinfo{person}{Andrew~J Younge}, \bibinfo{person}{Kulshrestha},
  \bibinfo{person}{Gregory~G. Pike}, \bibinfo{person}{Warren Smith},
  \bibinfo{person}{Jens Voeckler}, \bibinfo{person}{Renato~J. Figueiredo},
  \bibinfo{person}{Jose Fortes}, \bibinfo{person}{Kate Keahey}, {and}
  \bibinfo{person}{Ewa Deelman}.} \bibinfo{year}{2010}\natexlab{}.
\newblock \showarticletitle{{Design of the FutureGrid Experiment Management
  Framework}}. In \bibinfo{booktitle}{\emph{Proceedings of Gateway Computing
  Environments 2010 (GCE2010) at SC10}}. \bibinfo{publisher}{IEEE},
  \bibinfo{address}{New Orleans, LA}.
\newblock
\urldef\tempurl%
\url{https://doi.org/10.1109/GCE.2010.5676126}
\showDOI{\tempurl}


\bibitem[von Laszewski et~al\mbox{.}(2003b)]%
        {las-03-gridcomputing}
\bibfield{author}{\bibinfo{person}{Gregor von Laszewski},
  \bibinfo{person}{Jarek Gawor}, \bibinfo{person}{Sriram Krishnan}, {and}
  \bibinfo{person}{Keith Jackson}.} \bibinfo{year}{2003}\natexlab{b}.
\newblock \showarticletitle{{Commodity Grid Kits - Middleware for Building Grid
  Computing Environments}}.
\newblock In \bibinfo{booktitle}{\emph{{Grid Computing: Making the Global
  Infrastructure a Reality}}}, \bibfield{editor}{\bibinfo{person}{Fran Berman},
  \bibinfo{person}{Geoffrey Fox}, {and} \bibinfo{person}{Toney Hey}} (Eds.).
  \bibinfo{publisher}{Wiley}, \bibinfo{pages}{639--656}.
\newblock
\showISBNx{0-470-85319-0}
\urldef\tempurl%
\url{https://laszewski.github.io/papers/vonLaszewski-grid2002book.pdf}
\showURL{%
\tempurl}


\bibitem[von Laszewski et~al\mbox{.}(2002a)]%
        {las-02-javacog}
\bibfield{author}{\bibinfo{person}{Gregor von Laszewski},
  \bibinfo{person}{Jarek Gawor}, \bibinfo{person}{Peter Lane},
  \bibinfo{person}{Nell Rehn}, \bibinfo{person}{Mike Russell}, {and}
  \bibinfo{person}{Keith Jackson}.} \bibinfo{year}{2002}\natexlab{a}.
\newblock \showarticletitle{{Features of the Java Commodity Grid Kit}}.
\newblock \bibinfo{journal}{\emph{Concurrency and Computation: Practice and
  Experience}} \bibinfo{volume}{14}, \bibinfo{number}{13-15}
  (\bibinfo{year}{2002}), \bibinfo{pages}{1045--1055}.
\newblock
\showISSN{1532-0634}
\urldef\tempurl%
\url{https://doi.org/10.1002/cpe.674}
\showDOI{\tempurl}


\bibitem[von Laszewski et~al\mbox{.}(2002b)]%
        {las-02-infogram}
\bibfield{author}{\bibinfo{person}{Gregor von Laszewski},
  \bibinfo{person}{Jarek Gawor}, \bibinfo{person}{Carlos~J. Pe{\~n}a}, {and}
  \bibinfo{person}{Ian Foster}.} \bibinfo{year}{2002}\natexlab{b}.
\newblock \showarticletitle{{InfoGram: A Peer-to-Peer Information and Job
  Submission Service}}. In \bibinfo{booktitle}{\emph{{Proceedings of the 11th
  Symposium on High Performance Distributed Computing}}}
  \emph{(\bibinfo{series}{HPDC '02})}. \bibinfo{publisher}{IEEE Computer
  Society, Washington, DC, USA}, \bibinfo{address}{Edinbrough, U.K.},
  \bibinfo{pages}{333--342}.
\newblock
\showISBNx{0-7695-1686-6}
\urldef\tempurl%
\url{https://laszewski.github.io/papers/vonLaszewski-infogram.pdf}
\showURL{%
\tempurl}


\bibitem[von Laszewski et~al\mbox{.}(2004)]%
        {las-04-ftp-journal}
\bibfield{author}{\bibinfo{person}{Gregor von Laszewski},
  \bibinfo{person}{Jarek Gawor}, \bibinfo{person}{Pawel Plaszczak},
  \bibinfo{person}{Mike Hategan}, \bibinfo{person}{Kaizar Amin},
  \bibinfo{person}{Ravi Madduri}, {and} \bibinfo{person}{Scott Gose}.}
  \bibinfo{year}{2004}\natexlab{}.
\newblock \showarticletitle{{An Overview of Grid File Transfer Patterns and
  their Implementation in the Java CoG Kit}}.
\newblock \bibinfo{journal}{\emph{Journal of Neural Parallel and Scientific
  Computing}} \bibinfo{volume}{12}, \bibinfo{number}{3} (\bibinfo{date}{Sept.}
  \bibinfo{year}{2004}), \bibinfo{pages}{329--352}.
\newblock
\showISSN{1061-5369}
\urldef\tempurl%
\url{https://laszewski.github.io/papers/vonLaszewski-overview-gridftp.pdf}
\showURL{%
\tempurl}
\newblock
\shownote{Special Issue on Grid Computing}.


\bibitem[von Laszewski et~al\mbox{.}(2006a)]%
        {las-06-exp-a}
\bibfield{author}{\bibinfo{person}{Gregor von Laszewski},
  \bibinfo{person}{Christopher Grubbs}, \bibinfo{person}{Matthew Bone}, {and}
  \bibinfo{person}{David Angulo}.} \bibinfo{year}{2006}\natexlab{a}.
\newblock \showarticletitle{{The Java CoG Kit Experiment Manager}}. In
  \bibinfo{booktitle}{\emph{International Workshop on Grid Computing
  Environments 2006 in Conjunction with SC06}}.
\newblock
\urldef\tempurl%
\url{http://library.rit.edu/oajournals/index.php/gce/article/view/75/36}
\showURL{%
\tempurl}


\bibitem[von Laszewski et~al\mbox{.}(2006b)]%
        {las-06-workcoordination}
\bibfield{author}{\bibinfo{person}{Gregor von Laszewski},
  \bibinfo{person}{Mihael Hategan}, {and} \bibinfo{person}{Deepti Kodeboyina}.}
  \bibinfo{year}{2006}\natexlab{b}.
\newblock \showarticletitle{{Work coordination for Grid computing}}.
\newblock In \bibinfo{booktitle}{\emph{{Grid Technologies}}},
  \bibfield{editor}{\bibinfo{person}{M.P. Bekakos}, \bibinfo{person}{G.A.
  Gravvanis}, {and} \bibinfo{person}{H.R. Arabnia}} (Eds.).
  \bibinfo{series}{State-of-the-art in Science and Engineering},
  Vol.~\bibinfo{volume}{5}. \bibinfo{publisher}{Wit}.
\newblock
\showISBNx{1-84564-055-1}
\urldef\tempurl%
\url{https://laszewski.github.io/papers/vonLaszewski-work-coordination.pdf}
\showURL{%
\tempurl}


\bibitem[von Laszewski et~al\mbox{.}(2007b)]%
        {las-06-workflow-book}
\bibfield{author}{\bibinfo{person}{Gregor von Laszewski},
  \bibinfo{person}{Mihael Hategan}, {and} \bibinfo{person}{Depti Kodeboyina}.}
  \bibinfo{year}{2007}\natexlab{b}.
\newblock \showarticletitle{Grid Workflow with the Java CoG Kit}.
\newblock In \bibinfo{booktitle}{\emph{{Workflows for E-science: Scientific
  Workflows for Grids}}}, \bibfield{editor}{\bibinfo{person}{Ian~J. Taylor},
  \bibinfo{person}{Ewa Deelman}, \bibinfo{person}{Dennis~B. Gannon}, {and}
  \bibinfo{person}{Matthew Shields}} (Eds.).
  \bibinfo{publisher}{Springer-Verlag New York, Inc.},
  \bibinfo{address}{Secaucus, NJ, USA}.
\newblock
\showISBNx{1846285194}
\urldef\tempurl%
\url{https://laszewski.github.io/papers/vonLaszewski-workflow-book.pdf}
\showURL{%
\tempurl}


\bibitem[von Laszewski et~al\mbox{.}(2007c)]%
        {las07-workflow}
\bibfield{author}{\bibinfo{person}{Gregor von Laszewski},
  \bibinfo{person}{Mihael Hategan}, {and} \bibinfo{person}{Deepti Kodeboyina}.}
  \bibinfo{year}{2007}\natexlab{c}.
\newblock \bibinfo{booktitle}{\emph{Java CoG Kit Workflow}}.
\newblock \bibinfo{publisher}{Springer London}, \bibinfo{address}{London},
  \bibinfo{pages}{340--356}.
\newblock
\showISBNx{978-1-84628-757-2}
\urldef\tempurl%
\url{https://doi.org/10.1007/978-1-84628-757-2_21}
\showDOI{\tempurl}


\bibitem[von Laszewski et~al\mbox{.}(2023d)]%
        {las-2023-escience}
\bibfield{author}{\bibinfo{person}{Gregor von Laszewski}, \bibinfo{person}{J.P.
  J.P.~Fleischer}, \bibinfo{person}{Geoffrey~C. Fox}, \bibinfo{person}{Juri
  Papay}, \bibinfo{person}{Sam Jackson}, {and} \bibinfo{person}{Jeyan
  Thiyagalingam}.} \bibinfo{year}{2023}\natexlab{d}.
\newblock \showarticletitle{Templated Hybrid Reusable Computational Analytics
  Workflow Management with Cloudmesh, Applied to the Deep Learning MLCommons
  Cloudmask Application}. In \bibinfo{booktitle}{\emph{eScience'23}}. Second
  Workshop on Reproducible Workflows, Data, and Security (ReWorDS 2022),
  \bibinfo{address}{Limassol, Cyprus}.
\newblock
\urldef\tempurl%
\url{https://doi.org/10.1109/e-Science58273.2023.10254942}
\showDOI{\tempurl}


\bibitem[von Laszewski and Kodeboyina(2005)]%
        {las-05-workflowrepo}
\bibfield{author}{\bibinfo{person}{Gregor von Laszewski} {and}
  \bibinfo{person}{Deepti Kodeboyina}.} \bibinfo{year}{2005}\natexlab{}.
\newblock \showarticletitle{{A Repository Service for Grid Workflow
  Components}}. In \bibinfo{booktitle}{\emph{{Proceedings of the Joint
  International Conference on Autonomic and Autonomous Systems and
  International Conference on Networking and Services}}}
  \emph{(\bibinfo{series}{ICAS-ICNS '05})}. \bibinfo{publisher}{IEEE Computer
  Society, Washington, DC, USA}, \bibinfo{address}{Papeete, Tahiti, French
  Polynesia}, \bibinfo{pages}{84--}.
\newblock
\showISBNx{0-7695-2450-8}
\urldef\tempurl%
\url{https://laszewski.github.io/papers/vonLaszewski-workflow-repository.pdf}
\showURL{%
\tempurl}


\bibitem[von Laszewski et~al\mbox{.}(2012)]%
        {las-08-federated-cloud}
\bibfield{author}{\bibinfo{person}{Gregor von Laszewski},
  \bibinfo{person}{Hyungro Lee}, \bibinfo{person}{Javier Diaz},
  \bibinfo{person}{Fugang Wang}, \bibinfo{person}{Koji Tanaka},
  \bibinfo{person}{Shubhada Karavinkoppa}, \bibinfo{person}{Geoffrey~C. Fox},
  {and} \bibinfo{person}{Tom Furlani}.} \bibinfo{year}{2012}\natexlab{}.
\newblock \showarticletitle{{Design of an Accounting and Metric-based
  Cloud-shifting and Cloud-seeding Framework for Federated Clouds and
  Bare-metal Environments}}. In \bibinfo{booktitle}{\emph{Proceedings of the
  2012 Workshop on Cloud Services, Federation, and the 8th Open Cirrus Summit}}
  (San Jose, California, USA) \emph{(\bibinfo{series}{FederatedClouds '12})}.
  \bibinfo{publisher}{ACM}, \bibinfo{address}{New York, NY, USA},
  \bibinfo{pages}{25--32}.
\newblock
\showISBNx{978-1-4503-1754-2}
\urldef\tempurl%
\url{https://doi.org/10.1145/2378975.2378982}
\showDOI{\tempurl}


\bibitem[von Laszewski et~al\mbox{.}(1994)]%
        {las-94-ecwmf}
\bibfield{author}{\bibinfo{person}{Gregor von Laszewski}, \bibinfo{person}{Mike
  Seablom}, \bibinfo{person}{Milo Makivic}, \bibinfo{person}{Peter Lyster},
  {and} \bibinfo{person}{Sanya Ranka}.} \bibinfo{year}{1994}\natexlab{}.
\newblock \showarticletitle{{Design Issues for the Parallelization of an
  Optimal Interpolation Algorithm}}. In \bibinfo{booktitle}{\emph{{Coming of
  Age, Proceedings of the 4th Workshop on the Use of Parallel Processing in
  Atmospheric Science}}}, \bibfield{editor}{\bibinfo{person}{G.-R. Hoffman}
  {and} \bibinfo{person}{N.~Kreitz}} (Eds.). European Centre for Medium Weather
  Forecast, \bibinfo{publisher}{World Scientific}, \bibinfo{address}{Reading,
  UK}, \bibinfo{pages}{290--302}.
\newblock
\urldef\tempurl%
\url{https://laszewski.github.io/papers/vonLaszewski94-4dda-design.pdf}
\showURL{%
\tempurl}


\bibitem[von Laszewski et~al\mbox{.}(2019)]%
        {las-19-harc-comet}
\bibfield{author}{\bibinfo{person}{Gregor von Laszewski},
  \bibinfo{person}{Fugang Wang}, \bibinfo{person}{Geoffrey~C. Fox},
  \bibinfo{person}{Shawn Strande}, \bibinfo{person}{Christopher Irving},
  \bibinfo{person}{Trevor Cooper}, \bibinfo{person}{Dmitry Mishin}, {and}
  \bibinfo{person}{Michael~L. Norman}.} \bibinfo{year}{2019}\natexlab{}.
\newblock \showarticletitle{Human in the Loop Virtual Machine Management on
  Comet}. In \bibinfo{booktitle}{\emph{Humans in the Loop: Enabling and
  Facilitating Research on Cloud Computing}}. \bibinfo{address}{Chicago, IL,
  USA}.
\newblock
\showISBNx{978-1-4503-7279-4/19/07}
\urldef\tempurl%
\url{https://doi.org/10.1145/3355738.3355751}
\showDOI{\tempurl}


\bibitem[von Laszewski et~al\mbox{.}(2014)]%
        {las-14-bigdata}
\bibfield{author}{\bibinfo{person}{Gregor von Laszewski},
  \bibinfo{person}{Fugang Wang}, \bibinfo{person}{Hyungro Lee},
  \bibinfo{person}{Heng Chen}, {and} \bibinfo{person}{Geoffrey~C. Fox}.}
  \bibinfo{year}{2014}\natexlab{}.
\newblock \showarticletitle{Accessing multiple clouds with cloudmesh}. In
  \bibinfo{booktitle}{\emph{Proceedings of the 2014 ACM International Workshop
  on Software-Defined Ecosystems}} (Vancouver, BC, Canada)
  \emph{(\bibinfo{series}{BigSystem '14})}. \bibinfo{publisher}{Association for
  Computing Machinery}, \bibinfo{address}{New York, NY, USA},
  \bibinfo{pages}{21–28}.
\newblock
\showISBNx{9781450329095}
\urldef\tempurl%
\url{https://doi.org/10.1145/2609441.2609638}
\showDOI{\tempurl}


\bibitem[von Laszewski et~al\mbox{.}(2000b)]%
        {las-00-sbc}
\bibfield{author}{\bibinfo{person}{Gregor von Laszewski}, \bibinfo{person}{Mary
  Westbrook}, \bibinfo{person}{Ian Foster}, \bibinfo{person}{Edwin Westbrook},
  {and} \bibinfo{person}{Craig Barnes}.} \bibinfo{year}{2000}\natexlab{b}.
\newblock \showarticletitle{{Using Computational Grid Capabilities to Enhance
  the Ability of an X-Ray Source for Structural Biology}}.
\newblock \bibinfo{journal}{\emph{Cluster Computing}} \bibinfo{volume}{3},
  \bibinfo{number}{3} (\bibinfo{year}{2000}), \bibinfo{pages}{187--199}.
\newblock
\urldef\tempurl%
\url{https://doi.org/10.1023/A:1019036421819}
\showDOI{\tempurl}


\bibitem[von Laszewski et~al\mbox{.}(2009)]%
        {las-09-ccgrid}
\bibfield{author}{\bibinfo{person}{Gregor von Laszewski},
  \bibinfo{person}{Andrew Younge}, \bibinfo{person}{Xi He},
  \bibinfo{person}{Kumar Mahinthakumar}, {and} \bibinfo{person}{Lizhe Wang}.}
  \bibinfo{year}{2009}\natexlab{}.
\newblock \showarticletitle{{Experiment and Workflow Management Using Cyberaide
  Shell}}. In \bibinfo{booktitle}{\emph{4th International Workshop on Workflow
  Systems in e-Science (WSES 09) in conjunction with 9th IEEE International
  Symposium on Cluster Computing and the Grid}}. \bibinfo{publisher}{IEEE},
  \bibinfo{address}{Shanghai, China}, \bibinfo{pages}{568--573}.
\newblock
\urldef\tempurl%
\url{https://doi.org/10.1109/CCGRID.2009.66}
\showDOI{\tempurl}


\bibitem[vonLaszewski et~al\mbox{.}(2012)]%
        {las-12-fedcloud-proc}
\bibfield{editor}{\bibinfo{person}{Gregor vonLaszewski},
  \bibinfo{person}{Robert Grossman}, {and} \bibinfo{person}{Michael Kozuchand
  Rick McGeerand~Dejan Milojicic}} (Eds.). \bibinfo{year}{2012}\natexlab{}.
\newblock \bibinfo{booktitle}{\emph{{FederatedClouds '12: Proceedings of the
  2012 Workshop on Cloud Services, Federation, and the 8th Open Cirrus
  Summit}}} (San Jose, California, USA). \bibinfo{publisher}{ACM},
  \bibinfo{address}{New York, NY, USA}.
\newblock
\showISBNx{978-1-4503-1754-2}
\urldef\tempurl%
\url{http://dl.acm.org/citation.cfm?id=2378975&picked=prox&cfid=389635474&cftoken=32712991}
\showURL{%
\tempurl}


\bibitem[Wagner et~al\mbox{.}(2016)]%
        {las-16-virtcluster}
\bibfield{author}{\bibinfo{person}{Rick Wagner}, \bibinfo{person}{Philip
  Papadopoulos}, \bibinfo{person}{Dmitry Mishin}, \bibinfo{person}{Trevor
  Cooper}, \bibinfo{person}{Mahidhar Tatineti}, \bibinfo{person}{Gregor von
  Laszewski}, \bibinfo{person}{Fugang Wang}, {and} \bibinfo{person}{Geoffrey~C.
  Fox}.} \bibinfo{year}{2016}\natexlab{}.
\newblock \showarticletitle{User Managed Virtual Clusters in Comet}. In
  \bibinfo{booktitle}{\emph{Proceedings of the XSEDE16 Conference on Diversity,
  Big Data, and Science at Scale}}. \bibinfo{publisher}{ACM, New York, NY},
  \bibinfo{address}{Miami, USA}, \bibinfo{pages}{24:1--24:8}.
\newblock
\showISBNx{978-1-4503-4755-6}
\urldef\tempurl%
\url{https://doi.org/10.1145/2949550.2949555}
\showDOI{\tempurl}


\bibitem[Wilkinson et~al\mbox{.}(2016)]%
        {wilkinson2016fair}
\bibfield{author}{\bibinfo{person}{Mark~D. Wilkinson}, \bibinfo{person}{Michel
  Dumontier}, \bibinfo{person}{IJsbrand~Jan Aalbersberg},
  \bibinfo{person}{Gabrielle Appleton}, \bibinfo{person}{Myles Axton},
  \bibinfo{person}{Arie Baak}, \bibinfo{person}{Niklas Blomberg},
  \bibinfo{person}{Jan-Willem Boiten}, \bibinfo{person}{Luiz~Bonino da
  Silva~Santos}, \bibinfo{person}{Philip~E. Bourne}, \bibinfo{person}{Jildau
  Bouwman}, \bibinfo{person}{Anthony~J. Brookes}, \bibinfo{person}{Tim Clark},
  \bibinfo{person}{Merc{\`e} Crosas}, \bibinfo{person}{Ingrid Dillo},
  \bibinfo{person}{Olivier Dumon}, \bibinfo{person}{Scott Edmunds},
  \bibinfo{person}{Chris~T. Evelo}, \bibinfo{person}{Richard Finkers},
  \bibinfo{person}{Alejandra Gonzalez-Beltran}, \bibinfo{person}{Alasdair J.~G.
  Gray}, \bibinfo{person}{Paul Groth}, \bibinfo{person}{Carole Goble},
  \bibinfo{person}{Jeffrey~S. Grethe}, \bibinfo{person}{Jaap Heringa},
  \bibinfo{person}{Peter A.~C 't Hoen}, \bibinfo{person}{Rob Hooft},
  \bibinfo{person}{Tobias Kuhn}, \bibinfo{person}{Ruben Kok},
  \bibinfo{person}{Joost Kok}, \bibinfo{person}{Scott~J. Lusher},
  \bibinfo{person}{Maryann~E. Martone}, \bibinfo{person}{Albert Mons},
  \bibinfo{person}{Abel~L. Packer}, \bibinfo{person}{Bengt Persson},
  \bibinfo{person}{Philippe Rocca-Serra}, \bibinfo{person}{Marco Roos},
  \bibinfo{person}{Rene van Schaik}, \bibinfo{person}{Susanna-Assunta Sansone},
  \bibinfo{person}{Erik Schultes}, \bibinfo{person}{Thierry Sengstag},
  \bibinfo{person}{Ted Slater}, \bibinfo{person}{George Strawn},
  \bibinfo{person}{Morris~A. Swertz}, \bibinfo{person}{Mark Thompson},
  \bibinfo{person}{Johan van~der Lei}, \bibinfo{person}{Erik van Mulligen},
  \bibinfo{person}{Jan Velterop}, \bibinfo{person}{Andra Waagmeester},
  \bibinfo{person}{Peter Wittenburg}, \bibinfo{person}{Katherine Wolstencroft},
  \bibinfo{person}{Jun Zhao}, {and} \bibinfo{person}{Barend Mons}.}
  \bibinfo{year}{2016}\natexlab{}.
\newblock \showarticletitle{{The FAIR Guiding Principles for scientific data
  management and stewardship}}.
\newblock \bibinfo{journal}{\emph{Scientific Data}} \bibinfo{volume}{3},
  \bibinfo{number}{1} (\bibinfo{year}{2016}), \bibinfo{pages}{160018}.
\newblock
\showISBNx{2052-4463}
\urldef\tempurl%
\url{https://doi.org/10.1038/sdata.2016.18}
\showDOI{\tempurl}


\bibitem[Wilkinson et~al\mbox{.}(2022b)]%
        {caw2021-report}
\bibfield{author}{\bibinfo{person}{Sean Wilkinson}, \bibinfo{person}{Kathryn
  Knight}, \bibinfo{person}{Olga Kuchar}, \bibinfo{person}{Kshitij Mehta},
  \bibinfo{person}{Mallikarjun Shankar}, {and} \bibinfo{person}{Matthew Wolf}.}
  \bibinfo{year}{2022}\natexlab{b}.
\newblock \bibinfo{booktitle}{\emph{Official Report on the 2021 Computational
  and Autonomous Workflows Workshop (CAW 2021)}}.
\newblock \bibinfo{type}{{T}echnical {R}eport}. \bibinfo{institution}{Oak Ridge
  National Laboratory (ORNL), Oak Ridge, TN (United States)}.
\newblock
\urldef\tempurl%
\url{https://doi.org/10.2172/1862119}
\showDOI{\tempurl}


\bibitem[Wilkinson et~al\mbox{.}(2025)]%
        {wilkinson2025}
\bibfield{author}{\bibinfo{person}{Sean~R. Wilkinson}, \bibinfo{person}{Meznah
  Aloqalaa}, \bibinfo{person}{Khalid Belhajjame}, \bibinfo{person}{Michael~R.
  Crusoe}, \bibinfo{person}{Bruno de Paula~Kinoshita}, \bibinfo{person}{Luiz
  Gadelha}, \bibinfo{person}{Daniel Garijo}, \bibinfo{person}{Ove Johan~Ragnar
  Gustafsson}, \bibinfo{person}{Nick Juty}, \bibinfo{person}{Sehrish Kanwal},
  \bibinfo{person}{Farah~Zaib Khan}, \bibinfo{person}{Johannes K{\"o}ster},
  \bibinfo{person}{Karsten Peters-von Gehlen}, \bibinfo{person}{Line Pouchard},
  \bibinfo{person}{Randy~K. Rannow}, \bibinfo{person}{Stian Soiland-Reyes},
  \bibinfo{person}{Nicola Soranzo}, \bibinfo{person}{Shoaib Sufi},
  \bibinfo{person}{Ziheng Sun}, \bibinfo{person}{Baiba Vilne},
  \bibinfo{person}{Merridee~A. Wouters}, \bibinfo{person}{Denis Yuen}, {and}
  \bibinfo{person}{Carole Goble}.} \bibinfo{year}{2025}\natexlab{}.
\newblock \showarticletitle{Applying the {FAIR} Principles to computational
  workflows}.
\newblock \bibinfo{journal}{\emph{Scientific Data}} \bibinfo{volume}{12},
  \bibinfo{number}{1} (\bibinfo{year}{2025}), \bibinfo{pages}{328}.
\newblock
\showISBNx{2052-4463}
\urldef\tempurl%
\url{https://doi.org/10.1038/s41597-025-04451-9}
\showDOI{\tempurl}


\bibitem[Wilkinson et~al\mbox{.}(2022a)]%
        {wilkinson2022}
\bibfield{author}{\bibinfo{person}{Sean~R. Wilkinson}, \bibinfo{person}{Greg
  Eisenhauer}, \bibinfo{person}{Anuj~J. Kapadia}, \bibinfo{person}{Kathryn
  Knight}, \bibinfo{person}{Jeremy Logan}, \bibinfo{person}{Patrick Widener},
  {and} \bibinfo{person}{Matthew Wolf}.} \bibinfo{year}{2022}\natexlab{a}.
\newblock \showarticletitle{F*** workflows: when parts of {FAIR} are missing}.
  In \bibinfo{booktitle}{\emph{2022 {IEEE} 18th {International} {Conference} on
  e-{Science} (e-{Science})}}. \bibinfo{publisher}{IEEE},
  \bibinfo{address}{Salt Lake City, UT, USA}, \bibinfo{pages}{507--512}.
\newblock
\showISBNx{9781665461245}
\urldef\tempurl%
\url{https://doi.org/10.1109/eScience55777.2022.00090}
\showDOI{\tempurl}


\bibitem[Wilkinson et~al\mbox{.}(2022c)]%
        {wilkinson2022-2}
\bibfield{author}{\bibinfo{person}{Sean~R. Wilkinson}, \bibinfo{person}{Ketan
  Maheshwari}, {and} \bibinfo{person}{Rafael Ferreira~da Silva}.}
  \bibinfo{year}{2022}\natexlab{c}.
\newblock \showarticletitle{Unveiling User Behavior on {Summit} Login Nodes
  as a User}. In \bibinfo{booktitle}{\emph{Computational Science -- ICCS
  2022}}, \bibfield{editor}{\bibinfo{person}{Derek Groen},
  \bibinfo{person}{Cl{\'e}lia de~Mulatier}, \bibinfo{person}{Maciej Paszynski},
  \bibinfo{person}{Valeria~V. Krzhizhanovskaya}, \bibinfo{person}{Jack~J.
  Dongarra}, {and} \bibinfo{person}{Peter M.~A. Sloot}} (Eds.).
  \bibinfo{publisher}{Springer International Publishing},
  \bibinfo{address}{Cham}, \bibinfo{pages}{516--529}.
\newblock
\showISBNx{978-3-031-08751-6}
\urldef\tempurl%
\url{https://doi.org/10.1007/978-3-031-08751-6_37}
\showDOI{\tempurl}


\bibitem[Wilson et~al\mbox{.}(2021)]%
        {wilson2021}
\bibfield{author}{\bibinfo{person}{Eric Wilson}, \bibinfo{person}{John Vant},
  \bibinfo{person}{Jacob Layton}, \bibinfo{person}{Ryan Boyd},
  \bibinfo{person}{Hyungro Lee}, \bibinfo{person}{Matteo Turilli},
  \bibinfo{person}{Benjam{\'i}n Hern{\'a}ndez}, \bibinfo{person}{Sean
  Wilkinson}, \bibinfo{person}{Shantenu Jha}, \bibinfo{person}{Chitrak Gupta},
  \bibinfo{person}{Daipayan Sarkar}, {and} \bibinfo{person}{Abhishek
  Singharoy}.} \bibinfo{year}{2021}\natexlab{}.
\newblock \bibinfo{booktitle}{\emph{Large-Scale Molecular Dynamics Simulations
  of Cellular Compartments}}.
\newblock \bibinfo{publisher}{Springer US}, \bibinfo{address}{New York, NY},
  \bibinfo{pages}{335--356}.
\newblock
\showISBNx{978-1-0716-1394-8}
\urldef\tempurl%
\url{https://doi.org/10.1007/978-1-0716-1394-8_18}
\showDOI{\tempurl}


\bibitem[Wilson et~al\mbox{.}(2014)]%
        {software-carpentry}
\bibfield{author}{\bibinfo{person}{Greg Wilson}, \bibinfo{person}{D.~A.
  Aruliah}, \bibinfo{person}{C.~Titus Brown}, \bibinfo{person}{Neil P.~Chue
  Hong}, \bibinfo{person}{Matt Davis}, \bibinfo{person}{Richard~T. Guy},
  \bibinfo{person}{Steven H.~D. Haddock}, \bibinfo{person}{Kathryn~D. Huff},
  \bibinfo{person}{Ian~M. Mitchell}, \bibinfo{person}{Mark~D. Plumbley},
  \bibinfo{person}{Ben Waugh}, \bibinfo{person}{Ethan~P. White}, {and}
  \bibinfo{person}{Paul Wilson}.} \bibinfo{year}{2014}\natexlab{}.
\newblock \showarticletitle{Best Practices for Scientific Computing}.
\newblock \bibinfo{journal}{\emph{{PLoS} Biology}} \bibinfo{volume}{12},
  \bibinfo{number}{1} (\bibinfo{date}{jan} \bibinfo{year}{2014}),
  \bibinfo{pages}{e1001745}.
\newblock
\urldef\tempurl%
\url{https://doi.org/10.1371/journal.pbio.1001745}
\showDOI{\tempurl}


\bibitem[Workflow Language(2025)]%
        {workflow-list}
Workflow Language \bibinfo{year}{2025}\natexlab{}.
\newblock \bibinfo{title}{Existing Workflow systems}.
\newblock
\newblock
\urldef\tempurl%
\url{https://github.com/common-workflow-language/common-workflow-language/wiki/Existing-Workflow-systems}
\showURL{%
\tempurl}
\newblock
\shownote{[Online; accessed 03/01/2025]}.


\bibitem[Xu et~al\mbox{.}(2024)]%
        {xu2024surveyresourceefficientllmmultimodal}
\bibfield{author}{\bibinfo{person}{Mengwei Xu}, \bibinfo{person}{Wangsong Yin},
  \bibinfo{person}{Dongqi Cai}, \bibinfo{person}{Rongjie Yi},
  \bibinfo{person}{Daliang Xu}, \bibinfo{person}{Qipeng Wang},
  \bibinfo{person}{Bingyang Wu}, \bibinfo{person}{Yihao Zhao},
  \bibinfo{person}{Chen Yang}, \bibinfo{person}{Shihe Wang},
  \bibinfo{person}{Qiyang Zhang}, \bibinfo{person}{Zhenyan Lu},
  \bibinfo{person}{Li Zhang}, \bibinfo{person}{Shangguang Wang},
  \bibinfo{person}{Yuanchun Li}, \bibinfo{person}{Yunxin Liu},
  \bibinfo{person}{Xin Jin}, {and} \bibinfo{person}{Xuanzhe Liu}.}
  \bibinfo{year}{2024}\natexlab{}.
\newblock \showarticletitle{A Survey of Resource-efficient LLM and Multimodal
  Foundation Models}.
\newblock \bibinfo{journal}{\emph{arXiv}} (\bibinfo{year}{2024}).
\newblock
\urldef\tempurl%
\url{https://doi.org/10.48550/arXiv.2401.08092}
\showDOI{\tempurl}
\showeprint{2401.08092}
\newblock
\shownote{cs.LG}.


\bibitem[Zhao et~al\mbox{.}(2007)]%
        {las--7-swift}
\bibfield{author}{\bibinfo{person}{Yong Zhao}, \bibinfo{person}{M. Hategan},
  \bibinfo{person}{B. Clifford}, \bibinfo{person}{I. Foster},
  \bibinfo{person}{G. von Laszewski}, \bibinfo{person}{V. Nefedova},
  \bibinfo{person}{I. Raicu}, \bibinfo{person}{T. Stef-Praun}, {and}
  \bibinfo{person}{M. Wilde}.} \bibinfo{year}{2007}\natexlab{}.
\newblock \showarticletitle{{Swift: Fast, Reliable, Loosely Coupled Parallel
  Computation}}. In \bibinfo{booktitle}{\emph{Services, 2007 IEEE Congress
  on}}. \bibinfo{pages}{199--206}.
\newblock
\urldef\tempurl%
\url{https://doi.org/10.1109/SERVICES.2007.63}
\showDOI{\tempurl}


\end{thebibliography}

\end{document}